\documentclass[structabstract]{aa}  
\usepackage{txfonts}
\usepackage{graphicx,longtable,lscape,stfloats}
\usepackage{natbib}
\bibpunct{(}{)}{;}{a}{}{,} 

\newcommand{\ltsima} {$\; \buildrel < \over \sim \;$}  
\newcommand{\gtsima} {$\; \buildrel > \over \sim \;$}  
\newcommand{\lta} {\lower.5ex\hbox{\ltsima}}  
\newcommand{\gta} {\lower.5ex\hbox{\gtsima}}

\usepackage{color}
\usepackage[normalem]{ulem}
\usepackage{amstext}

\begin{document}

\title{Evolution of active region outflows throughout an active region 
  lifetime}
\subtitle{} \titlerunning{AR outflows}
\authorrunning{Zangrilli and Poletto}

\author{L.~Zangrilli\inst{1}
\and G.~Poletto\inst{2}}

\institute {Istituto Nazionale di Astrofisica (INAF) - Osservatorio 
Astrofisico di Torino, via Osservatorio 20, 10025 Pino Torinese, Italy
\and Istituto Nazionale di Astrofisica (INAF) - Osservatorio Astrofisico di 
Arcetri, largo Enrico Fermi 5, 50125 Firenze, Italy}

\offprints{luca.zangrilli@inaf.it}

 
\abstract
    {We have shown previously that SOHO/UVCS data allow us to detect active 
      region (AR) outflows at coronal altitudes higher than those reached by 
      other instrumentation. These outflows are thought to be a component of 
      the slow solar wind.}
    {Our purpose is to study the evolution of the outflows in the intermediate 
      corona from AR~8100, from the time the AR first forms until 
      it dissolves, after several transits at the solar limb.}
    {Data acquired by SOHO/UVCS at the time of the AR limb transits, at medium 
      latitudes and at altitudes ranging from 1.5 to $2.3~{\rm R_{\odot}}$, were 
      used to infer the physical properties of the outflows through the AR 
      evolution. To this end, we applied the Doppler dimming technique to UVCS 
      spectra. These spectra include the H~{\sc{i}}~Ly$\alpha$ line and the 
      O~{\sc vi} doublet lines at 1031.9 and 1037.6~\AA.}
    {Plasma speeds and electron densities of the outflows were inferred over 
      several rotations of the Sun. AR outflows are present in the newly born 
      AR and persist throughout the entire AR life. Moreover, we found two 
      types of outflows at different latitudes, both possibly originating in 
      the same negative polarity area of the AR. We also analyzed the behavior 
      of the Si~{\sc{xii}} 520~\AA\ line along the UVCS slit in an attempt to 
      reveal changes in the Si abundance when different regions are traversed. 
      Although we found some evidence for a Si enrichment in the AR outflows, 
      alternative interpretations are also plausible.}
    {Our results demonstrate that outflows from ARs are detectable in the 
      intermediate corona throughout the whole AR lifetime. This confirms that 
      outflows contribute to the slow wind.}

\maketitle

\section{Introduction}
\label{sec:intro}

The recent observations of outflows originating from the edges of active 
regions (ARs) contributed to identify one of the possible sources of the slow 
solar wind, whose origin has not yet been unambiguously identified. However, 
although they have been observed in many circumstances, the characteristics of 
upflows are not well known, nor is it established whether all reach the 
interplanetary space, or if there are motions that rise along one leg and 
descend along the other leg of a loop.

By using interplanetary scintillation analysis, \cite{Kojima1999a} found slow 
wind flows during the minimum of solar activity. These were associated with 
open magnetic field lines that apparently originated in one of the polarities 
of an AR. Observations with the TRACE satellite have sporadically shown 
outflows from ARs (\citealt{Winebarger2001a}). The so-called AR sources have 
been identified as slow wind contributors from the study of Ulysses and ACE in 
situ observations (see, e.g., \citealt{Neugebauer2002a}; 
\citealt{Liewer2004a}). According to the analysis made by 
\citealt{Neugebauer2002a}, the AR outflows can be comprised of different 
substreams, originating from multiple sites in the AR, and the distribution of 
open magnetic field lines can change with time.

More recently, continuous outflows from the edges of ARs have been identified 
through Hinode/XRT (\citealt{Sakao2007a}) and Hinode/EIS observations 
(\citealt{Harra2008a}, \citealt{Doschek2008a}). All these observations refer 
to low coronal levels. Analyses of Hinode/EIS line asymmetries 
(\citealt{Tian2011b}), combined with SDO/AIA space-time plots, suggested the 
occurrence of two velocity components: a background primary component 
(possibly blueshifted by $\approx 10~{\rm km~s^{-1}}$) and a secondary high 
speed ($\approx 100~{\rm km~s^{-1}}$) component, the latter only possibly 
reaching high coronal levels (\citealt{Tian2011b}). Furthermore, it is not 
clear whether the flows are continuous or quasi-periodic 
(\citealt{De_Pontieu2010a}, \citealt{Tian2011a}).

Several mechanisms have been proposed as drivers of the AR outflows, most of 
them involving magnetic reconnection (see, e.g., \citealt{Harra2008a}, 
\citealt{Marsch2008a}, \citealt{Del_Zanna2008a}). It has been pointed out by 
\cite{Baker2009a} that the strongest AR outflows originate from areas in the 
proximity of magnetic field configurations that are known as quasi-separatrix 
layers (QSLs; see, e.g., \citealt{Demoulin1996a}). These authors argued 
that the magnetic reconnection process, which operates in QSLs and separates 
AR loops from open magnetic field lines or larger-scale loops, is a possible 
mechanism underlying the AR outflows. A different mechanism for outflows, 
based on AR expansion, has been suggested by \cite{Murray2010a}.

It can be determined whether the outflows detected in the low corona reach the 
interplanetary medium by comparing the properties of plasma originating at the 
periphery of ARs and those of the in situ wind plasma, which are assumed to 
originate from that source. This approach avoids the problem of measuring 
plasma flows high up in the corona. The method of comparing the low corona 
plasma properties with the wind parameters measured in situ has been 
adopted by \cite{Ko2006a}, for example, who used SOHO/UVCS data to compare 
abundances inferred in the corona at the AR edges with abundances measured in 
the wind from ACE/SWICS. Analogously, \cite{Liewer2004a} analyzed ACE and 
Ulysses data and concluded that the ${\rm O}^{7+}/{\rm O}^{6+}$ ratio (a proxy 
for the electron temperature of the region where freezing-in occurs) is higher 
and more variable in the solar wind streams that are thought of as originating 
from ARs than the ratio detected from coronal hole (CH) sources.

An indirect evidence that low corona outflows extend into inteplanetary space 
has been proposed by \cite{Del_Zanna2011a} and \cite{Bradshaw2011a}, who 
compared Hinode/EIS observations with the radio noise storms mapped by the 
Nancay radioheliograph. In the scenario proposed by these authors, the 
observed noise storms originate in low-energy electron beams propagating into 
the high corona, where they are accelerated by interchange reconnection 
between closed and open loops in the low corona.

A multi-instrument approach in the study of AR outflows has been adopted by 
\cite{Vanninathan2015a}, who excluded the role of chromospheric jets as 
possible drivers and instead suggested that the magnetic field diffusion might 
be responsible for the upflows. Moreover, these authors obtained evidence of 
two outflow regimes from observing two velocity components that they ascribed 
to magnetic reconnection and a pressure-driven mechanism.

It is well established that the element abundances with a low first ionization 
potential (FIP), that is, lower than 10~eV, in the slow solar wind are 
significantly higher than those of the photosphere (see, e.g., 
\citealt{von_Steiger2000a}). The degree of enhancement of a particular element 
is commonly evaluated by the FIP bias factor, which is the ratio of the 
abundance of this element relative to that of a high-FIP (i.e., with a first 
ionization potential higher than 10~eV) reference element measured in the 
corona, with the same relative abundance as measured in the photosphere. By 
comparing the abundances determined by Hinode/EIS and SWICS, 
\cite{Brooks2011a} made a direct connection between the plasma flowing out 
from the edge of an AR with the slow wind detected in situ. These authors 
found that this AR outflow is characterized by values of the FIP bias of 
$\sim$~3-4, which distinguishes it from the bias derived from observations of 
the polar CH. These show values of $\sim$~1.0-1.1, as expected for 
the fast wind (see also \citealt{van_Driel-Gesztelyi2012a}, 
\citealt{Wang2012b}).

A recent full-Sun study, combining the analysis of coronal plasma composition, 
Doppler outflow speed mesurements, large-scale coronal magnetic field 
reconstruction, and in situ measurements of the solar wind at 1~AU, has been 
presented by \cite{Brooks2015a}. By analyzing Hinode/EIS and ACE/SWEPAM data, 
they produced a solar wind source map that confirmed the role of ARs as slow 
wind sources. The authors concluded that ARs appear to be the primary 
contributors to the slow wind flows.

Plasma flows at the intermediate corona levels, in the range from 1.5 to 
$2.5~{\rm R_{\odot}}$, have recently been detected by \cite{Zangrilli2012a}, who 
analyzed SOHO/UVCS coronal observations at the time when an AR transits at the 
solar limb. Applying the Doppler dimming (DD) technique to data that covered 
the 
NOAA~8124 AR complex on January 2, 1998, these authors showed that outflows 
were present in a narrow channel at the edge of a closed loop system, where, 
according to magnetic potential field extrapolations, bundles of open field 
lines connected the AR to the interplanetary space.

The temporal evolution of AR upflows has so far been studied in the relatively 
short time interval of a few days. \cite{Demoulin2013a} analyzed upflows from 
the edge of AR~10978 during its transit over the solar disk. They found that 
the upflows are characterized by a collimated and stationary component, and 
that their limb-to-limb evolution is dominated by projection effects onto the 
line of sight (LOS). Their results are also compatible with an origin of the 
upflows from reconnection along quasi-separatrix layers. \cite{Culhane2014a} 
followed the evolution of the plasma upflows from ARs over the days of its 
central meridian passage. The authors found a short timescale variability of 
the upflows and a strong stationary component and confirmed that AR upflows 
can contribute to the slow solar wind.

In the present work we address the neglected question of how plasma outflows 
evolve in the intermediate corona on a longer timescale by analyzing 
successive limb transits of an AR in a time interval of several months, from 
the time of its birth to the time of its death. We do not have to cope with 
projection effects on the solar disk because we only use limb observations and 
focus on possible changes that occurr in the time intervals from one limb 
transit to the next. Our purpose is to establish whether the deduced outflow 
speeds and electron densities are a function of the AR age or if they occur at 
typical speed and density levels, independent of the AR evolution. We also 
examine the location from which plasma originates to determine whether it 
changes with time or persists throughout the AR lifetime. A change in the AR 
outflow sources pattern might depend on a variation in the AR polarities 
orientation with time and on the AR magnetic field dispersal with the AR age, 
the latter perhaps leading to a unipolar field or to several low-lying closed 
loops with no escaping plasma. The evolution of the outflow properties might 
also depend on the activity level of the AR, as the occurrence of many flares 
or coronal mass ejections (CME) might indicate the likelihood of magnetic 
field lines opening. Finally, we analyze the brightness of the 520~\AA\ 
second-order line of the Si~{\sc{xii}} ion, which is detected in the 
O~{\sc{vi}} channel spectra, and discuss the origin of its changes along the 
UVCS slit.

The paper is organized as follows: in Sect.~\ref{sec:data} we describe the 
data, in Sect.~\ref{sec:analysis} we summarize the procedure by which we 
inferred the physical parameters of the AR and, in particular, those of the 
outflowing plasma. Section~\ref{sec:results} describes the results, which are 
discussed in Sect.~\ref{sec:discussion}, where a summary of the main results 
and the conclusions are also given.

\section{Data}
\label{sec:data}

\subsection{NOAA~8100 active region}

We here study NOAA~AR~8100, which is an interesting and fairly active region 
that has been the subject of several papers (see, e.g., 
\citealt{Yan2001a}, \citealt{Green2002a}, \citealt{Mandrini2004a}), which 
mainly addressed questions on its topology and the energy and helicity budget. 
Most of the literature refers to November 8, 1997, as the first limb passage of 
this region, when it was born as AR~8100 a few days before; however, the 
region had already been seen at the limb of the Sun on October 26-27 (Solar 
Geophysical Data n.~644 lists a flare observed at the Meudon Observatory from 
AR~8100, at S20, E75, on October 27, 1997) and made its last limb transit 
around February 28, 1998. In the following we refer to this active region as 
AR~8100, throughout its limb passages, although it was assigned different 
numbers at successive transits: it is called AR 8112, AR 8124, and AR 8142 in 
the second to fourth rotations, after which no number was assigned to the AR. 
The estimated number of flares and CMEs originating from the region is 
possibly higher than 50. The AR magnetic field exhibits a peculiar evolution 
as its polarities continued to rotate throughout the AR lifetime, changing 
from an $\approx {\rm E/W}$ to an $\approx {\rm N/S}$ orientation (see Fig.~6 
in \citealt{Mandrini2004a}). This behavior is very convenient for our 
analysis, allowing us to see loops both edge-on and face-on, respectively, 
thus facilitating our understanding of the 3D structure of the region.

\subsection{SOHO/UVCS data}
\label{subsec:sohouvcsdata}

SOHO/UVCS (see, e.g., \citealt{Domingo1995a}, \citealt{Kohl1995a}, 
\citealt{Kohl1997a}, \citealt{Kohl2006a}) routinely made synoptic observations 
of the Sun, acquiring spectra at eight different polar angles (PAs) every 45 
degrees, starting from the north pole. At every PA, the UVCS slit is set 
normal to the radial of the Sun, at a number of heliocentric distances that 
start at $1.5~{\rm R_\odot}$ and reach $3.5~{\rm R_\odot}$ at the equator, and 
lower 
altitudes at the other PAs, where weaker line intensities are expected. 
Typically we  used data taken at $PA=135~\deg$ and $PA=225~\deg$, 
corresponding to the latitude of $-45~\deg$ (southern hemisphere), SE and SW 
quadrant, respectively, for the east and west limb passages. The radial from 
the Sun center, which defines the PA, intercepts the spectrograph slit in an 
off-centered position. The resulting unequal distribution of the pixels along 
the spatial direction, relatively to the slit PA, implies a shorter coverage 
of the corona at low latitudes in the SE quadrant observations than in the SW. 
However, at the east limb the AR is marginally within the field of view of the 
observations taken with the slit at $PA = 90~\deg$ (and also at 135~$\deg$), 
which are made along the equatorial direction. This circumstance, and the need 
of a wider coverage of the corona at low latitudes, caused us to use both 
mid-latitudes and equatorial scans when analyzing the transits at the east 
solar limb.

SOHO/UVCS acquired data over the ten east and west limb transits of the AR, 
from the end of October 1997 to the end of February 1998. The passage at the 
west limb on January 2, 1998, has been analyzed by \cite{Zangrilli2012a}. We 
note in passing that since UVCS takes data only at the solar limb, when there 
is no direct way of knowing the magnetic field at the base of the region, any 
comparison between the UVCS structure and the magnetic field reconstructions 
(based on data collected throughout a solar rotation) is subject to some 
margin of uncertainty (see discussion in Sect.~\ref{sec:potential}).

\begin{figure}[t]
\centering
\includegraphics[width=7.5cm]{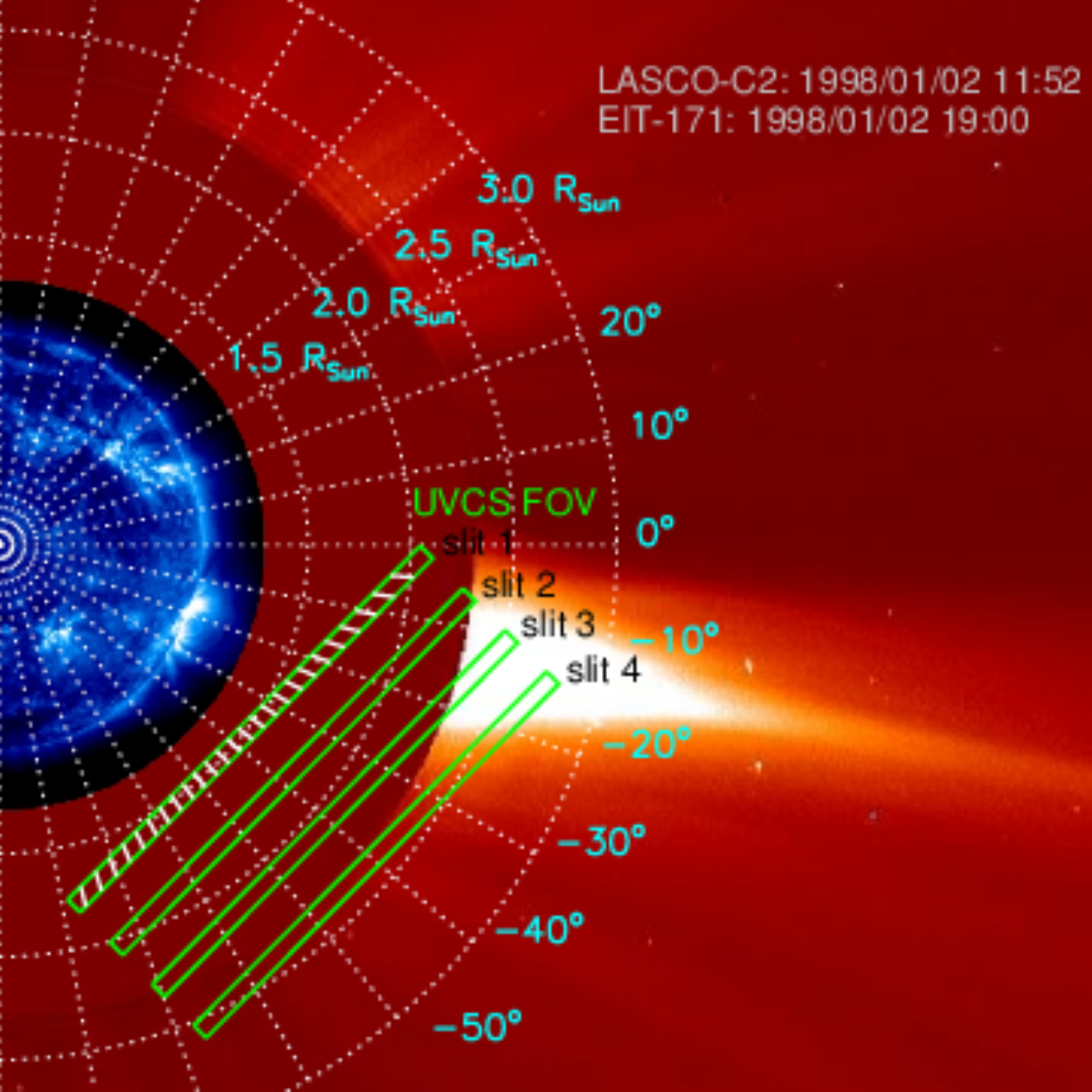}
\caption{Cartoon illustrating the configuration of the UVCS O~{\sc vi} channel 
  observations at the west limb: the UVCS field of view is shown in green as 
  the projection onto the plane of the sky of the slit at the four positions. 
  As an example of the data binning along the spatial direction, the slit at 
  $1.5~{\rm R_\odot}$ is divided into the three-degree wide angular sectors 
  within 
  which UVCS data have been summed. The slit positions at the two PAs at the 
  east limb are illustrated in the eastern panels of Fig.~\ref{fig:vmap}. As 
  context images, we show an image from EIT-171, taken on January 2, 1998 when 
  AR~8100 was at the west limb at a latitude of about $-20\deg$, and a 
  LASCO-C2 image, taken on the same day.}
\label{fig:slit}
\end{figure}

The field of view covered by UVCS ($40^\prime$ along the slit direction) is 
shown in Fig.~\ref{fig:slit}, in which the four slit positions and the data 
binning along the spatial direction are also given. The fixed edge of the UVCS 
slit, whose projections are represented in the cartoon as green rectangles, 
was set at 1.5, 1.8, 2.1 and $2.4~{\rm R_\odot}$. In the following, we refer to 
these slit positions as positions 1, 2, 3 and 4. Data are acquired on two 
consecutive days, when synoptic data are available (with the only exception of 
the sixth transit on January 2, 1998, owing to the lack of synoptic data over 
two consecutive days), and have been summed up over a three-degree wide 
sun-centered angular sector to increase the signal-to-noise ratio. The 
synoptic observations typically last only about 45 minutes over all four slit 
positions. This time is often not long enough to guarantee a good statistics, 
particularly at positions 3 and 4, and at relatively high latitudes. Hence, we 
summed data over two consecutive synoptics after ensuring, by inspecting 
SOHO/LASCO images (\citealt{Brueckner1995a}), that the spatial coronal 
structures along the slit were 
stable enough to allow the data combination. \cite{Zangrilli2012a} summed the 
spatial data over a number of pixels that was the same along the slit for all 
the four positions, spanning a decreasing latitude interval with distance. In 
the present study, the choice of a constant angular width of the bins at all 
slit positions allows partially compensating for the decreasing count rate in 
spectra with heliocentric distance and matching the observations at medium 
latitudes with those taken along the equatorial direction, which we used here.

UVCS data have been taken in the H~{\sc{i}}~Ly$\alpha$ and O~{\sc{vi}} 
channels, which have different characteristics: in the radial direction, the 
O~{\sc{vi}} channel slit width is $0.3~{\rm mm}$ , while the 
H~{\sc{i}}~Ly$\alpha$ channel slit width is $50~{\rm \mu m}$, thus covering 
different intervals of heliocentric distances, corresponding to 
$83^{\prime\prime}$ and $14^{\prime\prime}$ in the plane of the sky, respectively. 
Data have originally been acquired with a spatial resolution of 
$21^{\prime\prime}$ per bin in the plane of the sky, which has then been 
degraded by averaging data over three-degree wide sun-centered latitude 
sectors, as we discussed above.

Spectra acquired in the O~{\sc{vi}} channel include the doublet O~{\sc{vi}} 
lines at 1031.9 and 1037.6 \AA, the H~{\sc{i}}~Ly$\beta$ line at 1025.7 \AA, 
and the weaker line of the Si~{\sc{xii}} doublet at 520.7 \AA\ (in the second 
order). The spectral resolution of O~{\sc{vi}} data is $0.594$~\AA\ per bin. 
In the H~{\sc{i}}~Ly$\alpha$ channel the spectral resolution is $0.14$~\AA\ 
per bin. We point out that data are taken with exposure times that increase 
with the slit heliocentric altitude in an attempt to make counts at the 
largest distances accumulate long enough to become statistically significant. 
However, as the distance from the Sun increases along the slit as well (it is 
higher for pixels at the slit ends), there are cases when the exposure time is 
not long enough to provide significant data all along the slit, hence these 
positions were not considered in the study.

Data were analyzed using the UVCS - Data Analysis Software, which takes care 
of wavelength and radiometric calibration. Stray-light effects have been 
corrected through the semi-empirical relationship given by 
\cite{Cranmer2010a}. Line intensities were calculated after fitting the 
observed line profiles with a Gaussian and then integrating over wavelength.

\subsection{SOHO/EIT context images}

As context data, we used wavelet-enhanced images from SOHO/EIT 
(\citealt{Delaboudiniere1995a}) to infer 
the loop configurations in the low corona at the times when the AR crosses the 
limb of the Sun. EIT images were acquired in the 195 and 171~\AA\ wavebands, 
and, as far as possible, at the same dates as the SOHO/UVCS data we analyzed. 
The occurrence of weaker emission regions in EIT images might help us to 
identify the source regions of outflows, if any. Long-lived upflows have been 
recognized in Hinode/EIS analyses to correspond to low-density and 
low-electron temperature regions (e.g., \citealt{Doschek2008a}, 
\citealt{Del_Zanna2008a}, \citealt{Hara2008a}, \citealt{Demoulin2013a}). In 
particular, \cite{Doschek2008a} found evidence that low-intensity regions in 
ARs are significantly Doppler-shifted and are associated with open magnetic 
field lines that extend into the heliosphere; the estimated electron densities 
and temperatures in the outflows are considered rather low for an active 
region. We therefore expect to find the low-corona counterparts of the 
outflows we detect in the intermediate corona as weaker emission regions in 
EIT images.

At higher levels, we used potential field extrapolations that we obtained with 
the potential field source surface (PFSS) routine in the solar software (SSW) 
package (see, e.g., \citealt{Schrijver2001a}, \citealt{Schrijver2003a}). We 
used both EIT images and PFSS extrapolations as guidelines to infer 
where the magnetic field may be open and compared this with the outcome from 
UVCS data analysis. We point out that UVCS data are taken at altitudes high 
enough to be close to the source surface height (at least for the two highest 
positions of the slit), hence we expect the large-scale magnetic field there 
to be reasonably close to a potential field.

\section{Data analysis}
\label{sec:analysis}
\subsection{Doppler dimming analysis}
\label{sec:dimm}

The efficiency of the chromospheric radiation scattering process by a coronal 
ion is reduced when the plasma is moving, relative to the case of a static 
plasma, because the central wavelength of the scattering profile is Doppler 
shifted with respect to that of the line emitted from the solar disk. This is 
known as Doppler dimming (DD) effect, and its application as a diagnostic tool 
to measure the radial components of the outflow speeds in the corona has been 
extensively studied in the literature (see, e.g., \citealt{Hyder1970a}, 
\citealt{Withbroe1982a}, \citealt{Noci1987a}, \citealt{Noci1999a}).

In the following, we briefly summarize the method (based on the DD effect) by 
which we determined plasma outflow speeds and electron densities. A more 
complete description is provided in \cite{Zangrilli2002a} and 
\cite{Zangrilli2012a}. Briefly, we recall that the intensities of the 
H~{\sc{i}}~Ly$\alpha$ and O~{\sc{vi}} doublet lines depend on the plasma 
density, the outflow speed, and the element abundance; other parameters such 
as the plasma electron temperature and the ion kinetic temperatures along and 
across the magnetic field direction play a secondary role. The data we used 
are the intensities of the H~{\sc{i}}~Ly$\alpha$ line and the O~{\sc vi} 
doublet lines at 1031.9 and 1037.6~\AA, which, provided the temperatures are 
known by some other means, are expected to allow us to infer the three 
unknowns: electron density, outflow speed, and element abundance. This is 
indeed the case whenever protons and oxygen ions flow at the same speed; while 
in CHs this probably does not occur (see, e.g., 
\citealt{Cranmer2009a}), we made this assumption at intermediate latitudes, 
hence we have enough parameters to solve the problem.

The coronal UV lines form by collisional and radiative excitations (see, e.g., 
\citealt{Noci1999a}). To estimate the plasma speed, we need to know by how 
much the radiative component of the coronal line intensities has been dimmed. 
To this end, we have to calculate the intensity of the line on the basis of a 
coronal model, starting with trial values of the plasma speed and electron 
density, and integrating the calculated local emissivities along the LOS on 
the basis of the profiles of the physical parameters prescribed by the model. 
When the calculated intensity does not reproduce observations, the plasma 
speed and electron density are changed until the intensity converges toward 
the observed value, hence identifying the correct plasma outflow speed and 
density.

For the values assumed for the plasma temperatures, we adopted the electron 
temperature radial profile given by \cite{Gibson1999a} and the O~{\sc{vi}} 
kinetic temperature profile measured by \cite{Strachan2004a}, both obtained 
for a coronal streamer during a solar activity minimum. Parallel and normal 
(to the magnetic field) temperatures were assumed to have the same value. This 
is consistent with the results of \cite{Frazin2003a}, who claimed that kinetic 
temperature anisotropy in streamer regions, if any, starts above an altitude 
of 2.6~$R_\odot$. Neutral hydrogen kinetic temperatures were evaluated from our 
own data set of H~{\sc{i}}~Lyman~$\alpha$ profiles, taking into account 
corrections for the instrumental profile function and the spectrometer slit 
width.

A further parameter we did not discuss so far is the solar disk intensity, 
which affects the intensity of the coronal radiative line components. 
H~{\sc{i}}~Lyman~$\alpha$ disk intensities have been measured in the time 
interval we analyze by the SOLSTICE experiment on the UARS satellite (see 
\citealt{Woods2000a}). We point out that 
SOLSTICE measures the disk intensity from the same vantage point as Earth, 
while a coronal ion lying in the plane of the sky, directly above the AR, or 
in other words, in the most efficient geometrical scattering position, can be 
excited by a higher H~{\sc{i}}~Lyman~$\alpha$ flux originating from both the 
AR and the disk. A study of the AR contribution to the disk intensity has been 
made by \cite{Ko2002a}, who concluded that the AR contribution at the 
distances of the intermediate corona only amounts to a few percent of the 
total disk brightness. We took into account the small variation in the 
H~{\sc{i}}~Lyman~$\alpha$ disk brightness induced by the AR, adopting the 
value measured by SOLSTICE when the AR was at the central meridian, that is, 
six days before (for the west) or after (for the east) limb transit. Because 
there are no routine measurements of the O~{\sc{vi}} disk intensities, we used 
the values of 387.0 and $199.5~{(\rm erg/s/cm^2/sr)}$ for the 1031.9 and 
1037.6~\AA\ lines, respectively, derived from UVCS disk observations taken on 
December 4, 1996 (see \citealt{Zangrilli2002a}). The adopted line widths are 
taken from \cite{Warren1997a}.

\subsection{Si~{\sc{xii}} 520.7 \AA\ line analysis}
\label{sec:fip}

The Si~{\sc{xii}} doublet line at 520.7 \AA, which is detected in the 
second-order spectra of the O~{\sc{vi}} channel, offers us the opportunity of 
analyzing the behavior of a low-FIP element line intensity along the UVCS 
slit. As we mentioned in the Introduction, the low-FIP element abundances are 
enhanced in the slow wind with respect to fast wind abundances, which are 
approximately photospheric. Unfortunately, from the 520.7 \AA\ Si~{\sc xii} 
line intensity, $I({\rm Si~\textsc{xii}})$, and the electron density derived from the 
DD analysis alone, we cannot establish whether $I({\rm Si~\textsc{xii}})$ changes 
across the AR because of variations in the Si abundance or in the electron 
temperature, or as a result of a combination of these factors. We discuss our 
results in Sect.~\ref{sec:discussion}.

We know that the Si~{\sc{xii}}~520.7~\AA\ line is collisionally excited, and 
its intensity is given by (see, e.g., \citealt{Withbroe1982a}, 
\citealt{Noci1999a})

\begin{equation}
  I({\rm Si~\textsc{xii}})=\frac{h\nu_0}{4\pi}0.83A_{\rm Si}
  \int_{LOS} R_{\rm Si^{+11}}(T_{\rm e})n_{\rm e}^2C_{\rm coll}(T_{\rm e}) dx
\label{eq:sixii_1}
,\end{equation}

where the ${\rm Si^{+11}}$ ionization ratio, 
$R_{\rm Si^{+11}}=n_{\rm Si^{+11}}/n_{\rm Si}$, is dictated by the ionization balance 
assumption and is a function of the electron temperature, $T_{\rm e}$; 
$n_{\rm Si^{+11}}$ and $n_{\rm Si}$ are the ${\rm Si^{+11}}$ and silicon number 
density, respectively; $A_{\rm Si}=n_{\rm Si}/n_{\rm H}$ is the silicon abundance 
relative to hydrogen, $n_{\rm H}/n_{\rm e}=0.83$ is the hydrogen abundance 
relative to electrons, for a plasma composition with 10\% of helium, $n_e$ is 
the electron density; and $C_{\rm coll}$ is the collisional coefficient of the 
transition, which is a function of $T_{\rm e}$.

Equation~(\ref{eq:sixii_1}) can be written as

\begin{equation}
  I({\rm Si~\textsc{xii}})\propto [A_{\rm Si}] [n_{\rm e}^2] f({T_e}) L_{\rm n_e^2}
\label{eq:sixii_2}
,\end{equation}

{\noindent where} $[n_{\rm e}^2]$, the integration length scale $L_{\rm n_e^2}$ 
and $f({T_e})$ are usually evaluated around the temperature at which the line 
forms most strongly. In the following, we discard data acquired at the lowest 
slit position because the DD technique is more sensitive to the high-speed 
regime, and we discard data taken at slit positions 3 and 4 because counts in 
the line are often too few to guarantee a statistically significant result. 
Hence we use only data from slit position 2. For convenience, we write 
$I({\rm Si~\textsc{xii}})/[n_{\rm e}^2]$ below, which is proportional to 
$[A_{\rm Si}]f({T_e})$ when the integration length scale $L_{\rm n_e^2}$ is 
constant. We are interested in changes across the AR of the 
$I({\rm Si~\textsc{xii}})/[n_{\rm e}^2]$ quantity, therefore we choose to give its 
profile along the slit, in arbitrary units normalizing to its mean value over 
the altitude interval $-63$ to $-54~\deg$, avoiding uncertainties in the 
calibration of UVCS second-order spectra.

\subsection{Potential field extrapolation}
\label{sec:potential}

Inferring the coronal magnetic field topology allows us to identify the 
sources of the outflows detected by the DD analysis, and, 
possibly, to locate them at lower levels. The potential field source surface 
(PFSS) extrapolation technique (see, e.g., \citealt{Schrijver2001a}, 
\citealt{Schrijver2003a}) gives basic information on the magnetic field 
structure of the quiet corona. In particular, we used this approach to locate 
areas within the active region in which open magnetic field lines are rooted, 
assuming that the solar wind flows along open lines to eventually escape into 
interplanetary space. The PFSS package included in the SSW tree 
(\citealt{Schrijver2003a}) relies on a database of solar surface magnetic 
field models, calculated with a six-hour cadence, obtained from SOHO/MDI. 
These magnetic field models are given for any heliographic latitude and 
Carrington longitude, and allow magnetic field lines to also be extrapolated 
at the time ARs are at the solar limb, as in the case of UVCS observations.

We extrapolated magnetic field lines in a longitude interval of $\pm 10~\deg$ 
across the plane of the sky in a latitude interval spanning from the equator 
to the south pole, thus covering the whole extent of AR~8100. The lines were 
traced on a grid uniform in latitude and longitude, without setting any 
treshold for the magnetic flux at the solar surface. In the following, we 
roughly assume that open lines originating from latitudes higher than 
$-60~\deg$ belong to the polar CH region.

\begin{figure*}
  \centering
  \includegraphics[width=4.4cm]{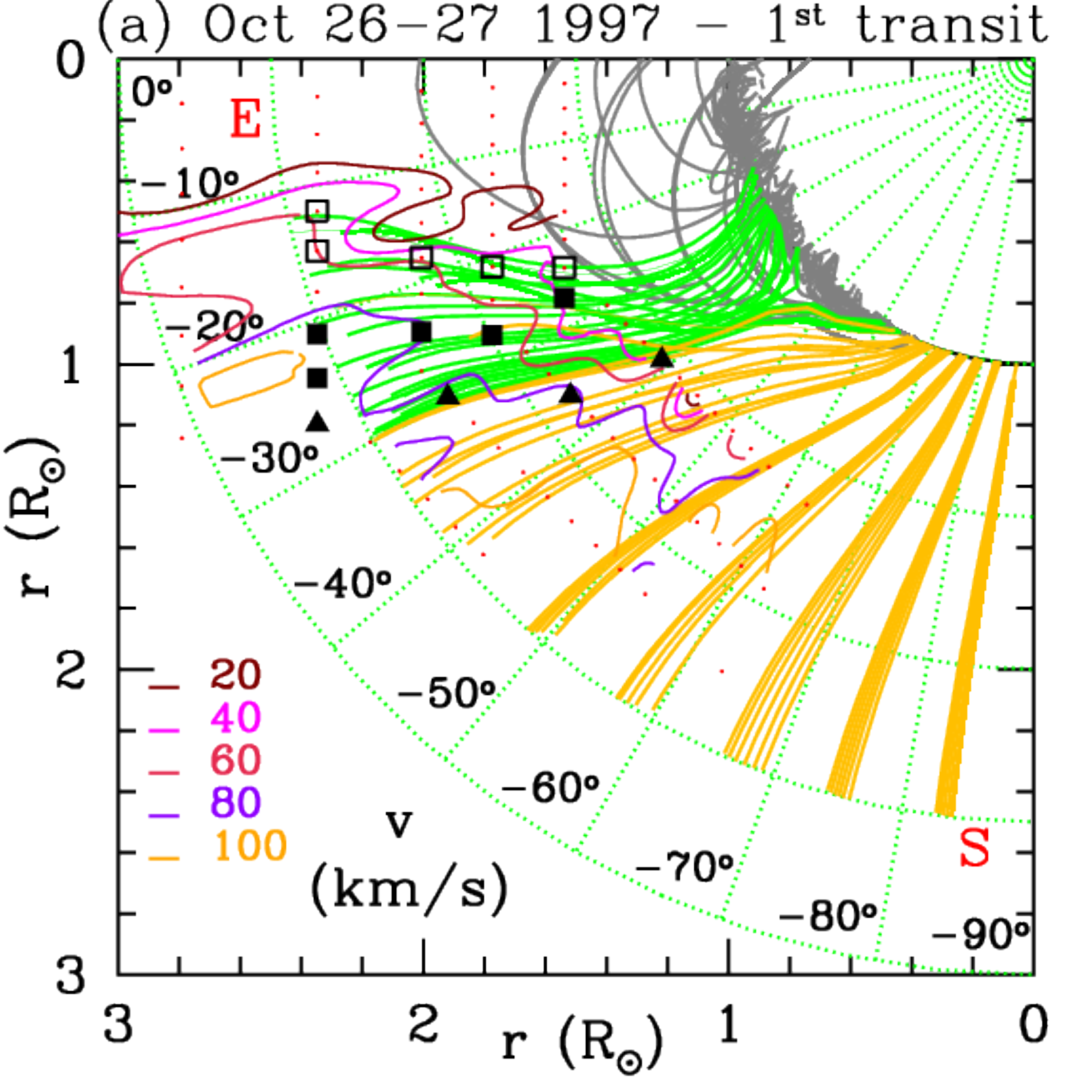}
  \includegraphics[width=4.4cm]{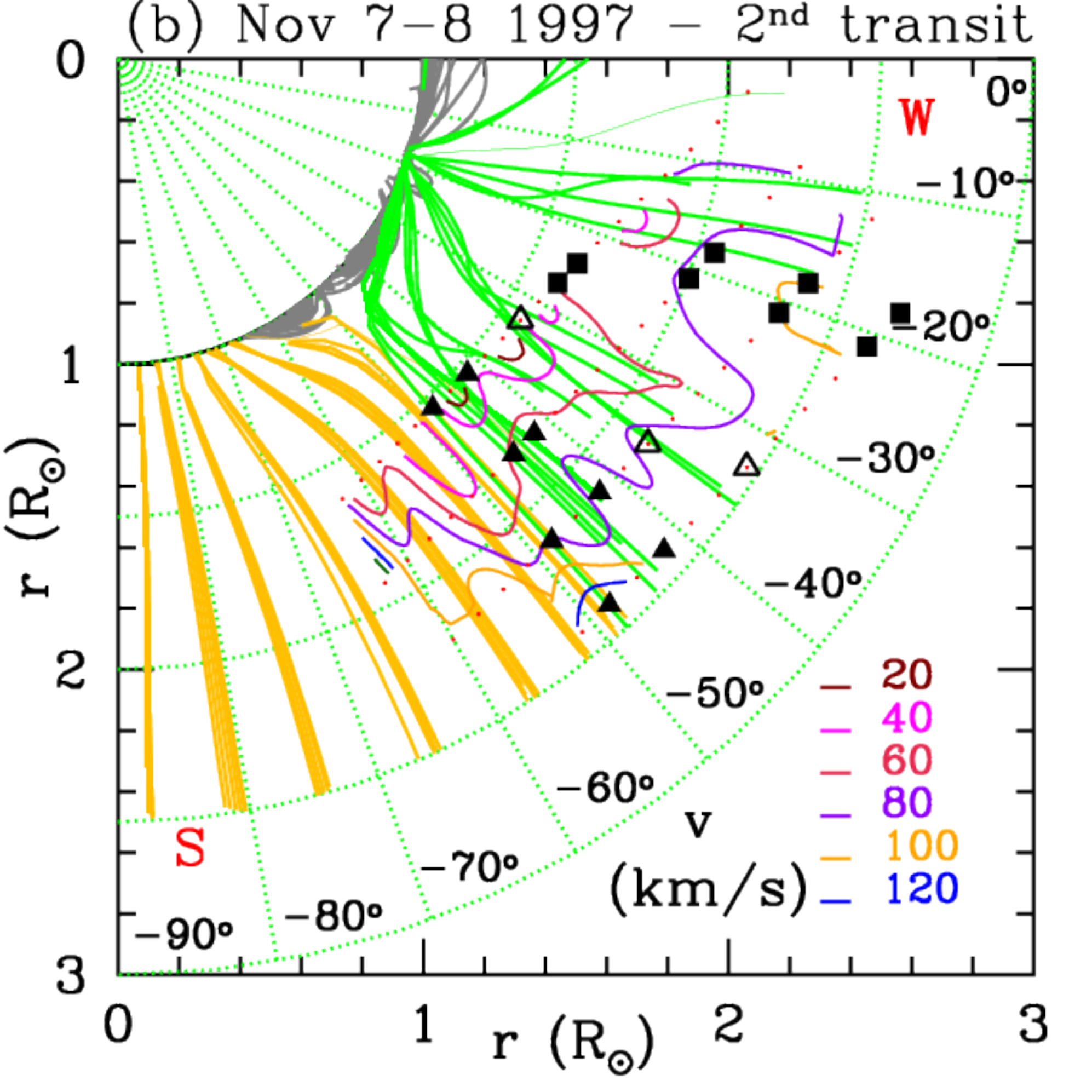}
  \includegraphics[width=4.4cm]{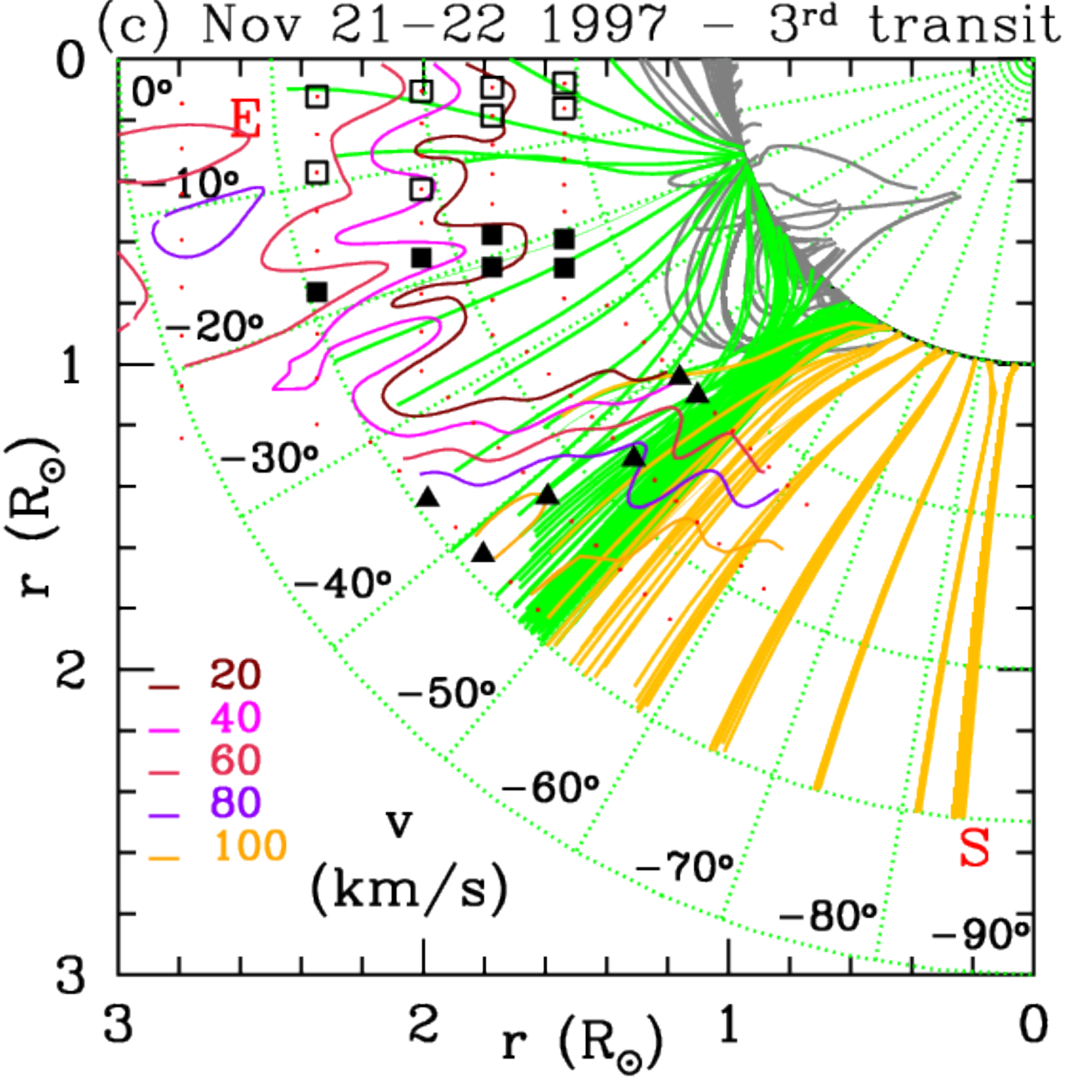}
  \includegraphics[width=4.4cm]{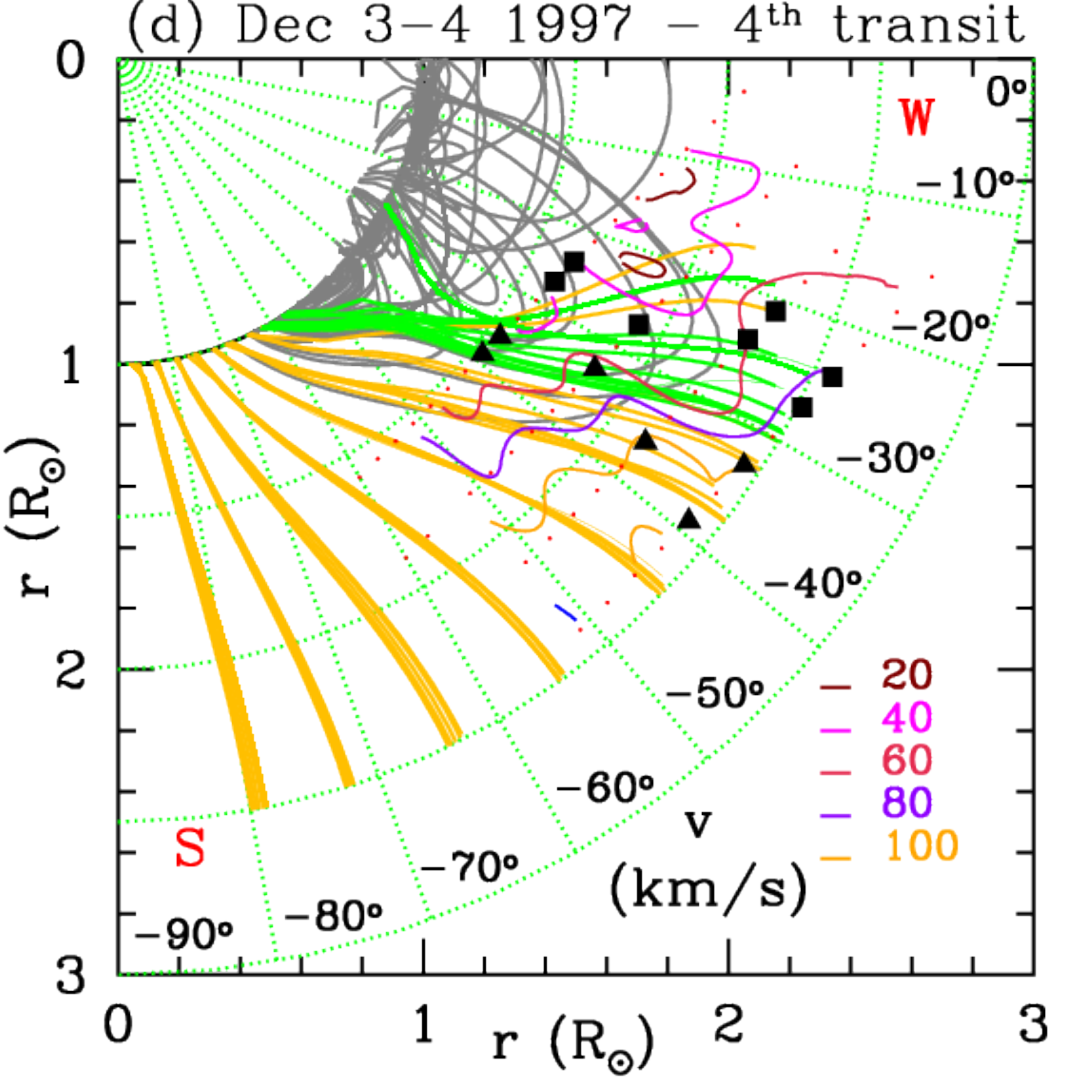}

  \includegraphics[width=4.4cm]{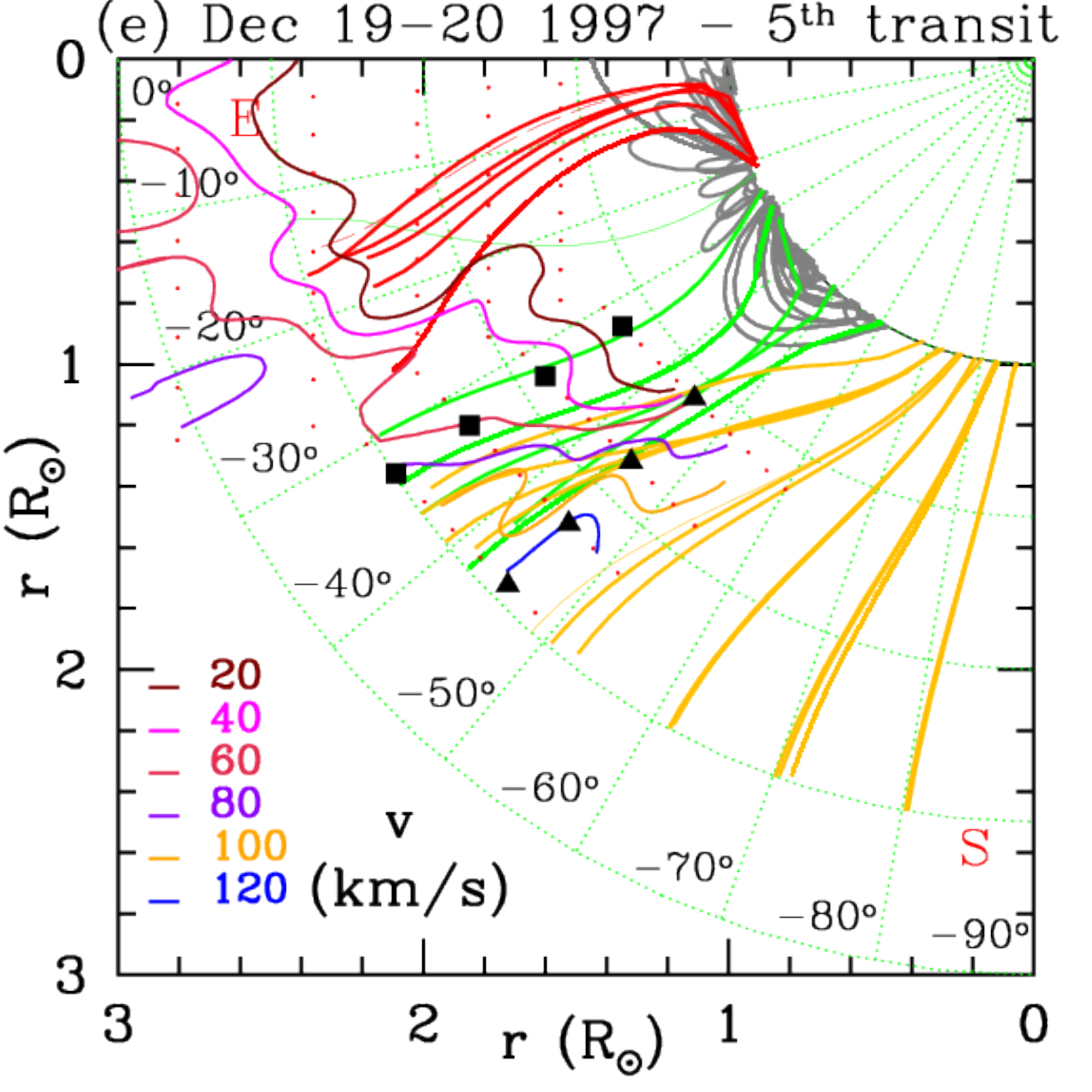}
  \includegraphics[width=4.4cm]{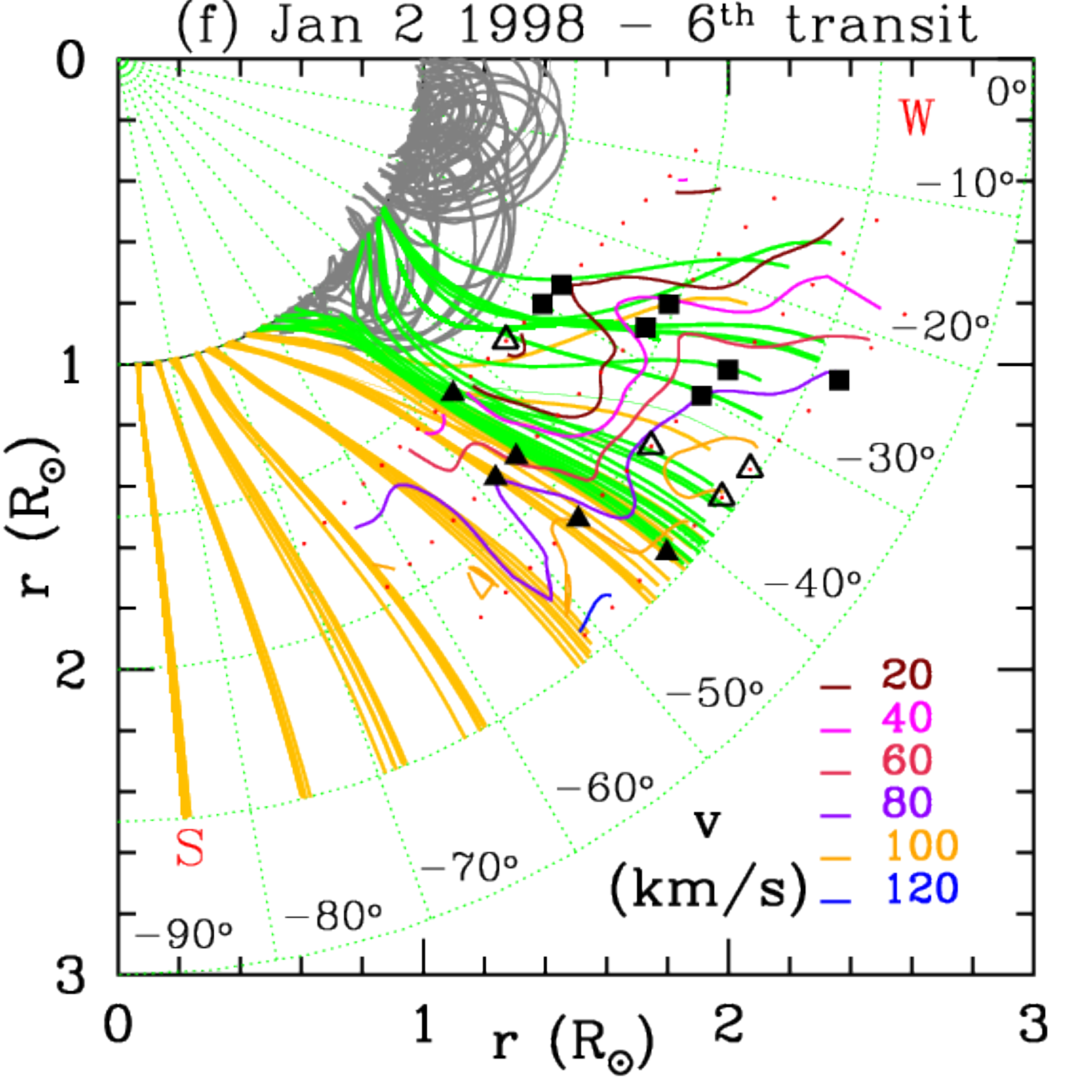}
  \includegraphics[width=4.4cm]{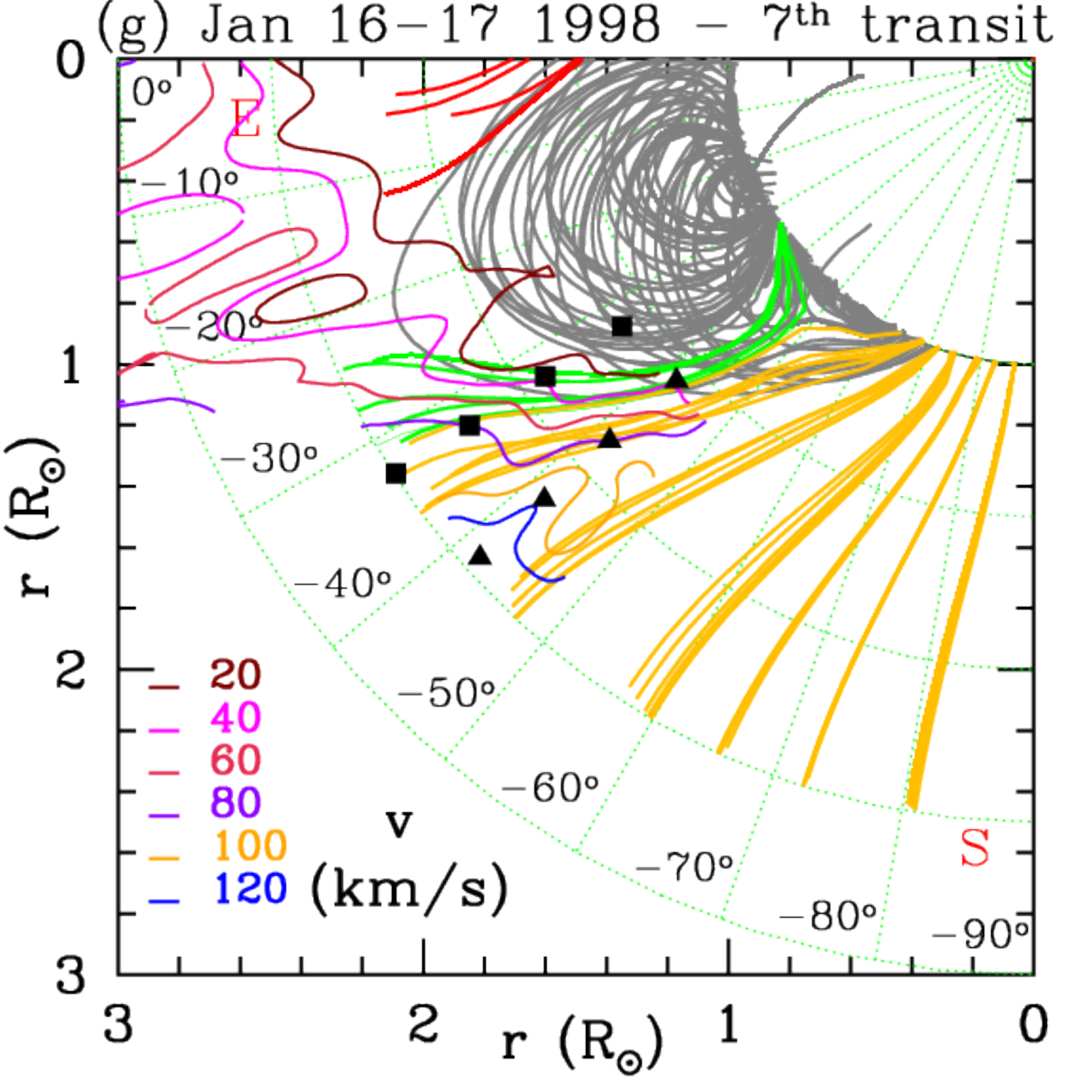}
  \includegraphics[width=4.4cm]{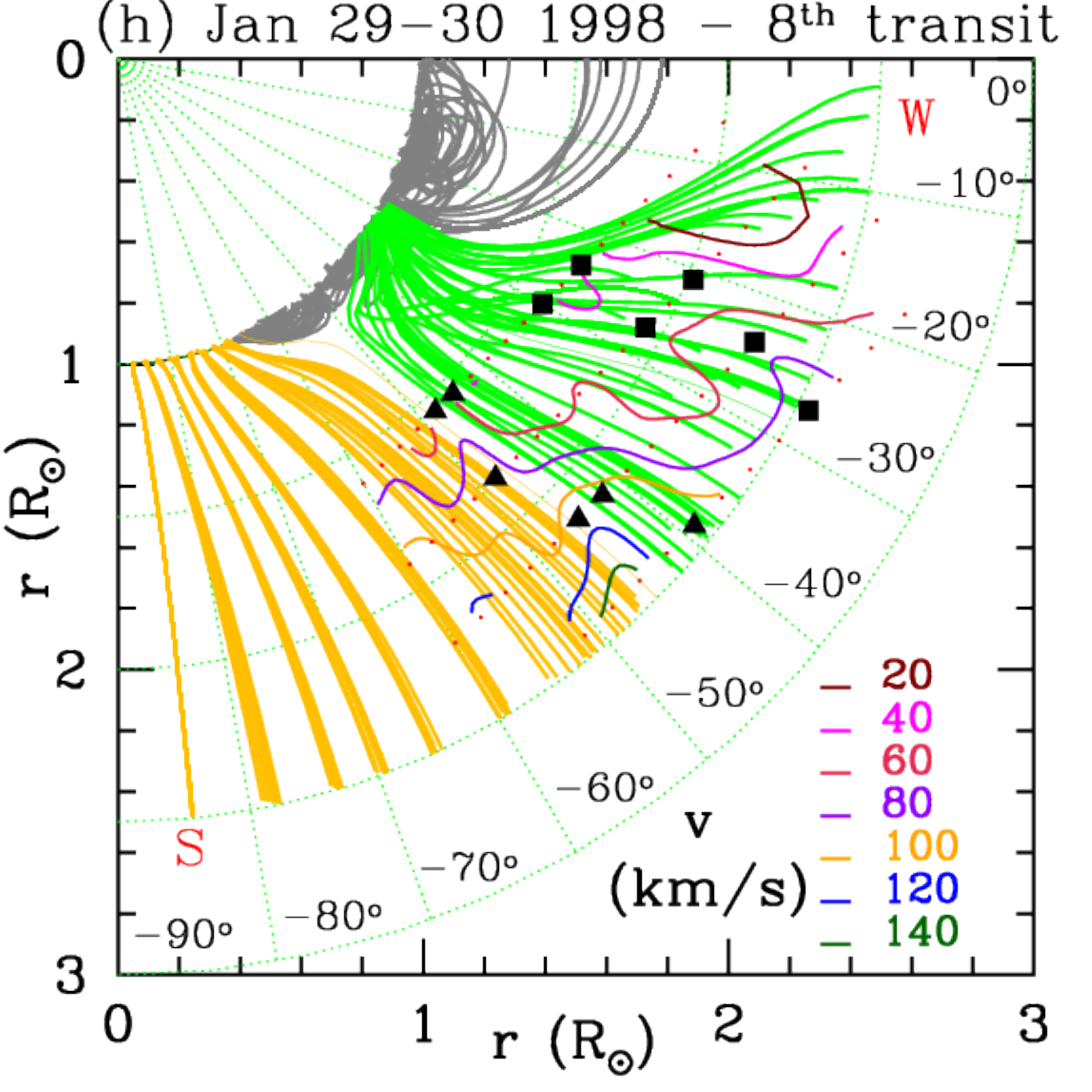}

  \includegraphics[width=4.4cm]{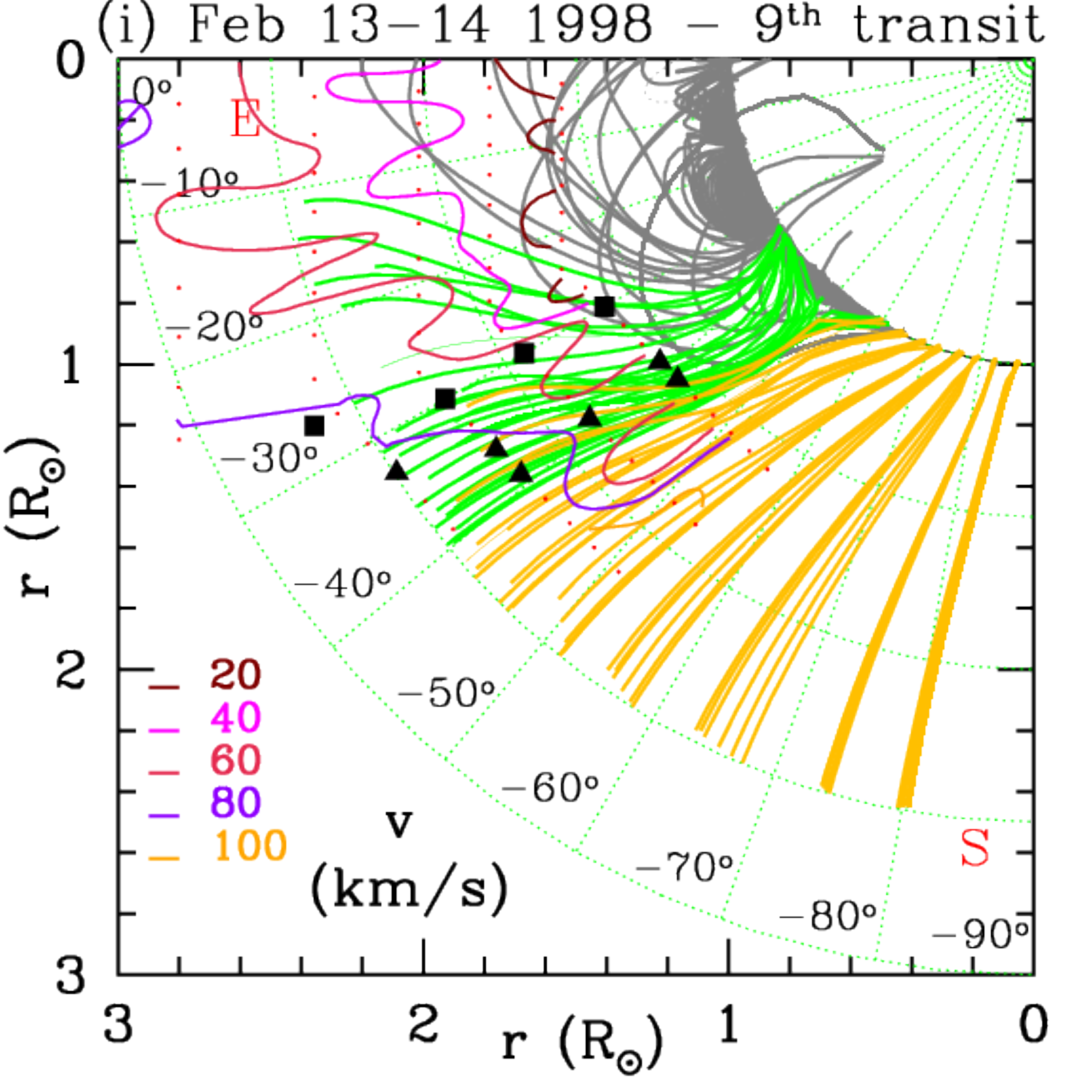}
  \includegraphics[width=4.4cm]{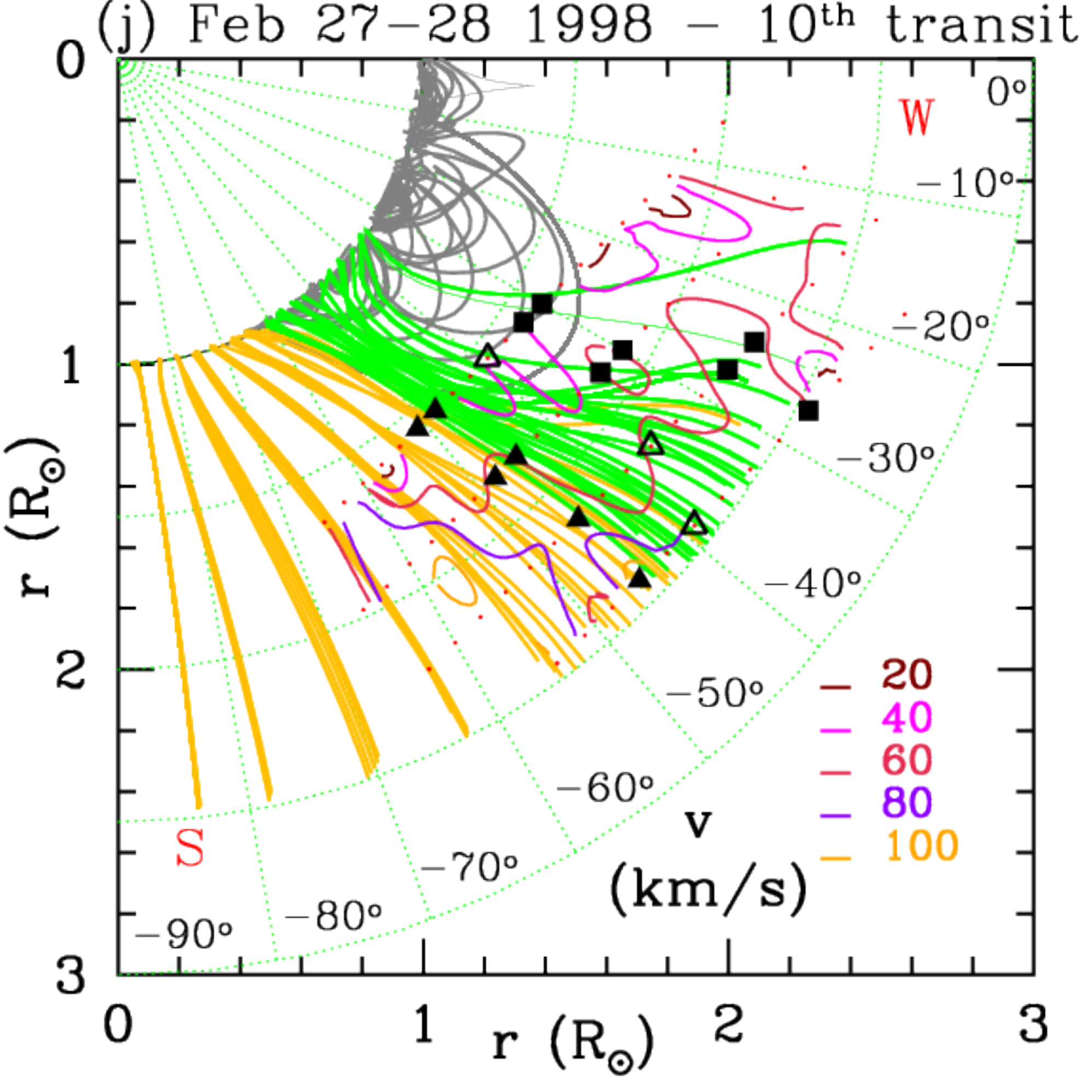}
  \caption{Isocontour maps of the outflow speed for the ten limb transits of 
    AR~8100. Open negative magnetic field lines originating from latitudes 
    lower (higher) than $-60~\deg$ are shown in green (orange). In panels e 
    and g positive open lines are also shown (red), originating from AR~8100 
    and from AR~8141 (in the northern emisphere), respectively. The UVCS slits 
    are shown as a sequence of red dots that represent the individual bins. 
    Full triangles (squares) identify data points within the intermediate- 
    (low-) latitude outflow channel candidates. Additional outflow channels 
    whose latitudes are slightly shifted equatorward with respect to the 
    latitudes at which intermediate- and low-latitude channels are found are 
    indicated with open triangles and squares.}
  \label{fig:vmap}
\end{figure*}

\section{Results}
\label{sec:results}

\begin{figure*}
  \centering
  \includegraphics[width=4.4cm]{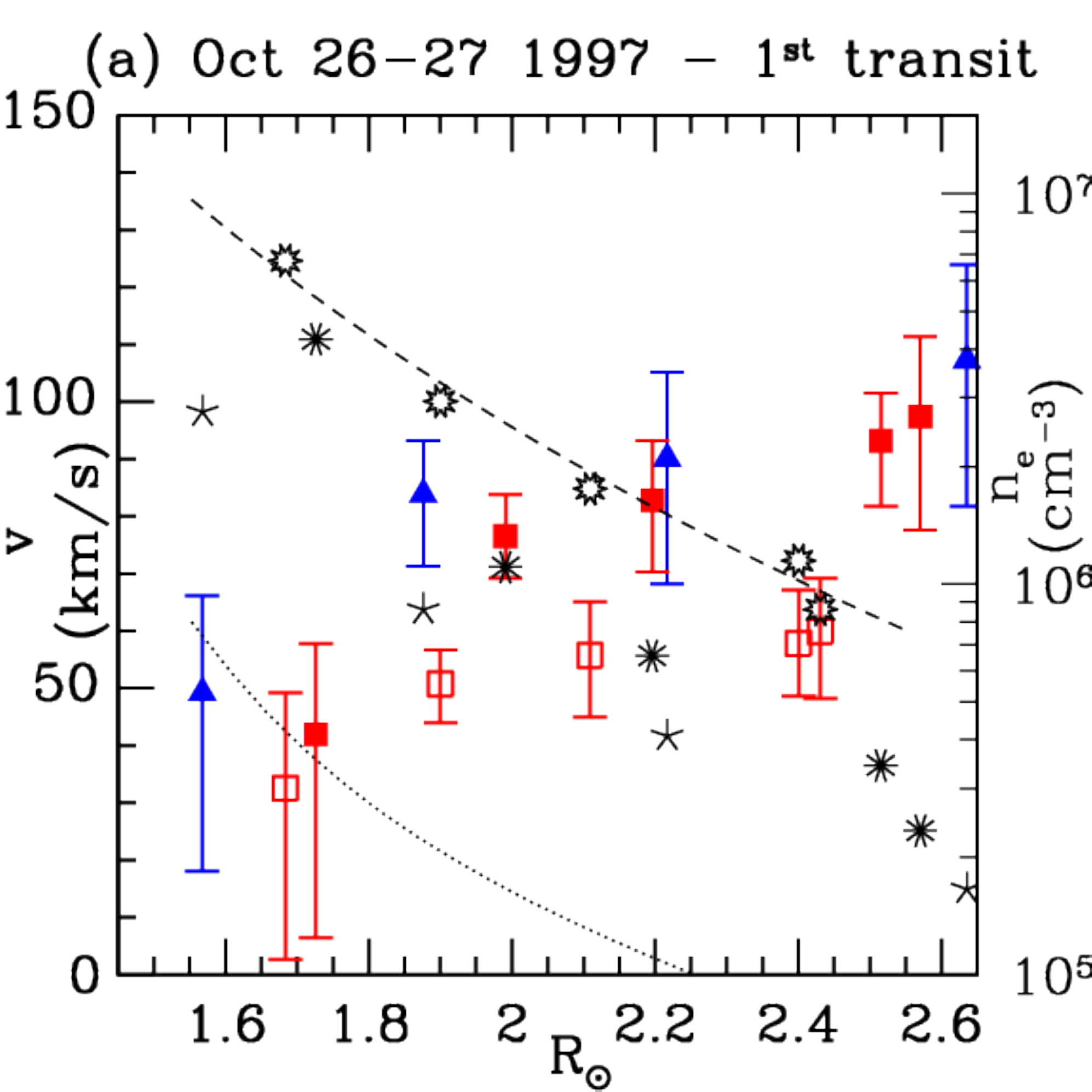}
  \includegraphics[width=4.4cm]{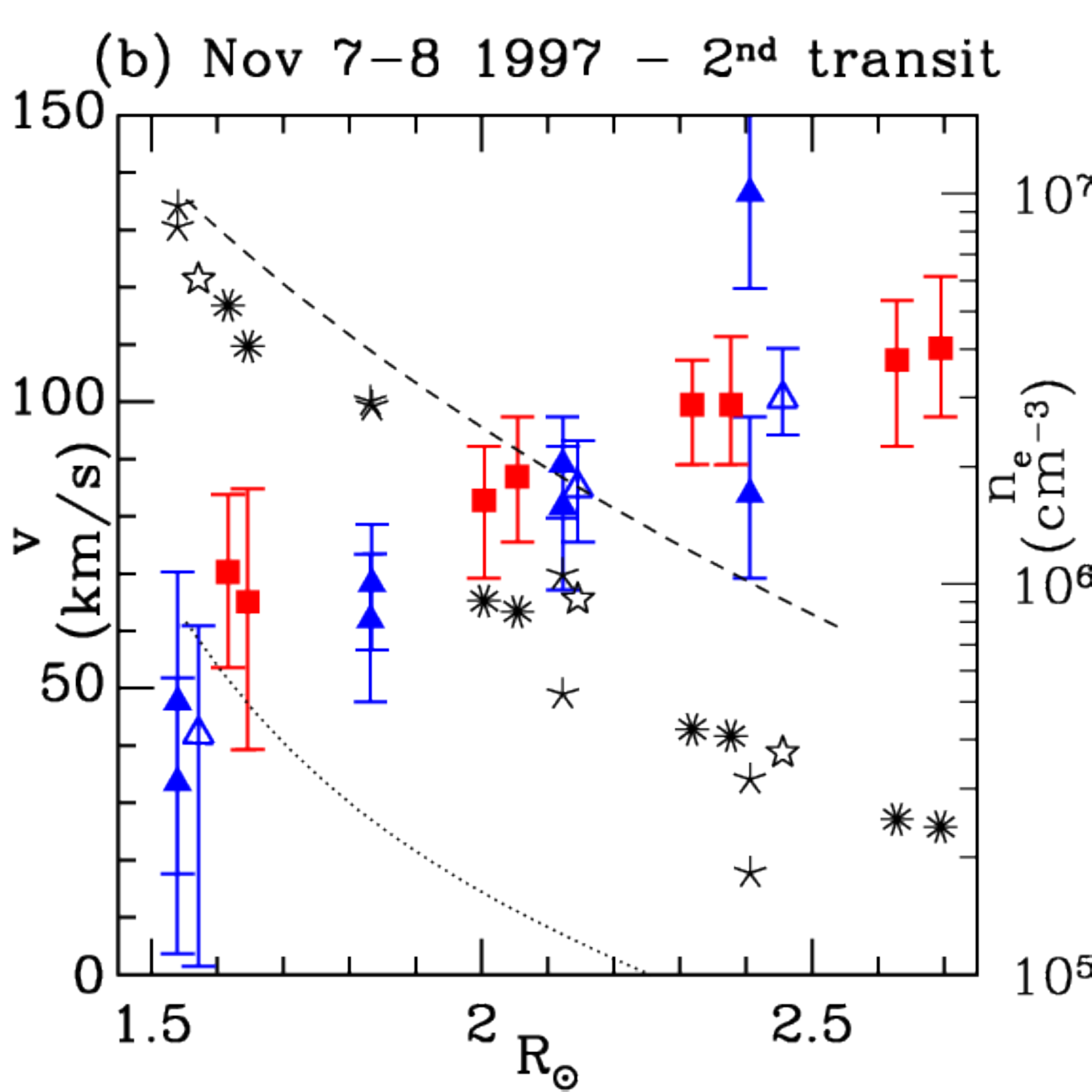}
  \includegraphics[width=4.4cm]{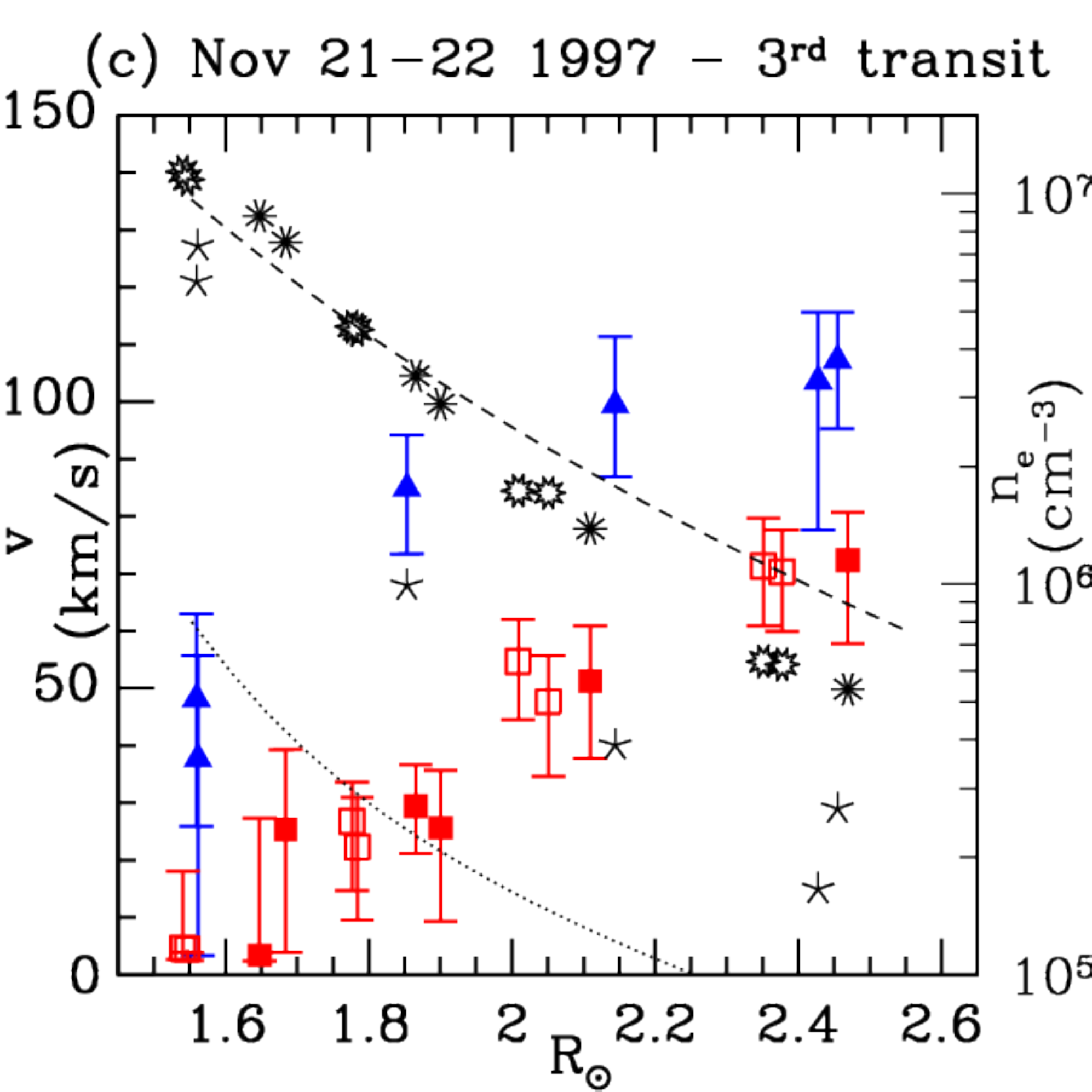}
  \includegraphics[width=4.4cm]{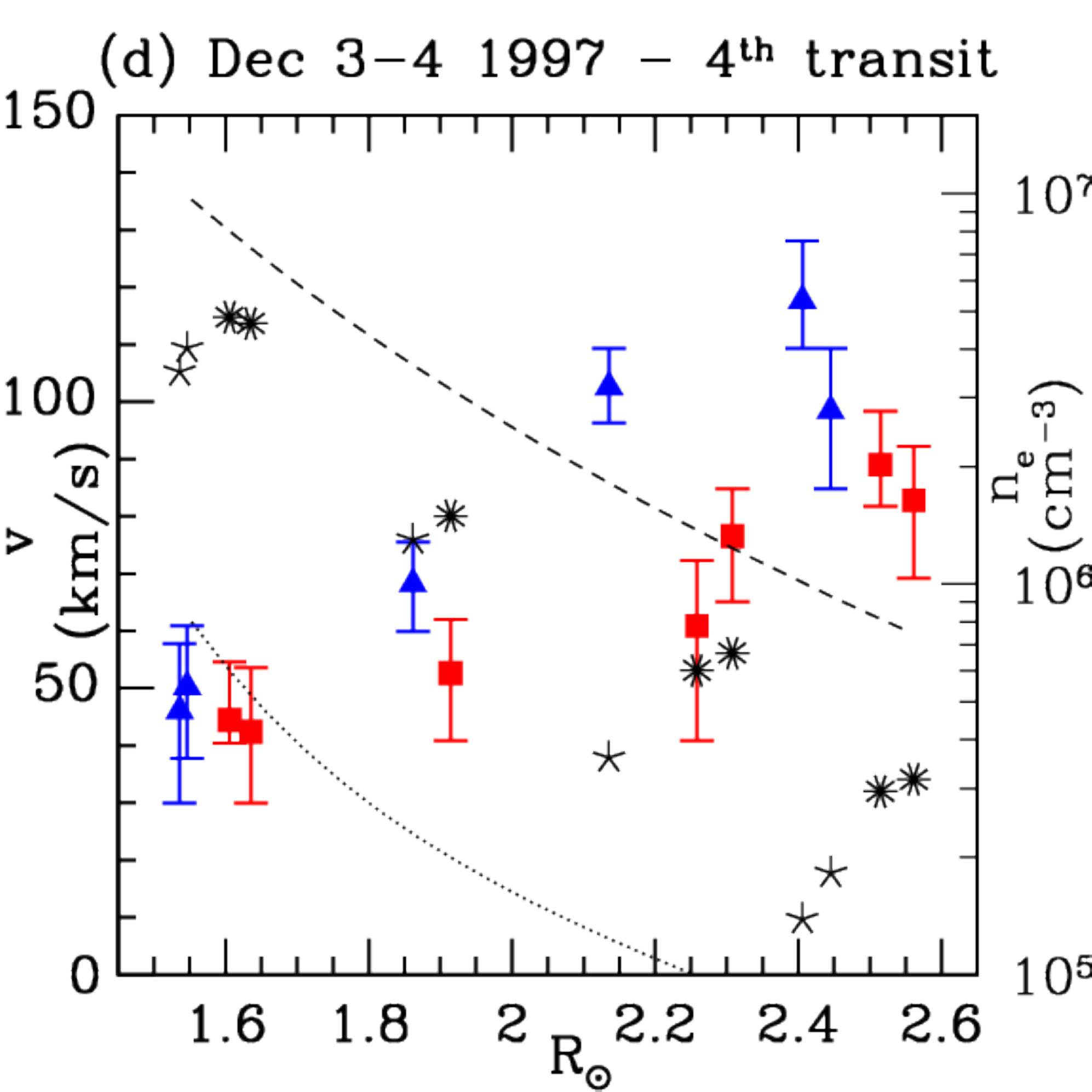}

  \includegraphics[width=4.4cm]{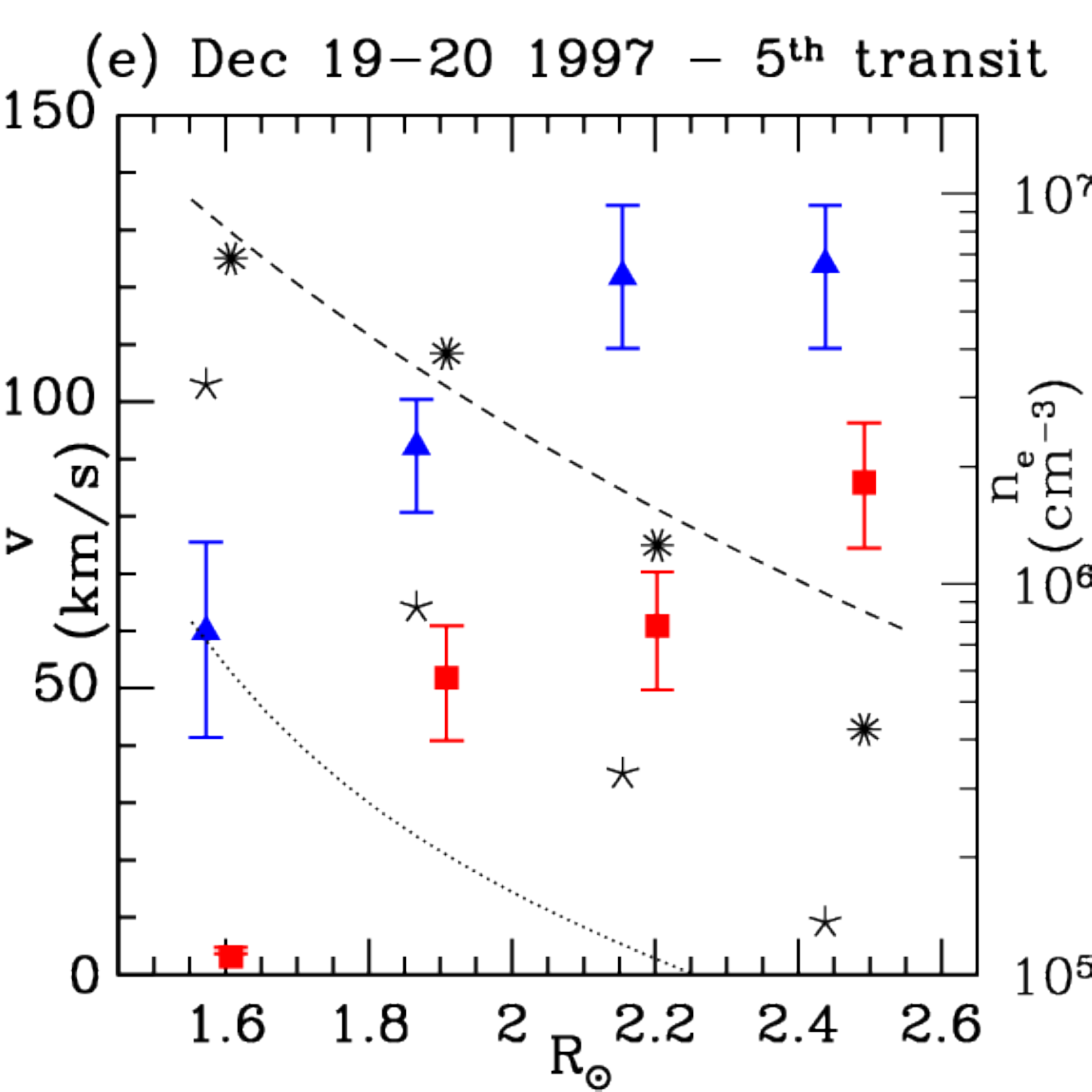}
  \includegraphics[width=4.4cm]{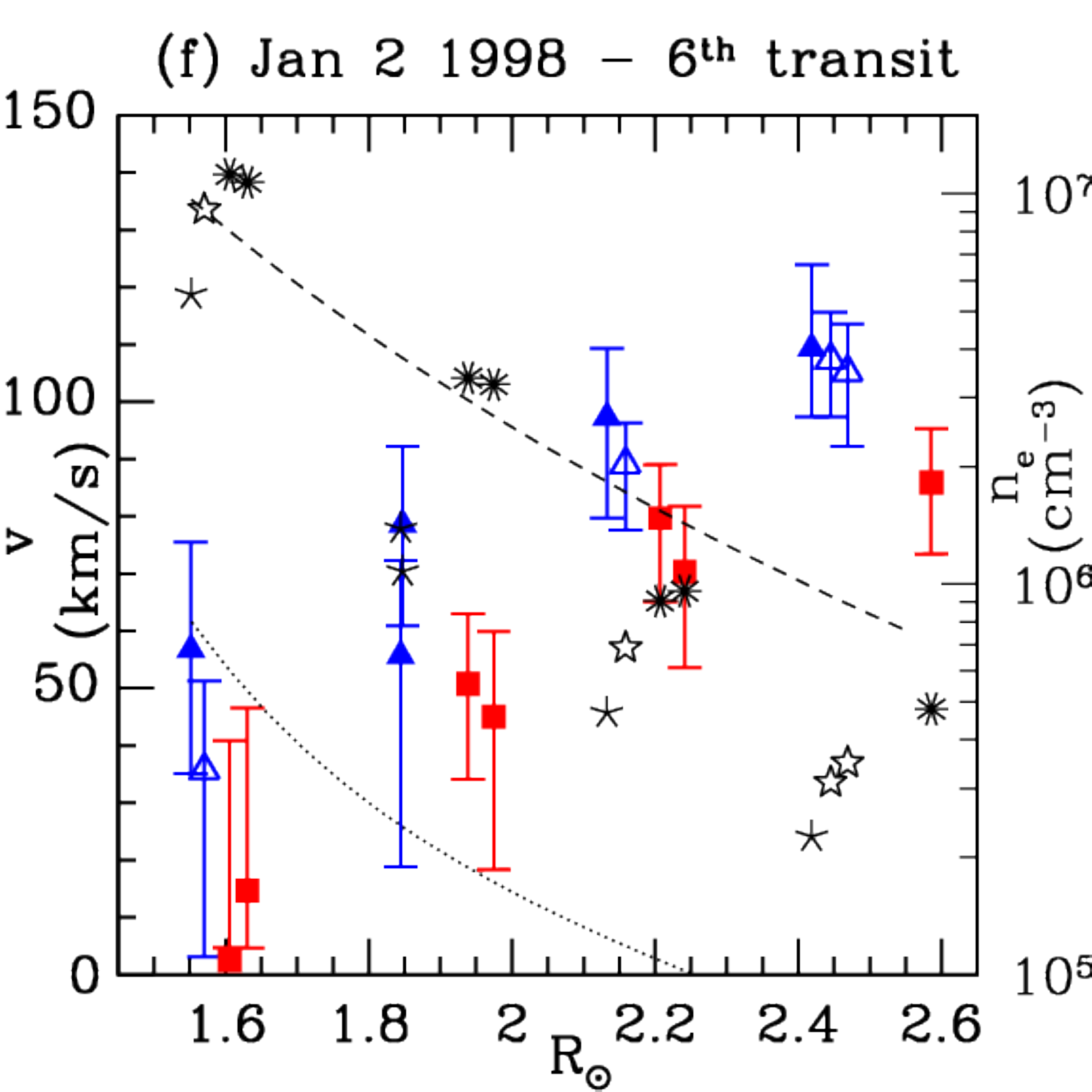}
  \includegraphics[width=4.4cm]{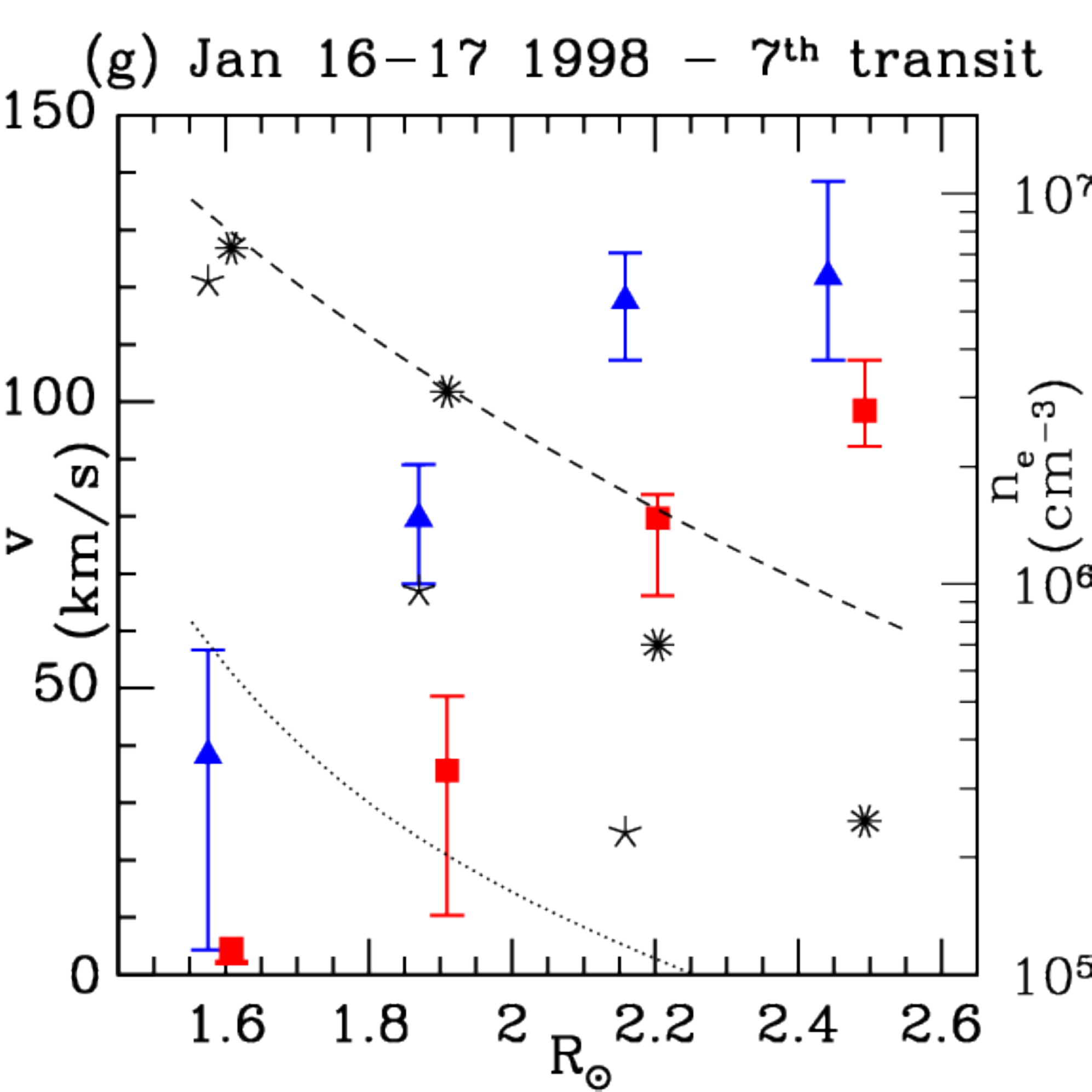}
  \includegraphics[width=4.4cm]{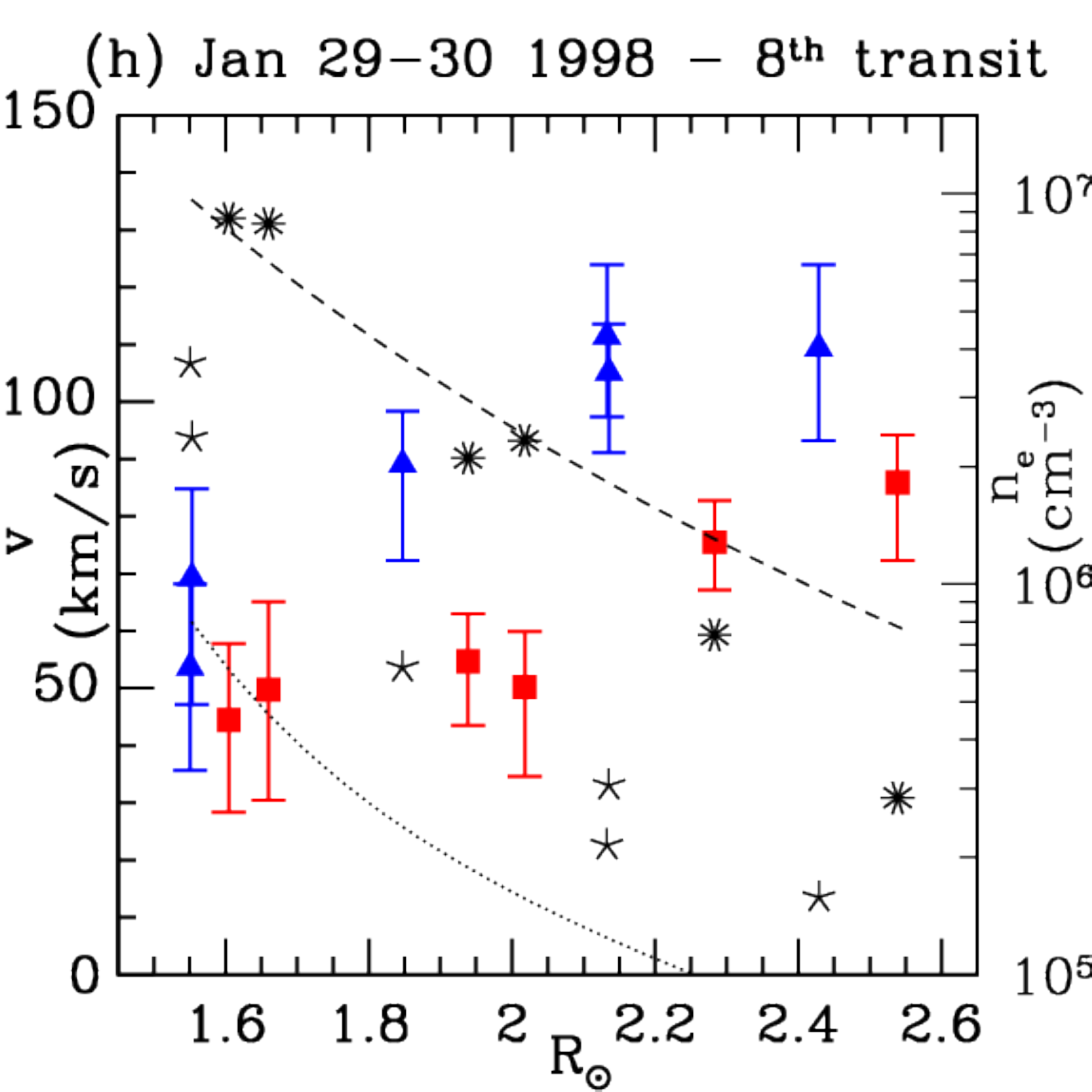}

  \includegraphics[width=4.4cm]{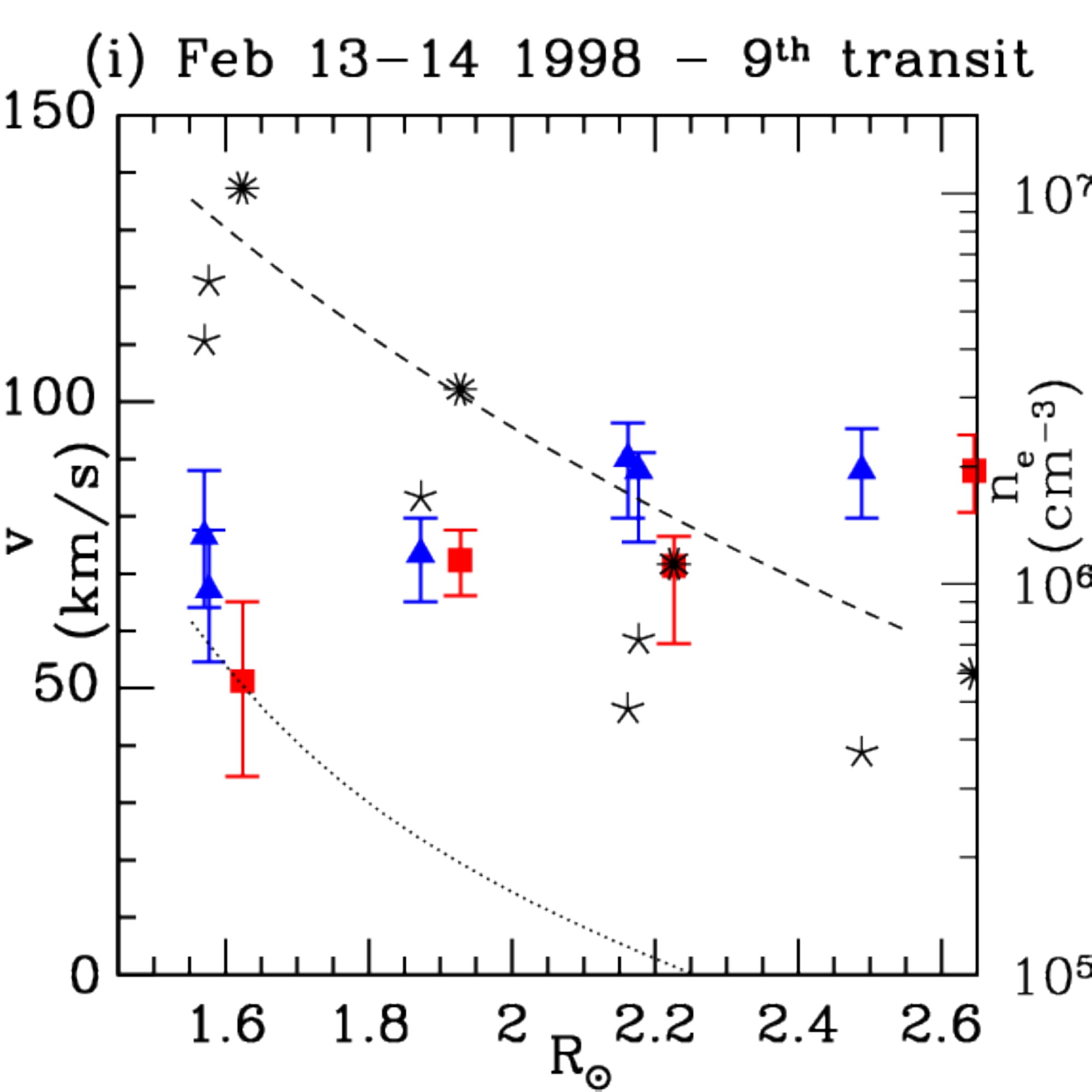}
  \includegraphics[width=4.4cm]{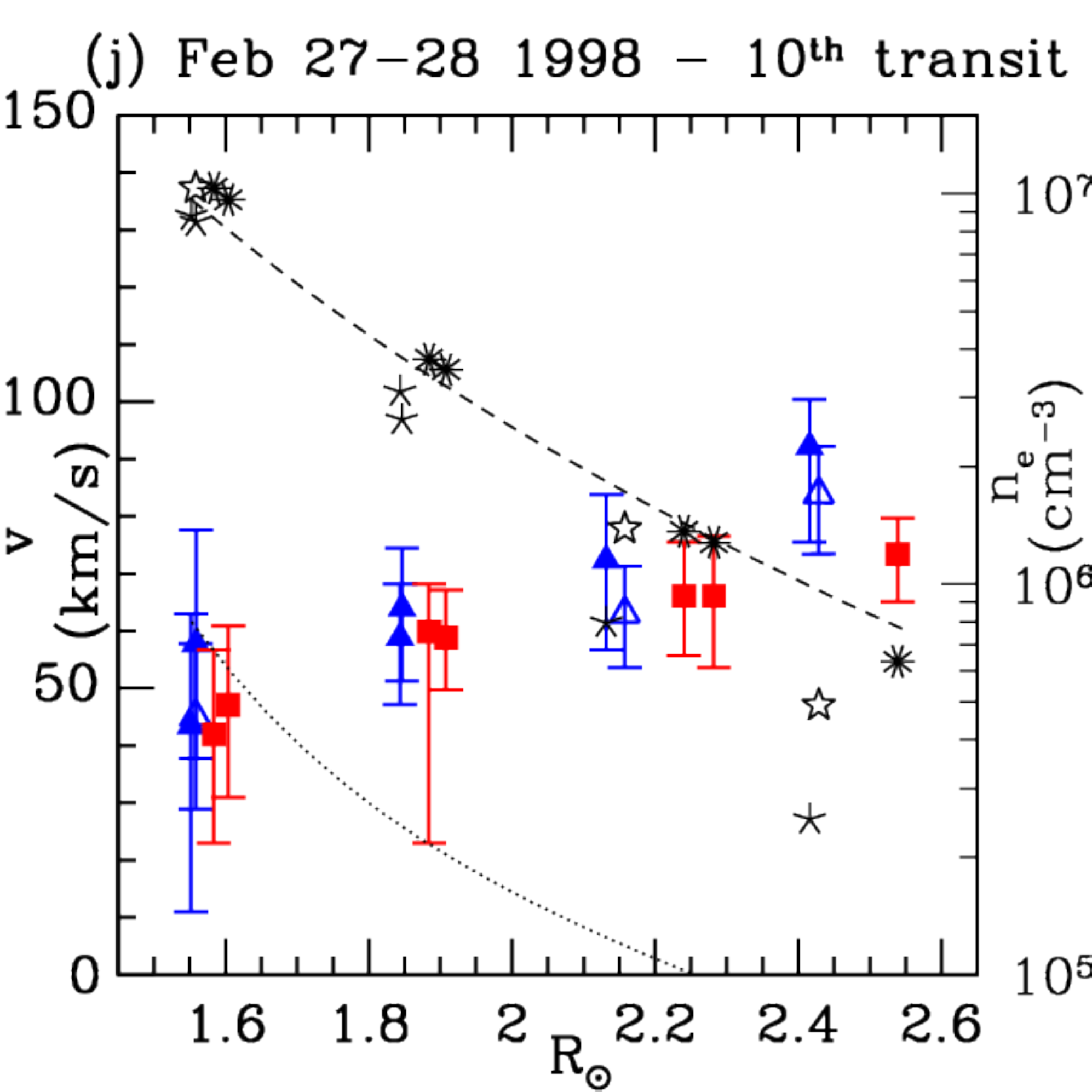}
  \caption{Outflow speed and electron density vs. distance for the ten limb 
    transits; the dotted and dashed curves are typical density profiles for a 
    CH and a streamer, respectively (see \citealt{Guhathakurta1999a} 
    and \citealt{Gibson1999a}). The solid blue triangles and the solid red 
    squares are the intermediate- and low-latitude outflow speeds, 
    respectively. The five- and ten-point stars represent the electron density 
    values for the intermediate- and low-latitude outflows, respectively. The 
    error bars in the electron density are not shown as they are barely 
    distinguishable in a logarithmic plot; see text for a discussion about the 
    error estimate for $v$ and $n_{\rm e}$. Analogously to 
    Fig.~\ref{fig:vmap}, open triangles and squares refer to the outflow 
    speeds in secondary intermediate- and low-latitude channels; the 
    relative electron densities are given as open ten- and five-point stars.}
  \label{fig:profv}
\end{figure*}

In this section we illustrate the results inferred from the DD technique 
described in Sect.~\ref{sec:dimm} and from the analysis of the 
Si~{\sc{xii}}~520.7~\AA\ line, starting from the first limb transit of 
AR~8100, which occurred on October 26-27, 1997, to the last one, on February 
27-28, 1998. The outflow speed results are shown as isocontour maps, obtained 
by interpolating the results of the DD analysis (see Fig.~\ref{fig:vmap}) as a 
function of latitude and heliocentric distance. Evidence for outflow channel 
candidates is provided by isocontour maps, dipping down, with respect to 
adjacent positions, toward lower altitudes. We required this to occur at 
several speed levels and at positions where open bundles of magnetic field 
lines, originating within the AR, are present. A further guide to identifying 
the AR outflow channels was obtained from the intensity ratio of the 
O~{\sc{vi}} doublet lines at 1031.9 and 1037.6~\AA, 
$I_{\rm O~{\textsc{VI}}-1031.9}/I_{\rm O~{\textsc{VI}}-1037.6}$. This diagnostic technique has 
been illustrated by many authors (see, e.g., \citealt{Noci1987a}, 
\citealt{Li1998a}), here it suffices to say that the intensity ratio is a 
decreasing function of the plasma speed, $v$, at least for 
$v < 100~{\rm km/s}$. 
Examples of the profiles of $I_{\rm O~{\textsc{VI}}-1031.9}/I_{\rm O~{\textsc{VI}}-1037.6}$ 
along the UVCS slit are discussed in Sects.~\ref{sec:second_t} and 
\ref{sec:third_t} at the times of the second and third AR limb transits. When 
the bundles of open lines clearly originated solely from the south CH, (i.e., 
at southern latitudes roughly higher than $60~\deg$), we discarded the channel 
candidate from our analysis. The occurrence of open lines that project onto 
the same area and originate from both the CH and the AR, as is often the case 
in channels at intermediate latitude, is discussed in 
Sect.~\ref{sec:discussion}.

The speeds and electron densities inferred from the DD analysis that belong to 
outflow channel candidates, which are tentatively classified as low-latitude 
and intermediate-latitude channels, are also given in separate diagrams (see 
Fig.~\ref{fig:profv}). The error bars were estimated from the total counts 
statistics in the line intensites. The main source of error in the values 
inferred for the outflow speed and electron density is the statistical 
uncertainty in the intensity line ratio 
$I_{\rm O~{\textsc{VI}}-1031.9}/I_{\rm O~{\textsc{VI}}-1037.6}$ (the H~{\sc{i}}~Lyman~$\alpha$ 
intensity is affected by a negligible error). We estimated the upper and lower 
limits in $v$ and $n_{\rm e}$ by setting the value of the ratio 
$I_{\rm O~{\textsc{VI}}-1031.9}/I_{\rm O~{\textsc{VI}}-1037.6}$ to its highest and lowest value 
in the DD analysis, according to the Poissonian statistics of the intensities 
of individual lines. Owing to the non-linear relationship between the oxygen 
doublet intensity ratio and the outflow speed (see, e.g., 
\citealt{Noci1987a}), values inferred from the DD study that are lower than 
about 50~km/s, which are obtained mostly at low heliocentric distances, tend 
to be more uncertain than those relative to higher speed regimes, in spite of 
the more favorable count statistics. Typical outflow speed and electron 
density errors range within the intervals of 10-40 percent, and 2-10 percent, 
respectively. For a thorough discussion on the influence of DD model 
parameters on the final results we refer to \cite{Zangrilli2012a}.

\begin{figure}[h]
  \centering
  \includegraphics[width=4.cm]{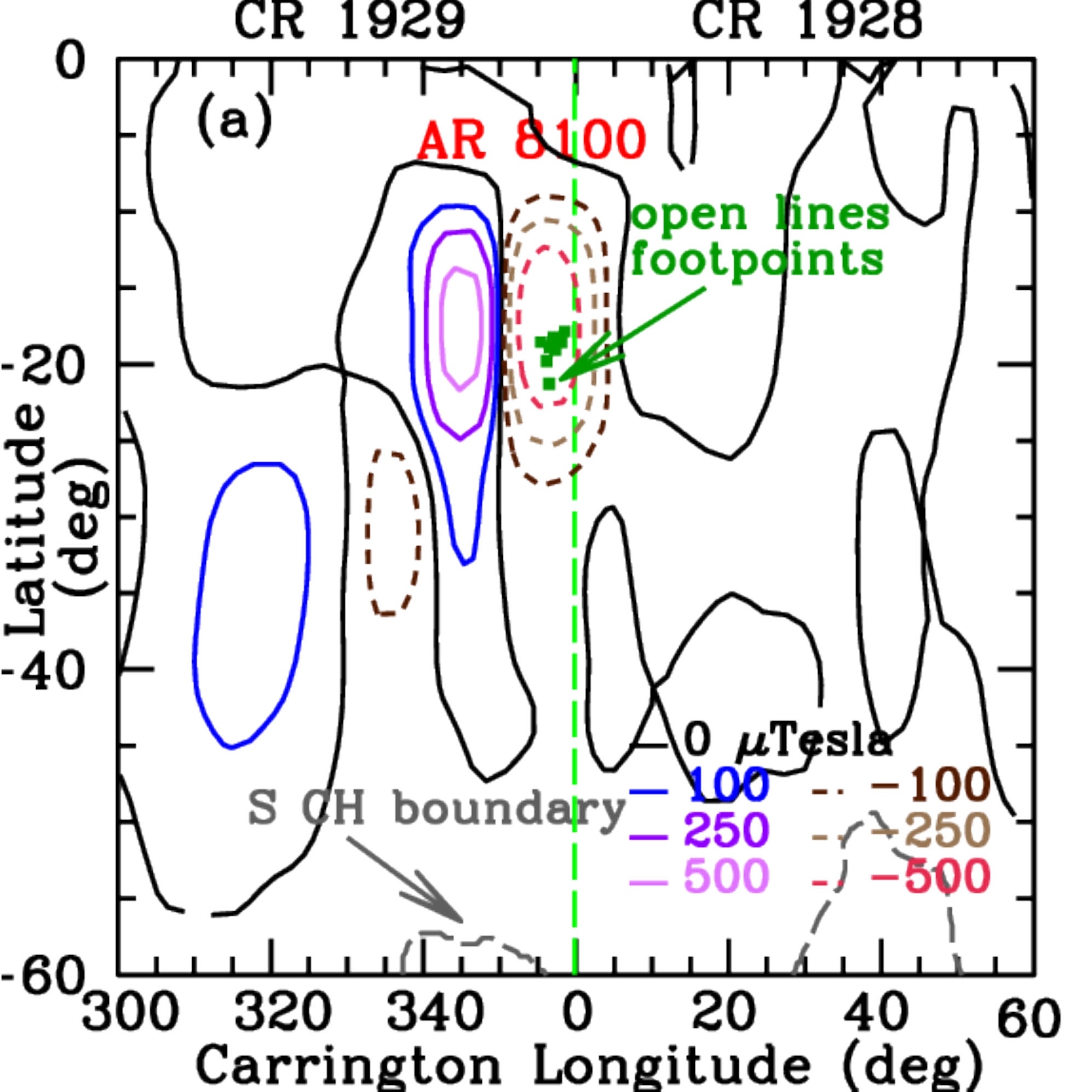}
  \includegraphics[width=4.cm]{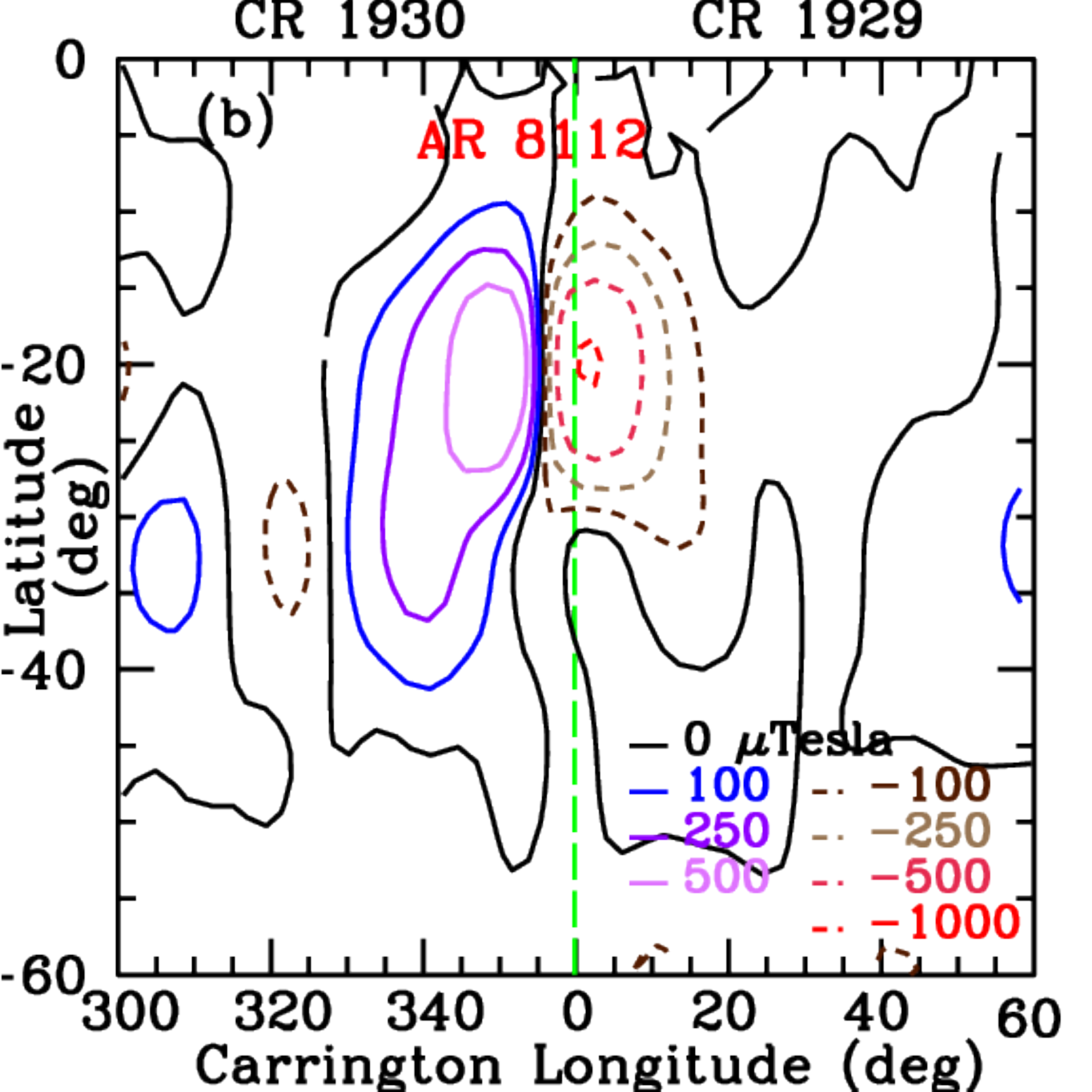}

  \includegraphics[width=4.cm]{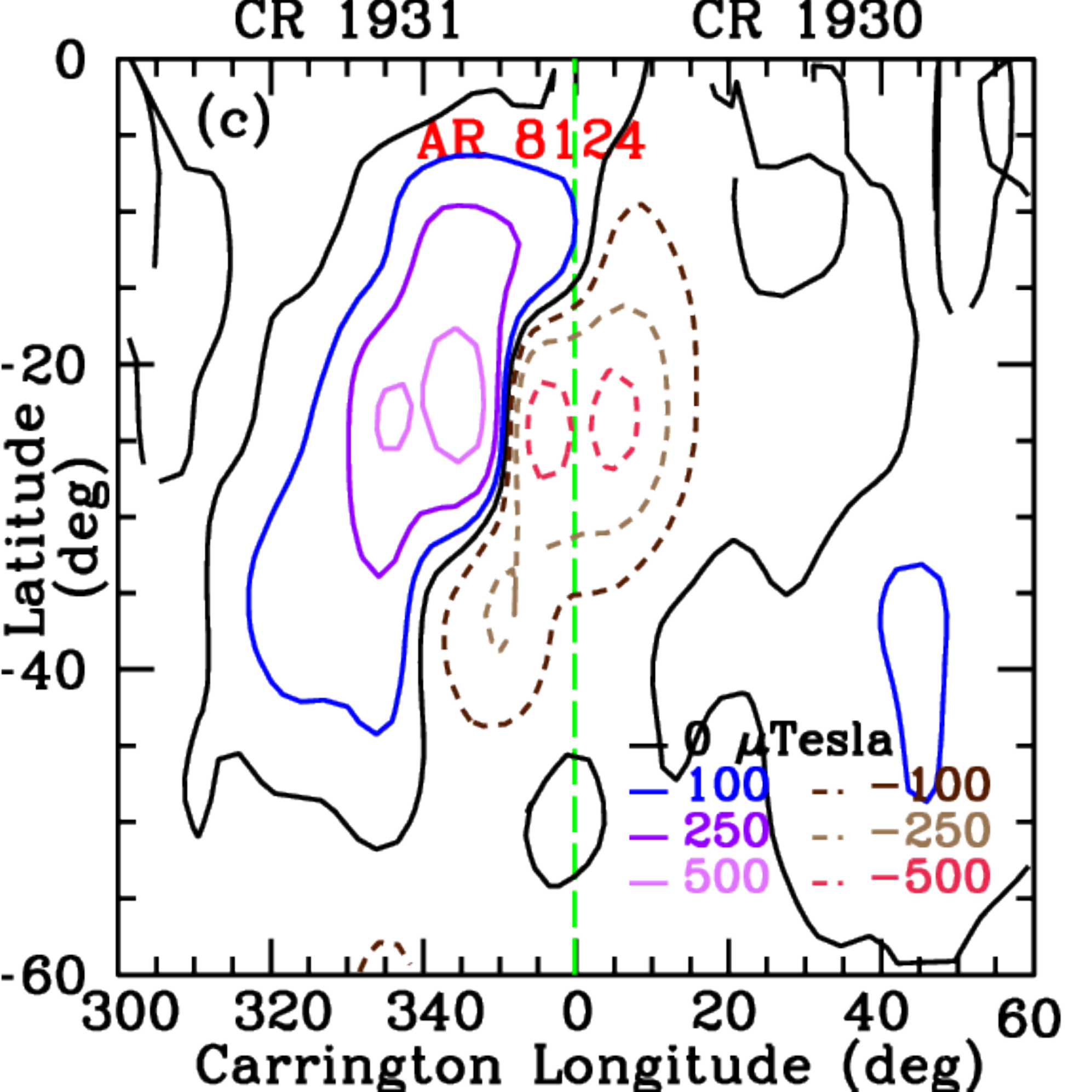}
  \includegraphics[width=4.cm]{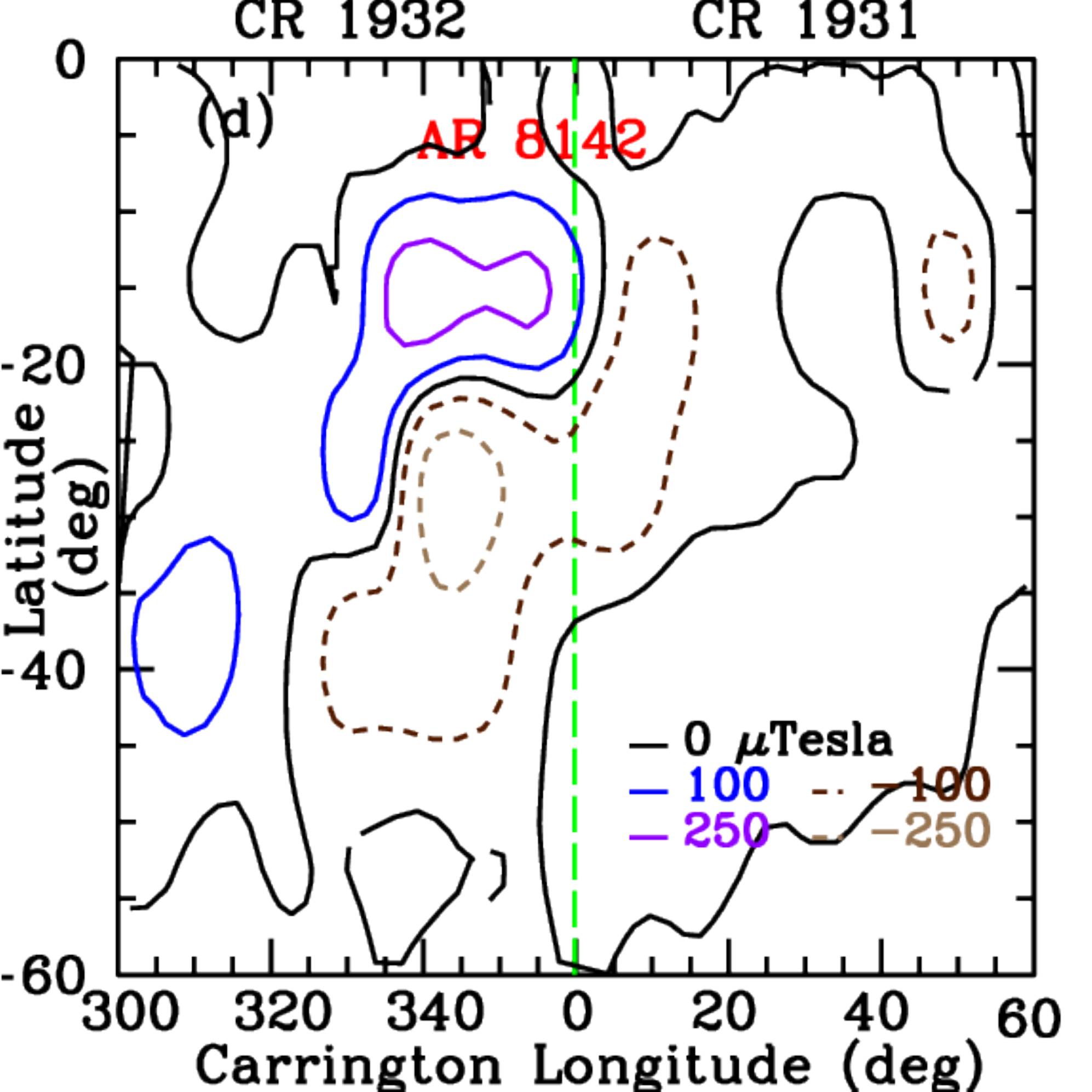}

  \includegraphics[width=4.cm]{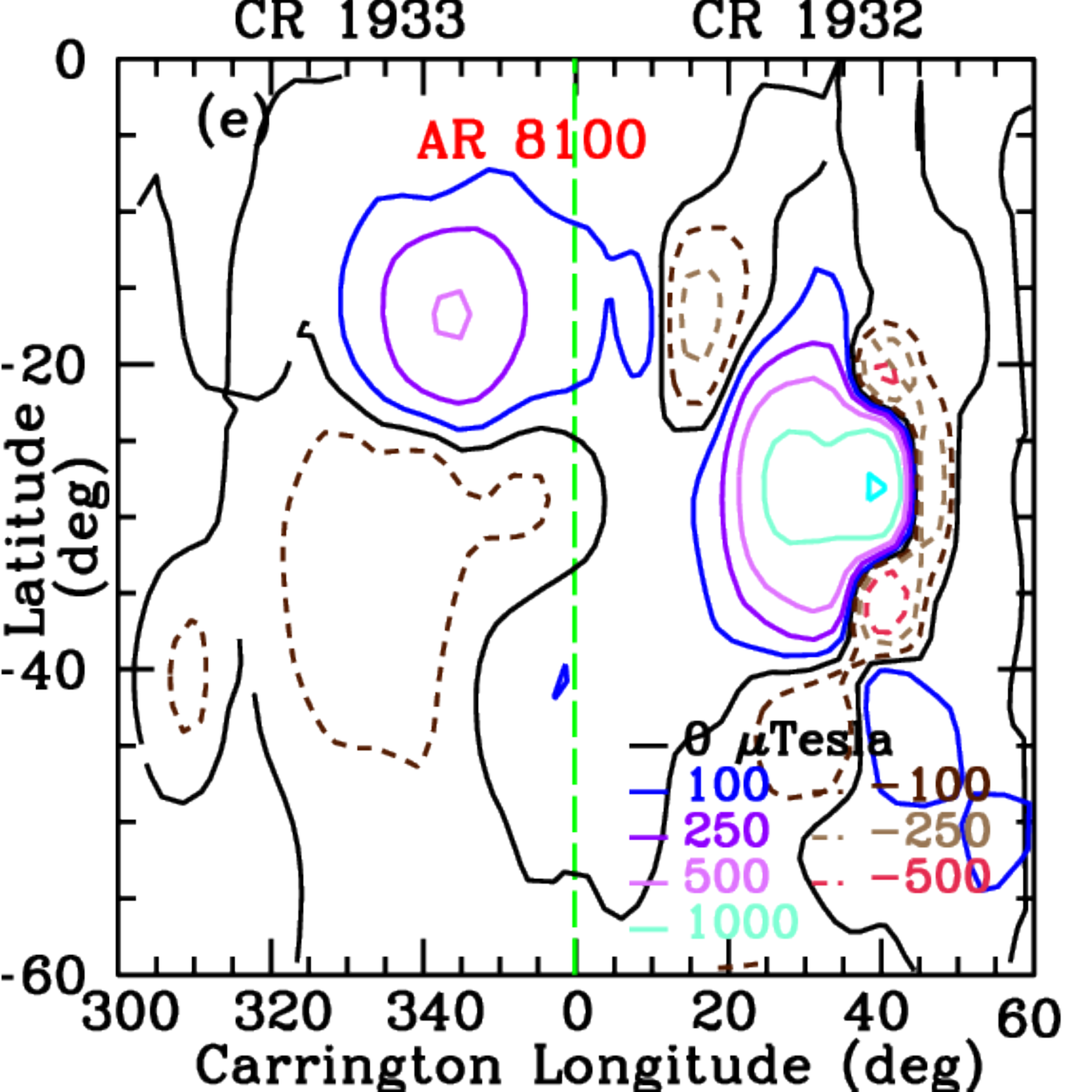}
  \caption{Photospheric magnetic field distribution around AR~8100, taken 
    during Carrington rotations from 1929 to 1933, adapted from the synoptic 
    maps provided by the Wilcox Solar Observatory. The neutral field line is 
    drawn as a black solid line, while positive and negative contours are 
    drawn as colored solid and dashed lines. The greenish squares in panel (a) 
    are the footpoints of the open magnetic field lines. The CH boundaries are 
    based on He~{\sc{I}}~10830\AA\ line observations from the Kitt Peak 
    telescope and are shown as a gray long-dashed contour.}
  \label{fig:wilcox}
\end{figure}

\begin{figure}[h]
  \centering
  \includegraphics[width=4.cm]{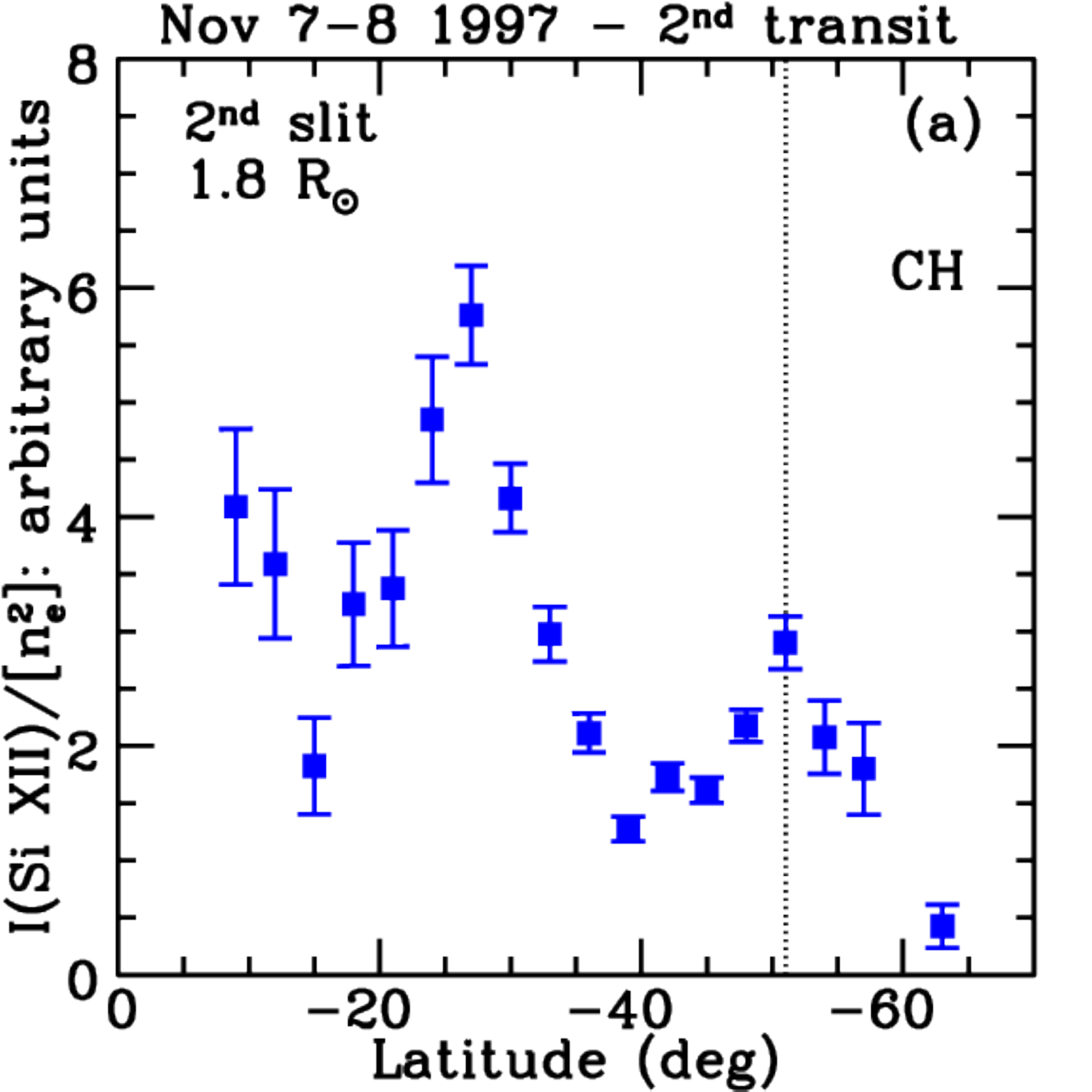}
  \includegraphics[width=4.cm]{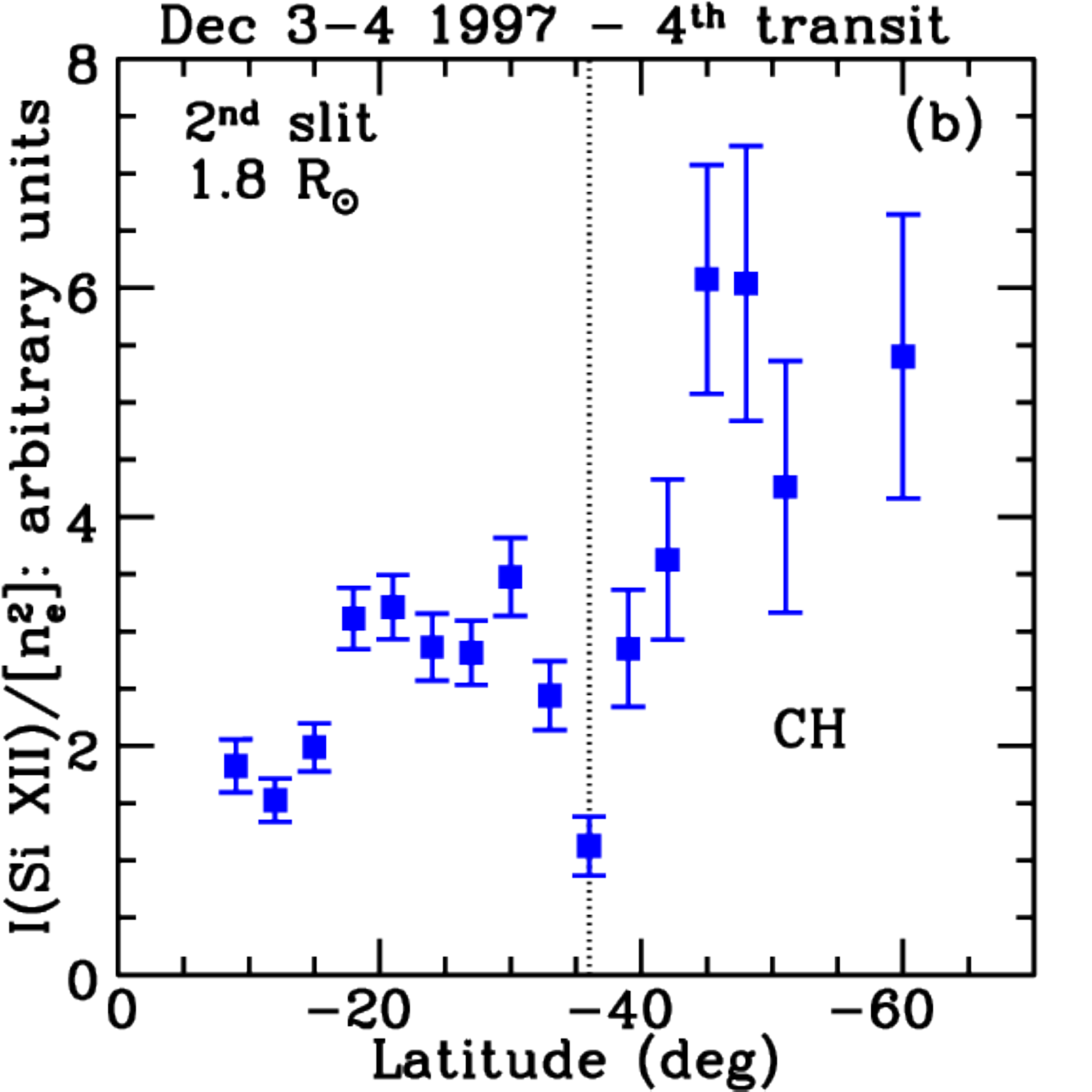}

  \includegraphics[width=4.cm]{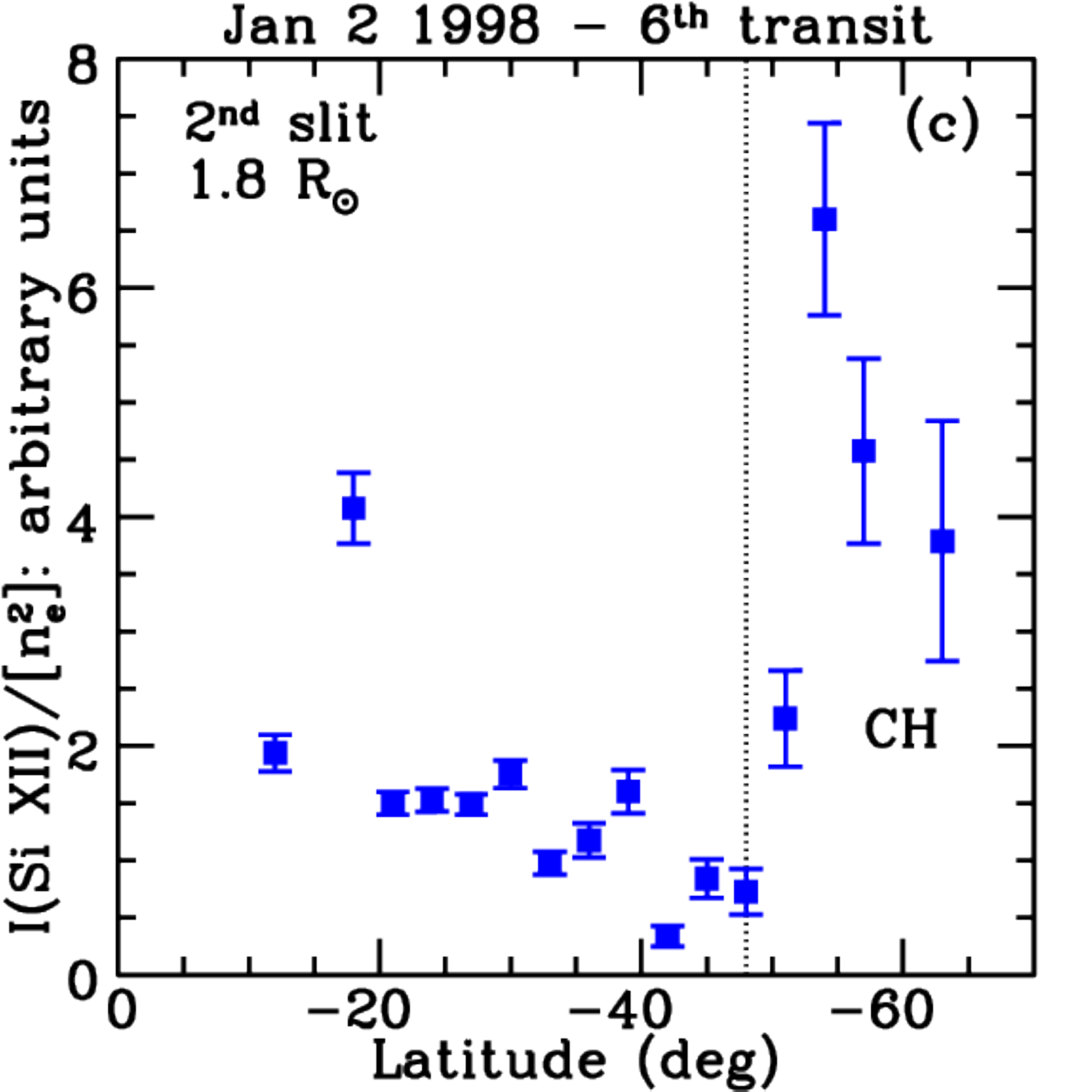}
  \includegraphics[width=4.cm]{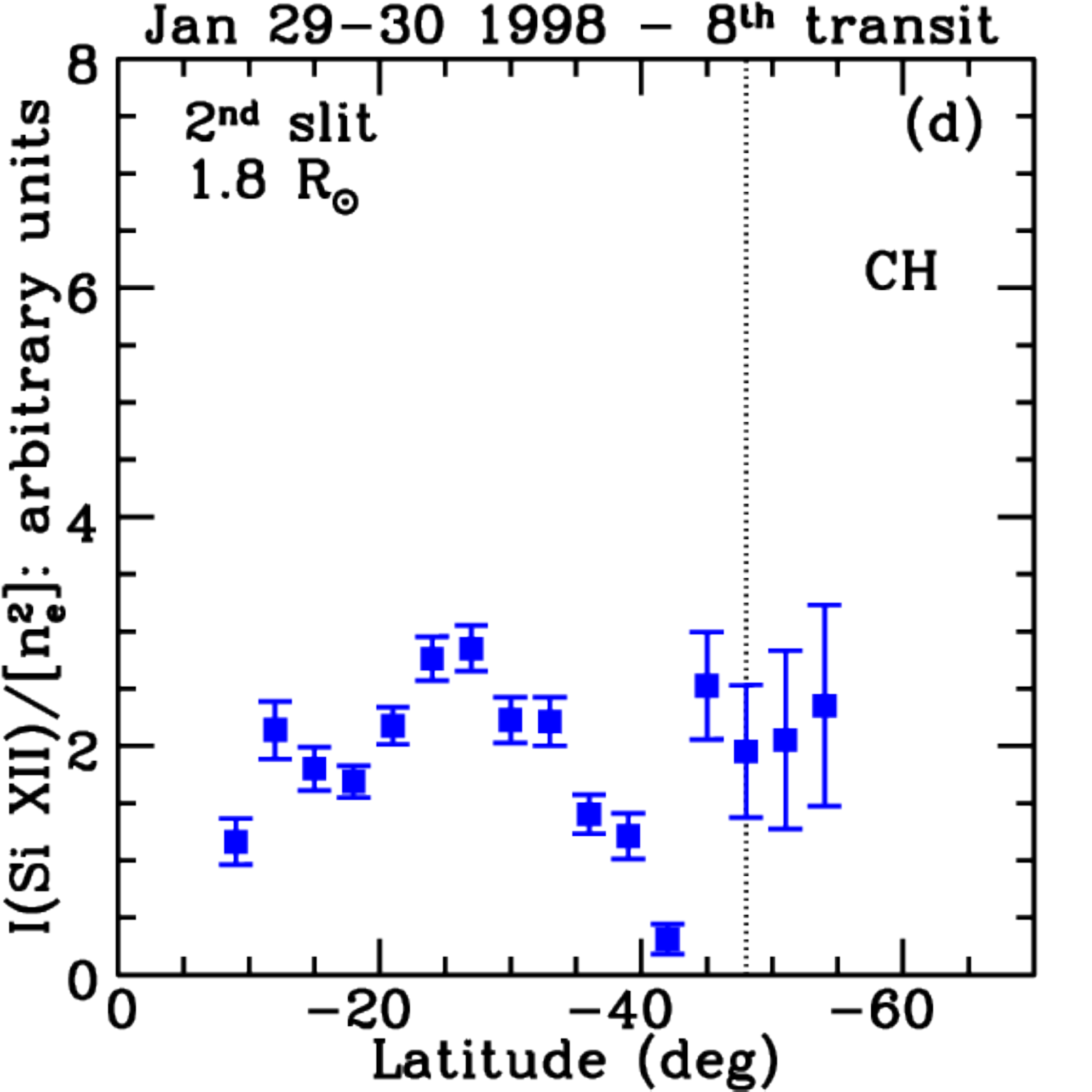}

  \includegraphics[width=4.cm]{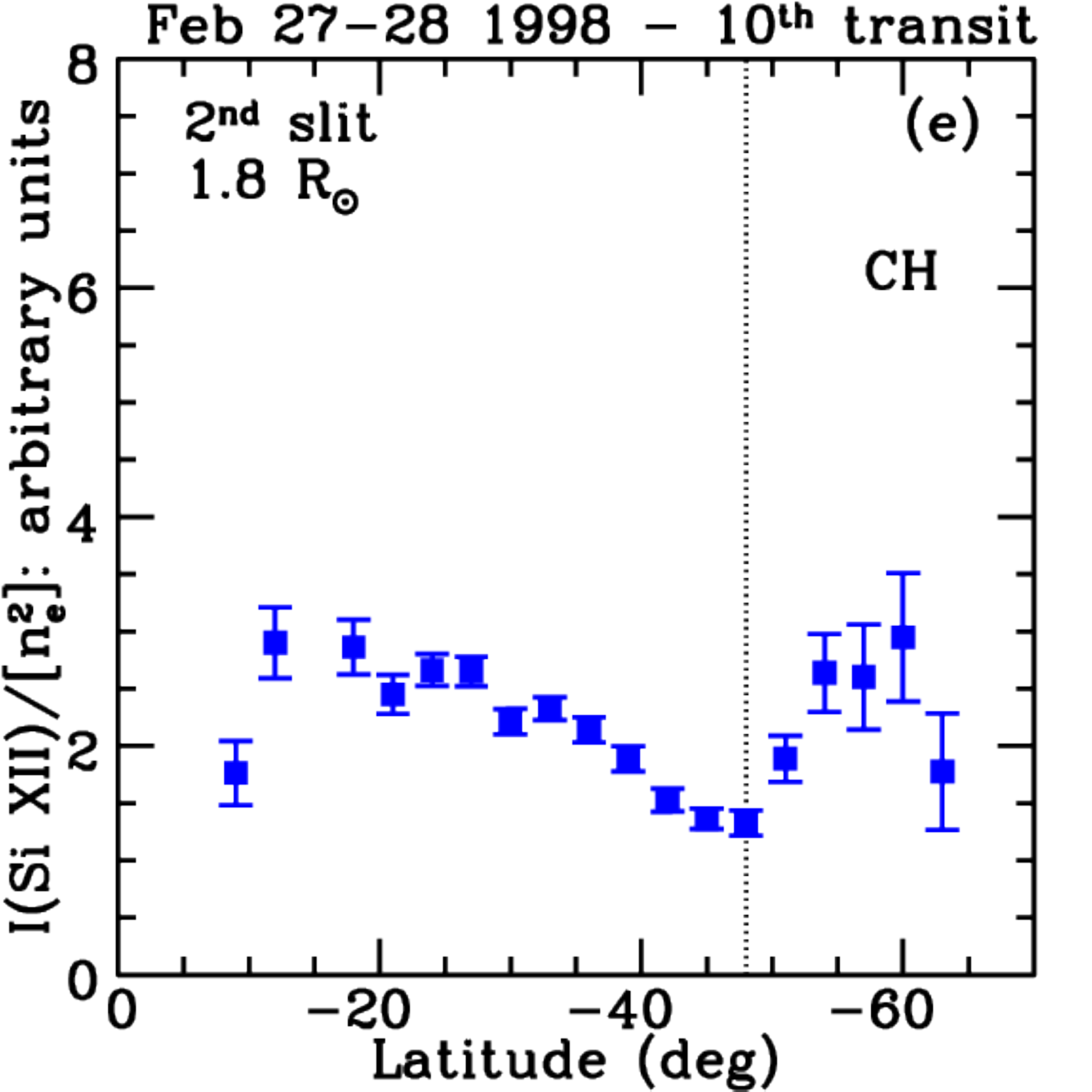}
  \caption{Profile along the slit of the normalized (see Sect.~\ref{sec:fip}) 
    quantity $I({\rm Si~\textsc{xii}})/[n_{\rm e}^2]$, for the limb transits 2, 4, 6, 8 
    and 10. The equator corresponds to $0~\deg$, the south pole to $-90~\deg$. 
    The vertical dotted line defines the boundary with the south polar CH.}
  \label{fig:fip}
\end{figure}

The contour maps of Fig.~\ref{fig:wilcox} describe the distribution and the 
evolution of the surface (photospheric) magnetic field strength of AR~8100 and 
the change in the orientation of the field polarities during Carrington 
rotations (CR) from 1929 to 1933. The data of these magnetograms are taken 
from the synoptic charts published by the Wilcox Solar Observatory 
(http://wso.stanford.edu/synopticl.html). The CH boundaries are also shown in 
Fig.~\ref{fig:wilcox}a; data are taken from the synoptic maps of the 
He~{\sc{I}}~10830\AA\ line equivalent width of the Kitt Peak telescope of the 
National Solar Observatory (ftp://vso.nso.edu/kpvt/synoptic/).

We show in Fig.~\ref{fig:profv} the outflow speeds and densities for every 
limb transit of the AR and in Fig.~\ref{fig:fip} the profiles of 
$I({\rm Si~\textsc{xii}})/[n_{\rm e}^2]$ along the slit (set at position 2) for the AR 
west limb transits. Features for each limb passage are illustrated in 
individual subsection and discussed in Sect.~\ref{sec:discussion}.

\begin{figure}[h!]
  \centering
  \includegraphics[width=4cm]{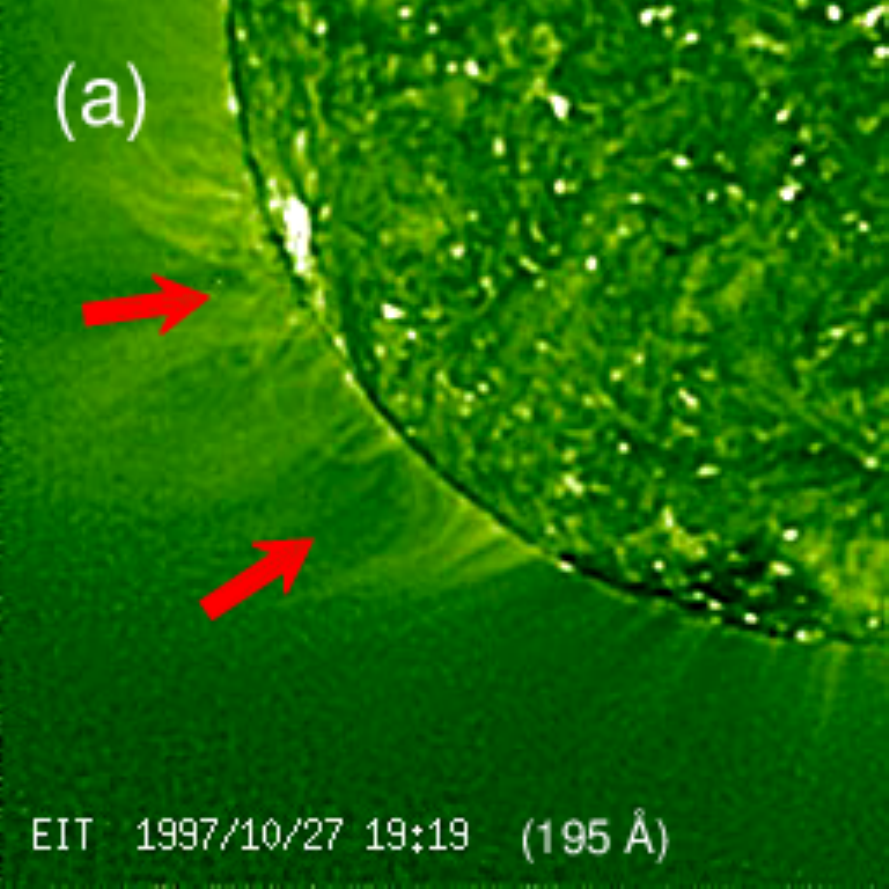}
  \includegraphics[width=4cm]{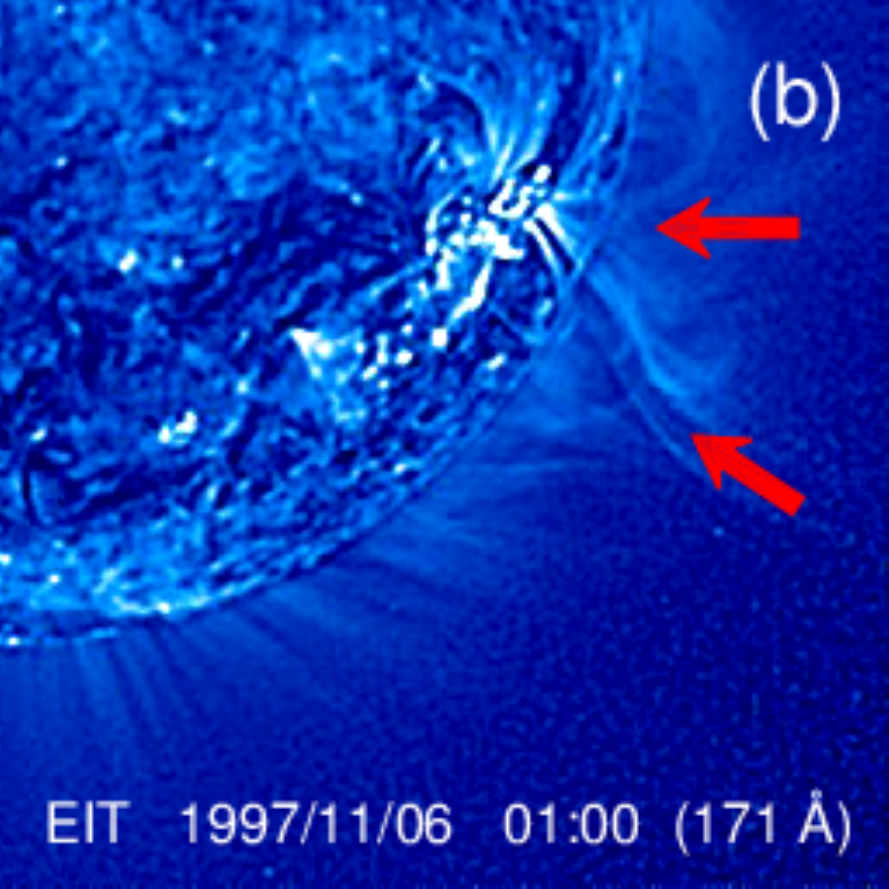}

  \includegraphics[width=4cm]{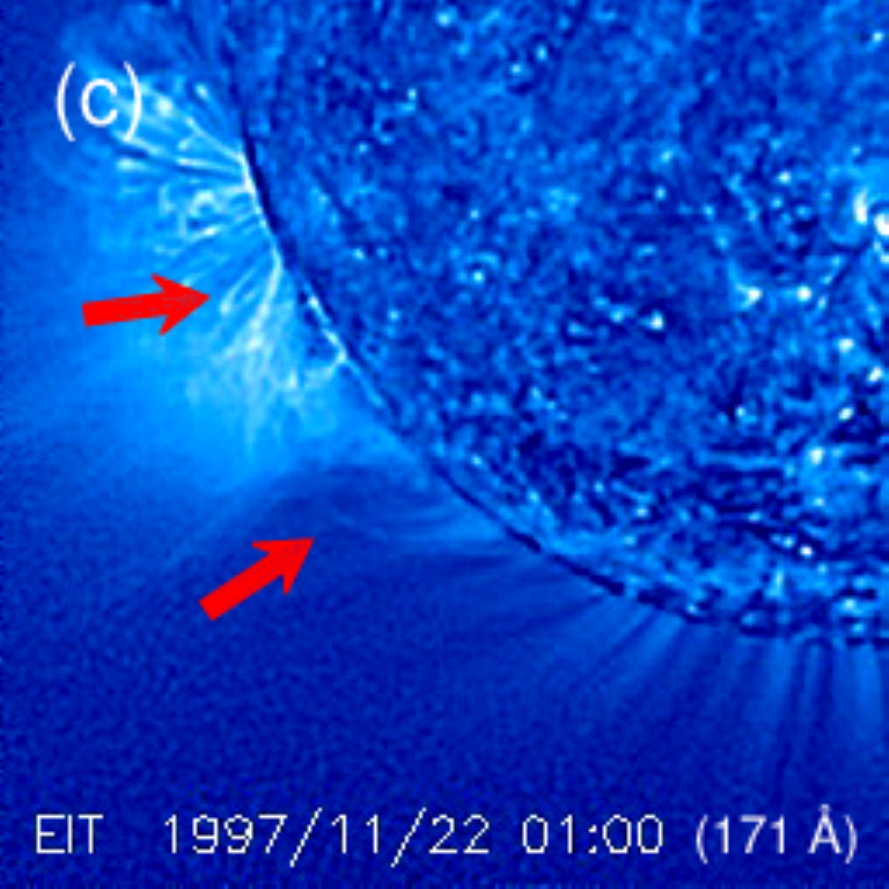}
  \includegraphics[width=4cm]{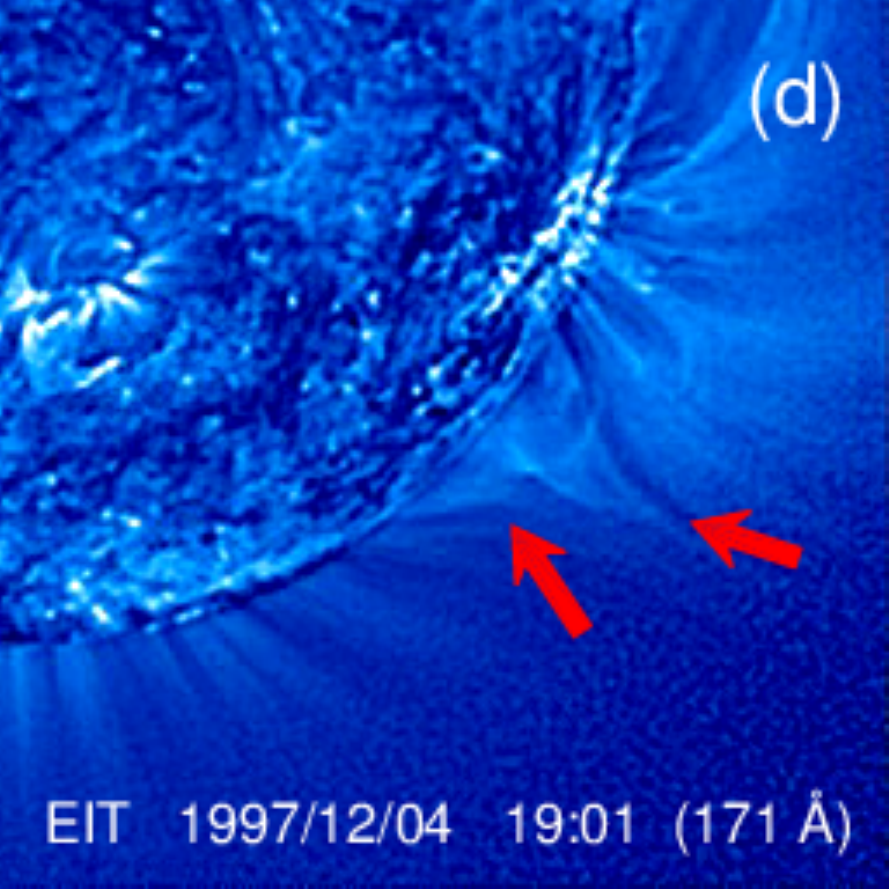}

  \includegraphics[width=4cm]{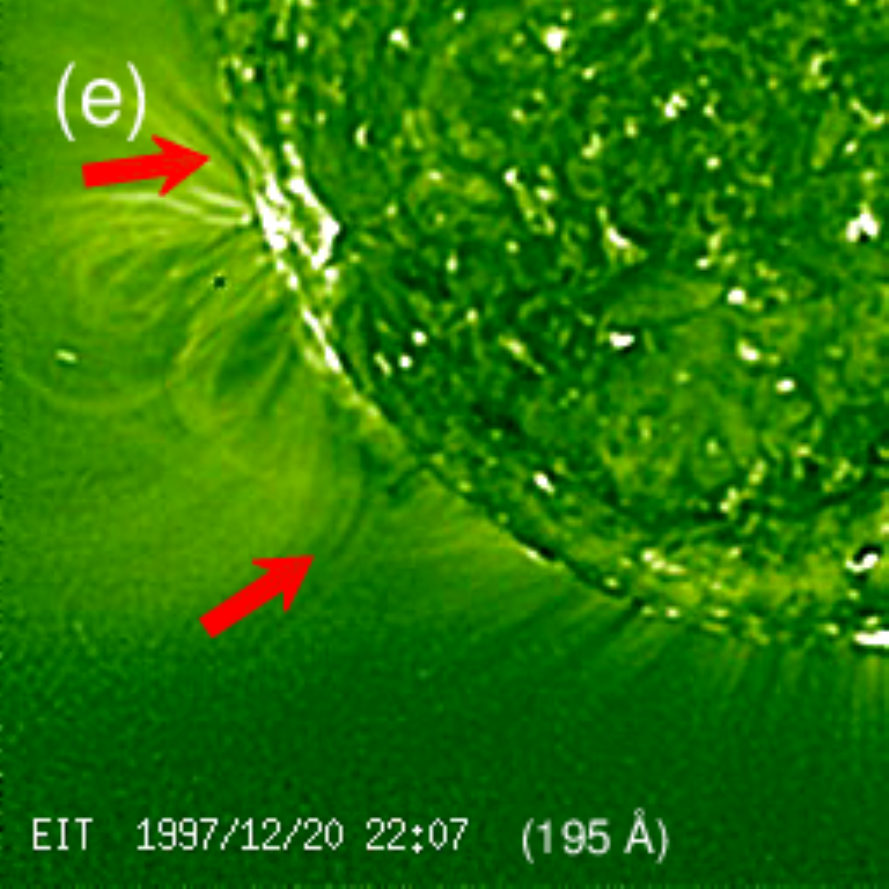}
  \includegraphics[width=4cm]{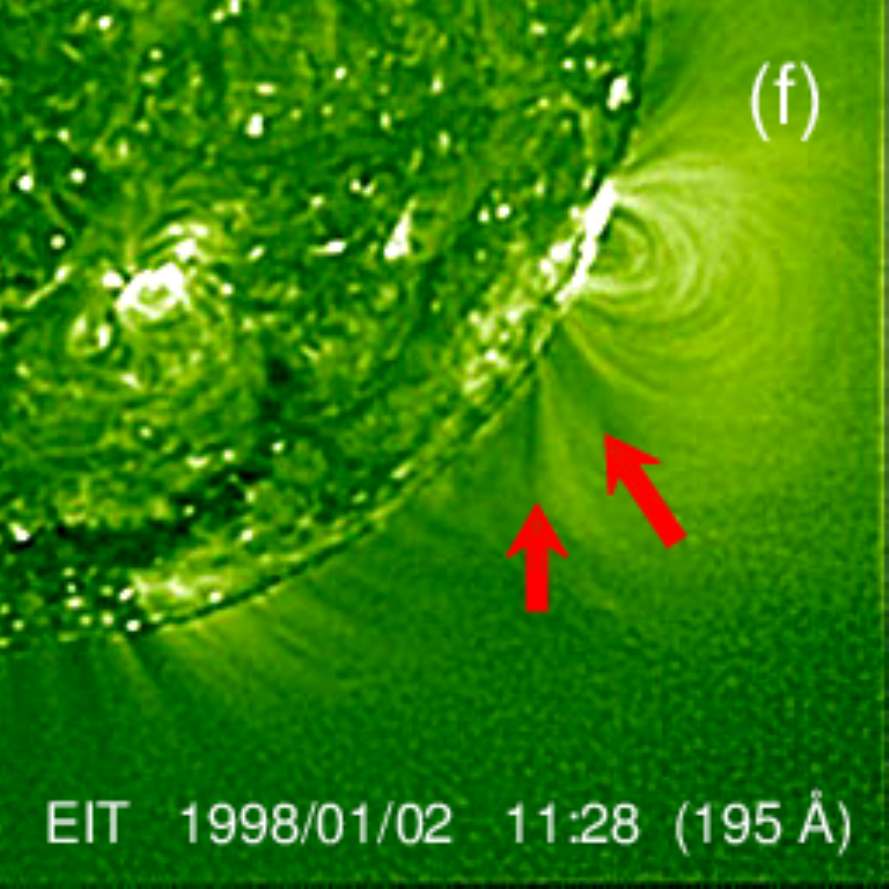}

  \includegraphics[width=4cm]{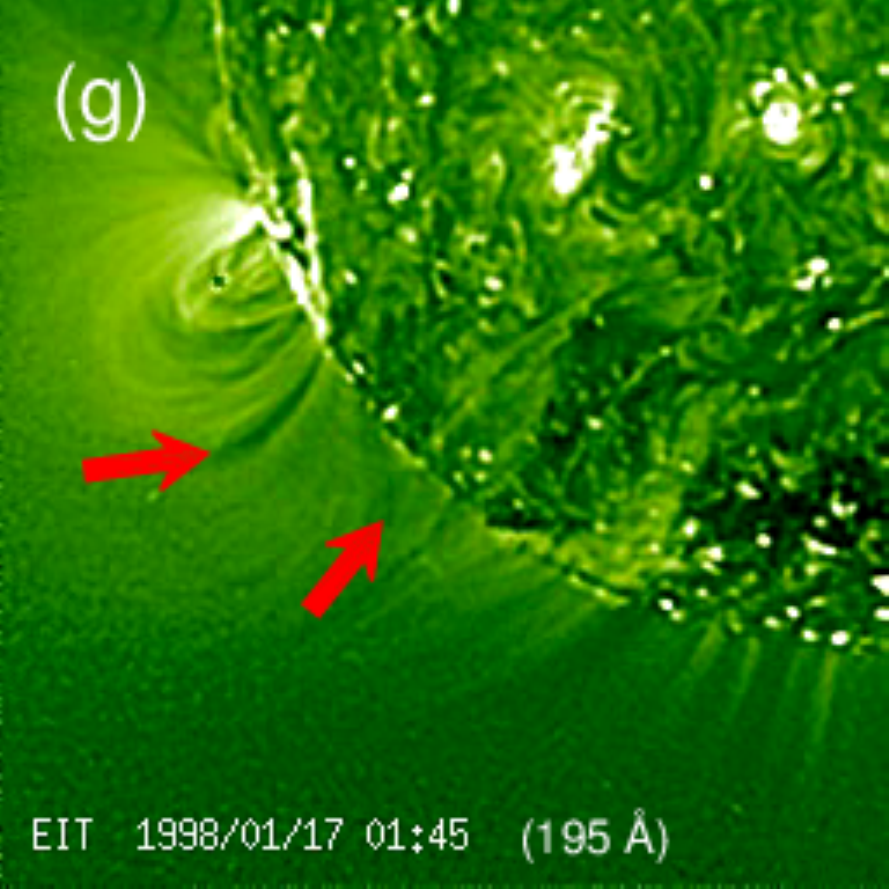}
  \includegraphics[width=4cm]{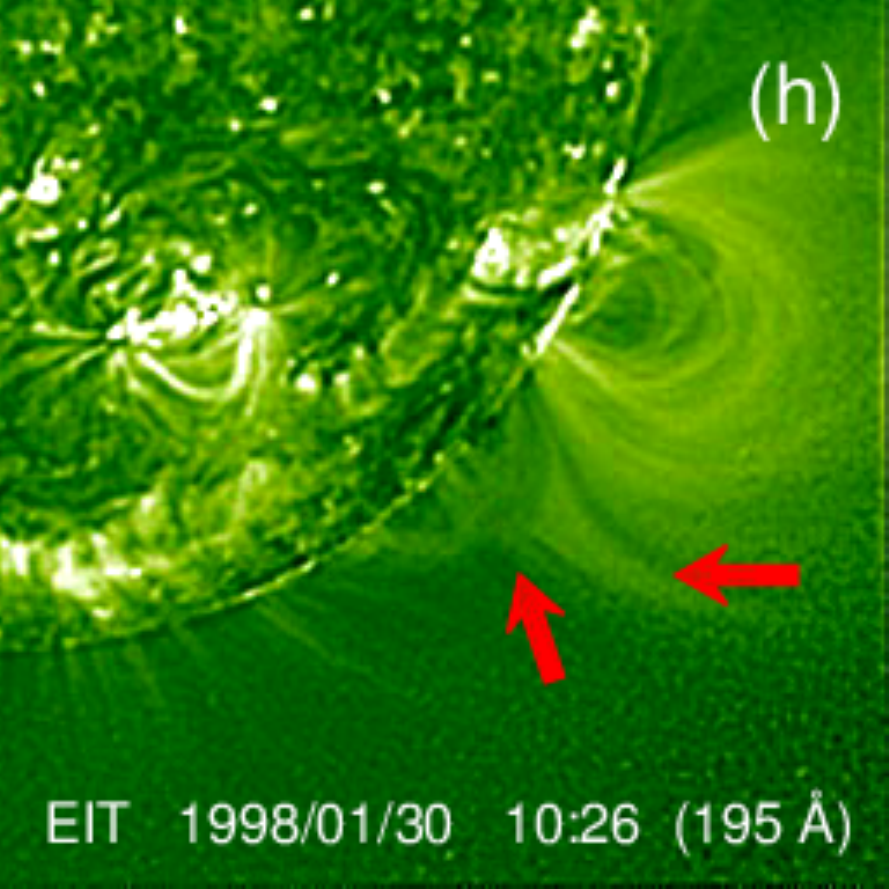}

  \includegraphics[width=4cm]{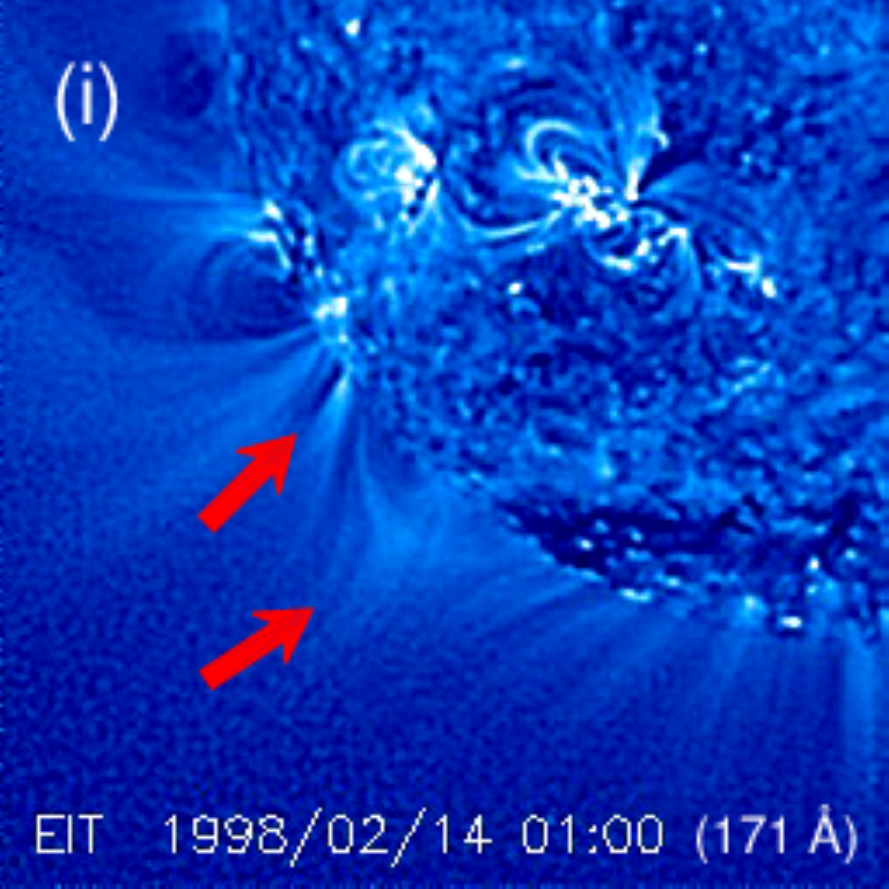}
  \includegraphics[width=4cm]{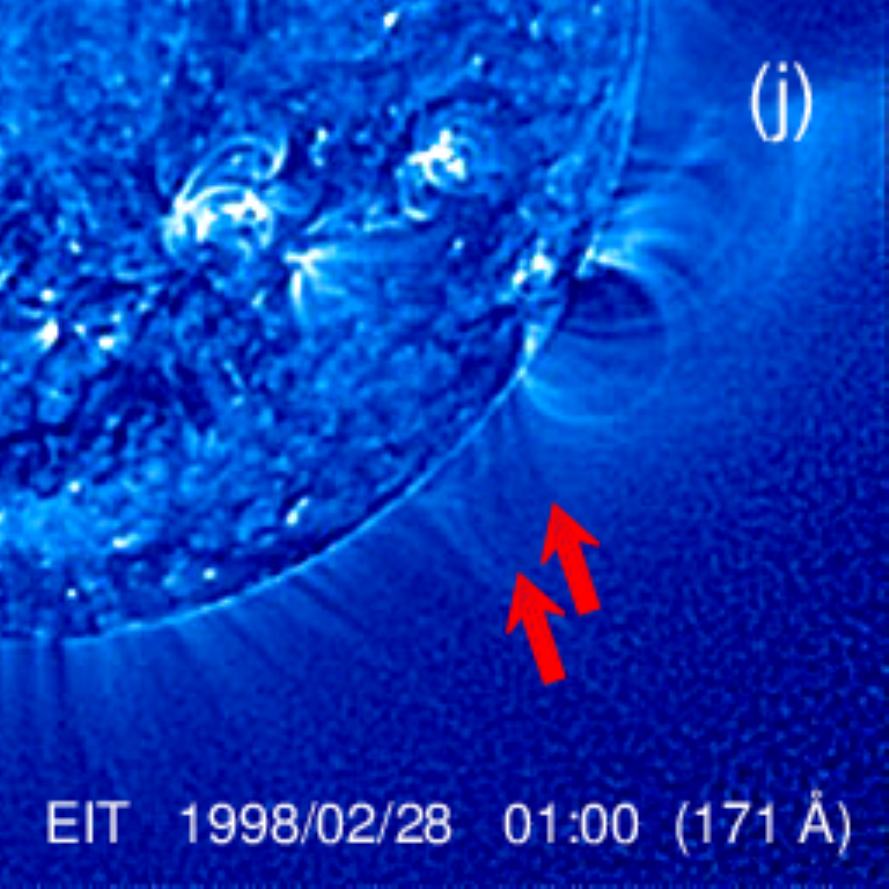}
  \caption{EIT 171\AA\ and 195\AA\ wavelet-enhanced images, acquired during 
    the AR limb transits. In the panels the red arrows point to weak emission 
    regions, which we identify as the low corona counterparts of the low- and 
    intermediate-latitude outflow channels that we infer from the DD analysis 
    of UVCS data. We considered EIT 195\AA\ images when the 171\AA\ channel 
    data were not available. For the second limb transit (panel b), we had to 
    consider images taken on November 6 because no EIT data were available for 
    November 7 and 8.}
  \label{fig:tEIT}
\end{figure}

\subsection{First limb transit}

The results from the DD analysis of the first limb transit, which occurred in 
the SE quadrant on October 26 and 27, 1997, are shown in Fig.~\ref{fig:vmap}a. 
Red dots indicate the location of the UVCS data used in our modeling; magnetic 
field lines obtained from PFSS extrapolations are labeled with different 
colors, gray (closed field lines), green (open), and orange (open, but 
originating from southern latitudes greater than $-60~\deg$ that pertain to 
the south polar CH. Hints of three outflow channels appear in the intermediate 
corona at latitudes $-15$, $-25$, and $-35~\deg$; they are 
identified in panel (a) of Fig.~\ref{fig:vmap} by dips in the contour levels 
of the speed map that occur at different altitudes where bundles of open lines 
are found. They originate in a loop system and in the boundary of the 
active~region and the coronal~hole (AR-CH). Their positions correspond to 
local minima in the $I_{\rm O~{\textsc{VI}}-1031.9}/I_{\rm O~{\textsc{VI}}-1037.6}$ profiles. We 
classified two of them as low-latitude and intermediate-latitude channels: 
these are identified by full squares and full triangles, respectively in the 
map of Fig.~\ref{fig:vmap}a. A third outflow is indicated by open squares and 
apparently bifurcates from the low-latitude outflow at $1.7~{\rm R_\odot}$ to 
then extend closer to the east equator. We note that the intermediate-latitude 
channel is also partially associated with open lines originating from the 
south polar CH. These lines project onto the same areas as those rooted in the 
AR.

Figure~\ref{fig:profv}a shows the speed profiles of the three outflow channels 
we found from the DD analysis. The low-latitude speed profile (solid squares) 
possibly attains values lower than those of the intermediate-latitude profile 
(solid triangles), particularly at distances smaller than $2~{\rm R_\odot}$, 
while at higher altitudes the two profiles reach similar values. However, the 
relatively large errors do not allow us to draw any clear conclusion about 
possible differences in the speed regime between the two outflows. The 
secondary branch of the low-latitude channel (open squares) continues at 
relatively low speed values, also at distances larger than $2~{\rm R_\odot}$.

The low-percentage errors in the calculated electron densities allow us to 
state that the three density profiles (see Fig.~\ref{fig:profv}a) are 
different: in particular, two profiles are intermediate between typical 
streamer and CH profiles, and the third one, pertaining to the 
equatorial region, is more similar to typical streamer values. Moreover, the 
highest latitude channel appears to show a mixture of plasma from the AR and 
the polar CH.

The panels of Fig.~\ref{fig:tEIT} show the west and east side of AR~8100, as 
the AR moves from the east to the west limb passage. Moreover, the gradual 
rotation of the AR polarities occasionally brings different parts of its 
structure into the foreground. These parts appear to consist of three loop 
families. The first system is closest to the equator and consists of slanted 
loops, which are seen neither exactly face-on nor edge-on during the first 
transit. The second family appears at medium latitudes as a fainter and 
apparently higher reaching system than the first. It gradually brightens with 
time and evolves to a face-on orientation owing to the AR rotation, clearly 
revealing a fan-loop structure (see Fig.~\ref{fig:tEIT}e). The third family is 
located at higher southern latitudes and consists of several bright features 
whose geometry is initially not easy to interpret. They are connected in the 
N-S direction.

After the solar rotation dragged the AR close to the central meridian, where 
better magnetic field data could be acquired, multiple episodes of flux 
emergence 
were observed. The AR produced some 35 flares and 25 CMEs at this location 
during its first transit (see \citealt{Green2002a}). Although the high 
variability makes any comparison between extreme ultraviolet (EUV) morphology 
and isocontour maps rather uncertain, we searched EIT images for areas of 
weaker EUV emission within the UVCS FOV because long-lived upflows have been 
recognized from Hinode/EIS analyses to correspond to low-intensity regions 
(e.g., \citealt{Doschek2008a}, \citealt{Del_Zanna2008a}, \citealt{Hara2008a}, 
\citealt{Demoulin2013a}). The EIT~195~\AA\ image in Fig.~\ref{fig:tEIT}a 
clearly shows the occurrence of a weaker emission area in between the first 
and second loop system (latitude $\approx -24~\deg$) and, analogously, of a 
wide gap at the southern boundary of the second system at about $-43~\deg$ 
latitude. These positions nicely correspond to the region from which open 
magnetic field line bundles originate and are then superposed onto the 
channels in the outflow map. The positions of the weaker emission areas that 
we identify in the EIT images of the ten AR limb transits, and that we claim 
to be the low corona counterparts of the outflows we detected in the 
intermediate corona with the DD analysis, are indicated with red arrows in the 
panels of Fig.~\ref{fig:tEIT}.

The clues we collected in the first AR 8100 limb transit that are in favor of 
the occurrence of outflow channels in regions of open field lines and where 
the EIT images show low-density areas, would be corroborated if the following 
limb transits showed a similar behavior. We therefore analyzed successive limb 
passages of AR~8100 to provide further support of the evidence provided so 
far.

\subsection{Second limb transit}
\label{sec:second_t}

During the second AR limb transit, AR~8100 was fairly active: the shearing 
motions have been highlighted by \cite{Green2002a}, and according to the LASCO 
CME catalog (http://cdaw.gsfc.nasa.gov/CME\_list/), on November 8 a CME 
spanning 76 degrees occurred at $PA=271~\deg$ at 08:59~UT, and two CMEs 
occurred on the previous day at PAs $270~\deg$ and $278~\deg$, respectively. 
This shows that the magnetic configuration of the region experienced 
disruptions, although no event is recorded on November 9. We mention that a 
X9.4 flare occurred on November 6 at 11:49~UT, followed by a halo CME at 
12:10~UT, whose source has been identified in the AR~8100 site.

Analogously to the first transit, the outflow speed map in 
Fig.~\ref{fig:vmap}b is characterized by three systems of channels above the 
AR at latitudes $-20$, $-35$, and $-45~\deg$; two of them appear to belong to 
the intermediate-latitude (solid and open triangles) and the third to the 
low-latitude family (solid squares).

\begin{figure*}
  \centering
  \includegraphics[width=8cm]{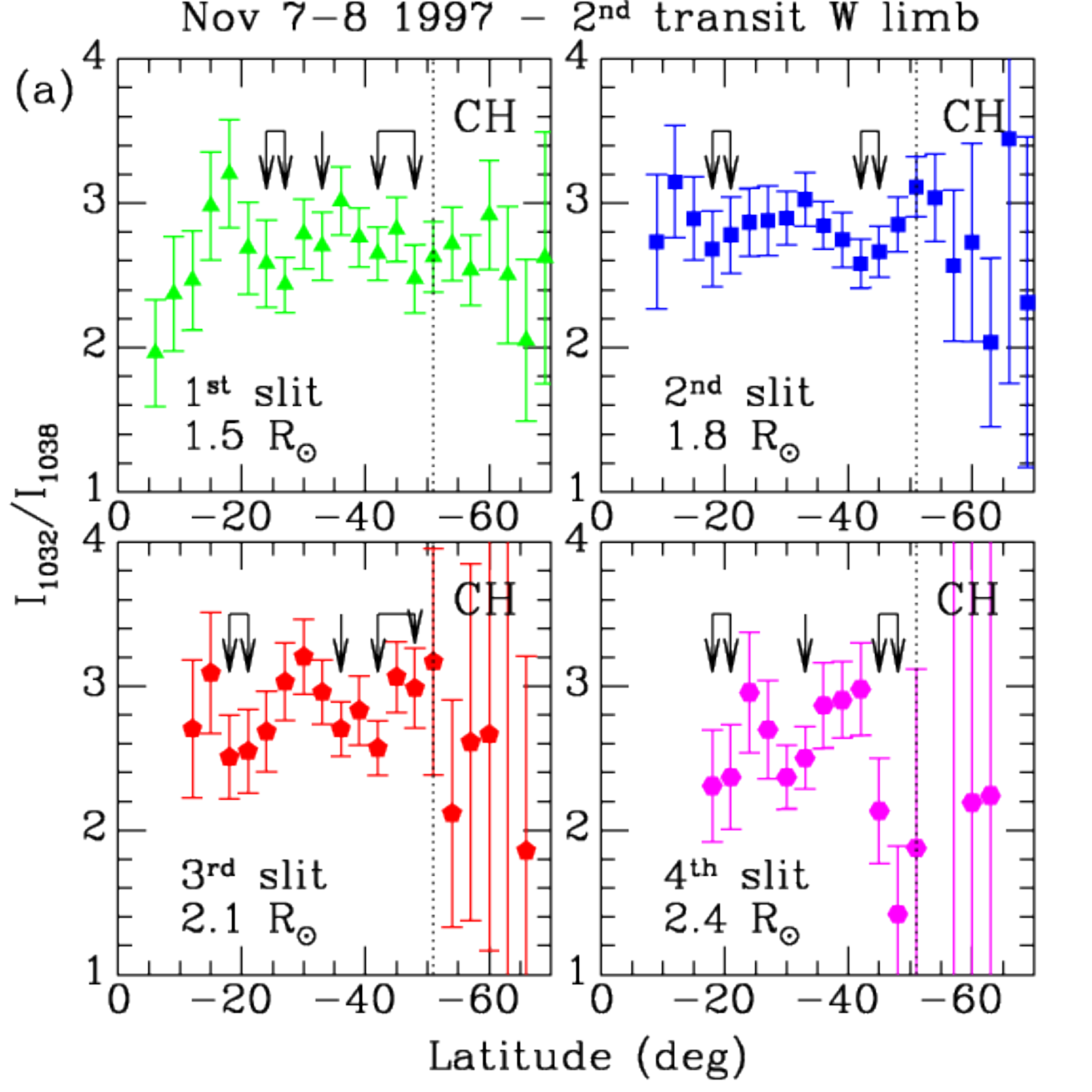}
  \includegraphics[width=8cm]{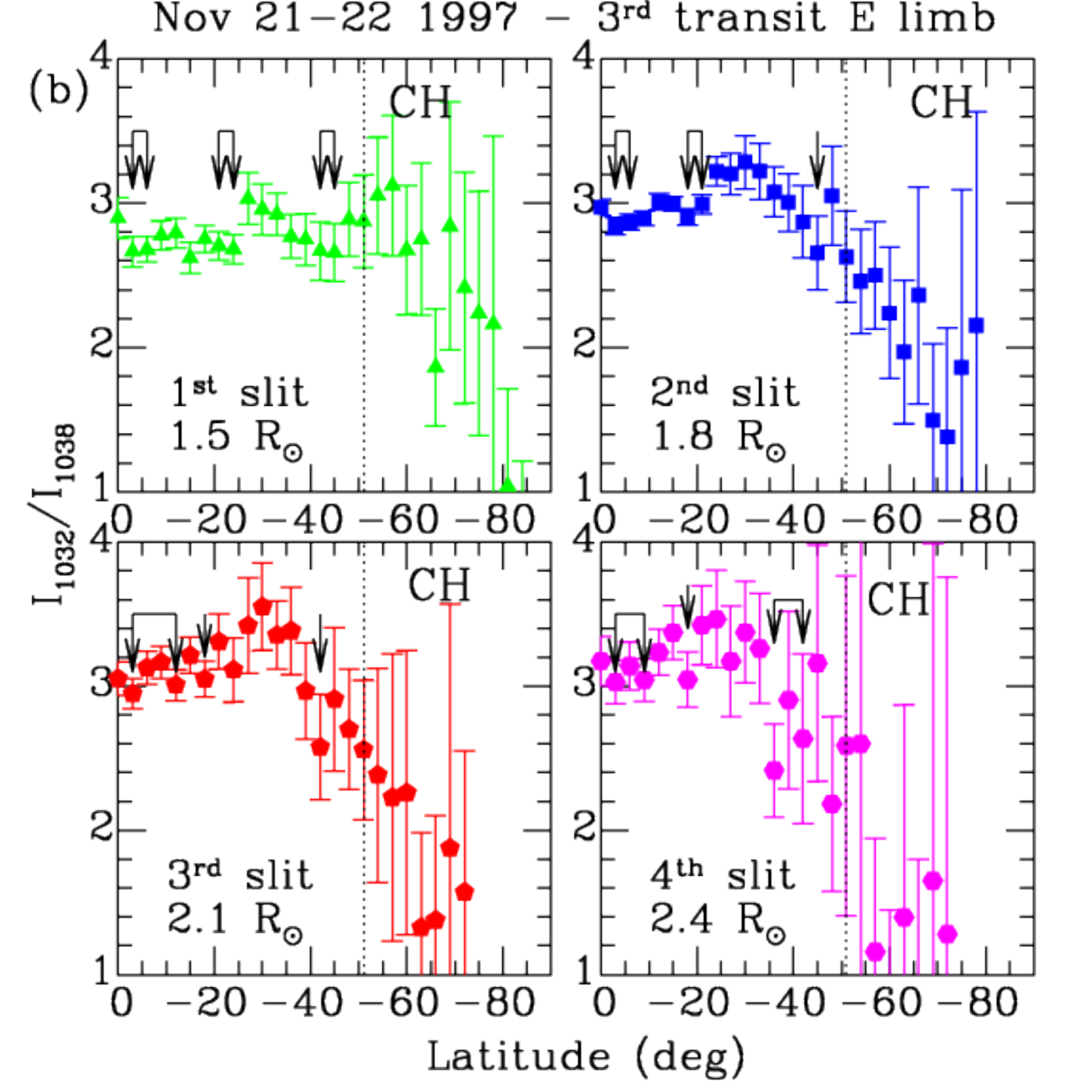}
  \caption{Profiles of the O~{\sc{vi}} doublet intensity ratio along the UVCS 
    slit at four slit positions from data taken at $PA=225~\deg$ 
    (panel a), and $PA=135~\deg$ and $PA=90~\deg$ (panel b) for the second and 
    third limb transit, respectively. The positions of the AR outflows are 
    shown by arrows, and the vertical dotted line defines the boundary with 
    the south polar CH.}
  \label{fig:ovirat}
\end{figure*}

Figure~\ref{fig:profv}b shows the outflow speed and electron density profiles 
of the low- and intermediate-latitude channels: the properties of the outflows 
appear indistinguishable in outflow speed and electron density. 
Figure~\ref{fig:ovirat}a shows the profile of the O~{\sc vi} doublet line 
intensity ratio, $I_{\rm O~{\textsc{VI}}-1031.9}/I_{\rm O~{\textsc{VI}}-1037.6}$, along the UVCS slit at 
the time of the second limb transit of the AR in the four different positions 
of the UVCS slit at $PA=235~\deg$. It is obvious from the plot that the 
O~{\sc vi} line ratio is lowest at the high outflow channels.

The magnetic PFSS extrapolation is characterized by three loop systems with 
open field lines originating far apart on the east side (see 
Fig.~\ref{fig:vmap}b). The open magnetic field lines from the AR site define 
reasonably well the outflow lanes we found on the basis of the criteria we 
established. The most relevant difference with respect to the first transit 
concerns the topology of the open magnetic field lines, which clearly appear 
to originate from a common region inside the AR, as shown in 
Fig.~\ref{fig:vmap}b. The AR magnetic field map for CRs 1928/1929 is also 
presented in Fig.~\ref{fig:wilcox}a, where all the footpoints of the open 
field lines (greenish squares, corresponding to the green open lines of 
Fig.~\ref{fig:vmap}b) lie within the same portion of negative polarity of the 
AR and are far from the AR-CH boundary (gray long-dashed contour).

The configuration of the AR at its second west limb passage is presented in 
Fig.~\ref{fig:tEIT}b, which shows a wavelet-enhanced EIT~171~\AA\ image taken 
on November 6, 1997, one day before the time interval we considered in the DD 
analysis, because no wavelet EIT images are available on November 7-8. 
Figure~\ref{fig:tEIT}b clearly shows a dark lane at the southern latitudes 
$\approx$ $-40$ and $-25~\deg$ (both marked with red arrows), matching the 
positions of the outflow channels and of the open field lines in the PFSS 
extrapolation (see Fig.~\ref{fig:vmap}b).

\begin{figure}
  \centering
  \includegraphics[width=8cm]{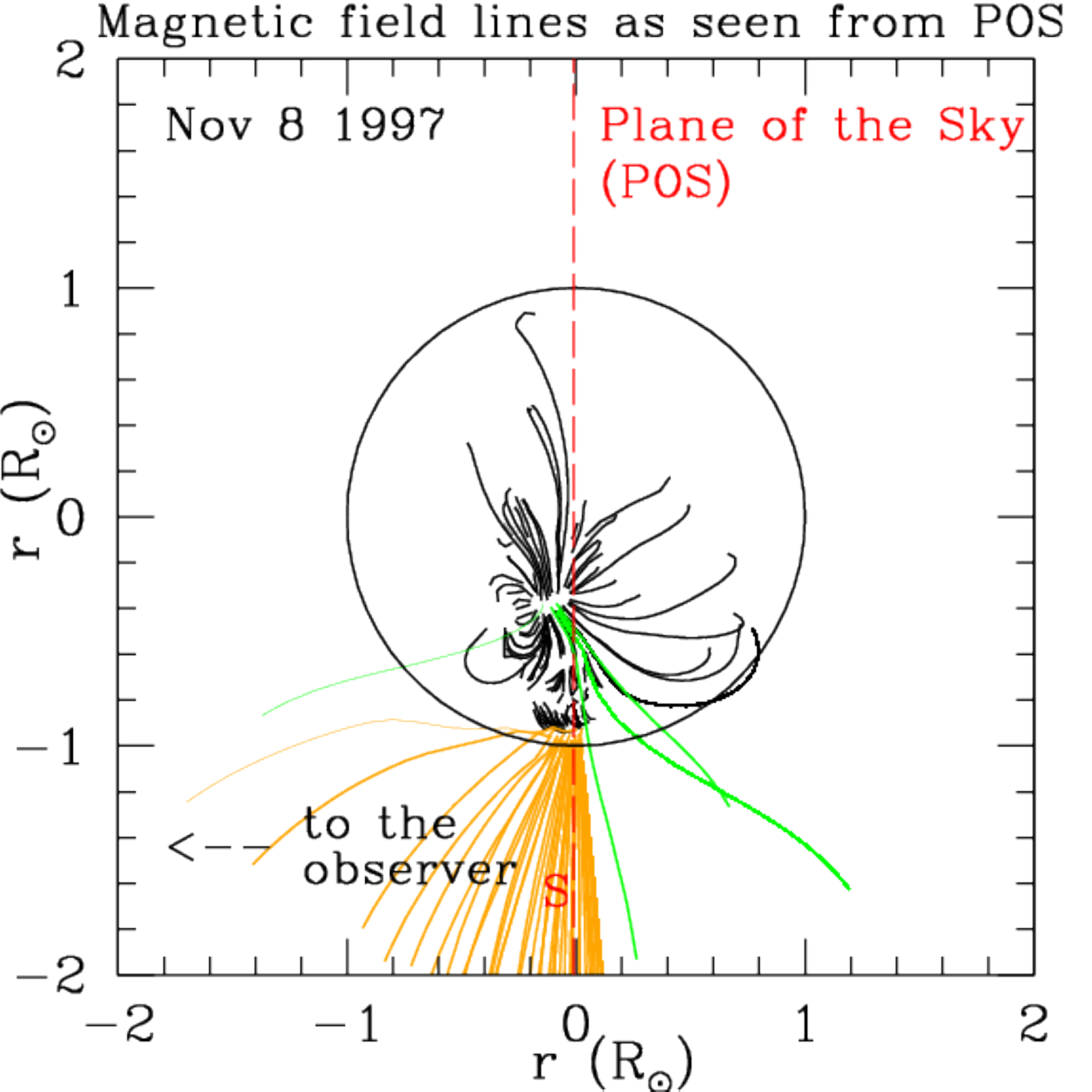}
  \caption{PFSS extrapolation above the AR at the second limb transit, as seen 
    from an observer directly above the AR, on the west limb of the Sun. The 
    arrow points to the position of UVCS at the time of observations.}
  \label{fig:t02_side_view}
\end{figure}

A different view of the field line geometry is shown in 
Fig.~\ref{fig:t02_side_view}, from a vantage point on the west limb of the 
Sun, that is, above the AR at the time it crosses the west limb (the view is 
rotated by $90~\deg$ with respect to that of Fig.~\ref{fig:vmap}b), where the 
green open lines originate from the AR, and the orange lines from the CH. The 
west AR side is characterized by a fan-loop structure, while from the east 
side the closed lines connect the main polarity to the network magnetic field. 
The bundle of open field lines that have been shown to overlie the outflow 
channels inferred from the DD analysis in corona originates in between these 
two loop structures. Recent studies revealed that fan-loop structures are 
often associated with quasi-separatrix layers (QSLs; see, e.g., 
\citealt{Schrijver2010a}). \cite{Baker2009a} identified the QSLs that separate 
closed field lines, where the magnetic field connectivity experiences a 
drastic change, as regions from which AR outflows may originate. Although 
detailed calculations are required to find the possible location of QSLs in 
the AR magnetic field topology, the magnetic field structure shown in 
Fig.~\ref{fig:t02_side_view} indeed suggests the occurrence of QSLs at the 
border separating the open magnetic field lines from the fan-loop system, 
hence enforcing our claim of having identified AR outflows in the intermediate 
corona.

Panel a of Fig.~\ref{fig:fip} shows the profile of 
$I({\rm Si~\textsc{xii}})/[n_{\rm e}^2]$ across the UVCS slit at the time of the second 
limb transit of the AR and an altitude of $1.8~{\rm R_\odot}$: the profile shows 
two peaks at latitudes $\approx -25$ and $\approx -50~\deg$, which coincide 
with the positions of the two main outflows identified in the DD analysis (see 
Fig.~\ref{fig:vmap}b). Under the hypothesis that the electron temperature, 
hence the function $f(T_{\rm e})$, is constant along the UVCS slit (see the 
discussion in Sect.~\ref{sec:fip}), the variation in the ratio 
$I({\rm Si~\textsc{xii}})/[n_{\rm e}^2]$ can be explained by invoking a silicon 
enrichment in the coronal plasma at the positions of the two peaks, which is 
where we detected slow wind outflows. This interpretation is also valid when 
the temperature in the channel is lower than the temperature of adjacent 
regions because $f(T_{\rm e,channel}) < f(T_{\rm e,AR})$ (according to the 
ionization balance calculations given in the literature, see, e.g., 
\citealt{Arnaud1985a}, we have the maximum of Si~{\sc xii} fraction for 
$T_{\rm e}\approx 2\times 10^6~{\rm K}$, which is above the electron 
temperature regime we expect at these altitudes; see the discussion in 
Sect.~\ref{sec:discussion}) and a higher Si abundance would be required to 
account for our results. This means that the plot would be consistent with the 
enrichment of the low FIP elements found by other authors in AR outflows, as 
mentioned in Sect.~\ref{sec:intro}. At the altitude we examine, however, the 
Si~{\sc xii} ion is very likey frozen-in, and we have no means to know the 
temperatures at which this phenomenon occurs in the channels and in the AR. 
Because (see Sect.~\ref{sec:fip}) the Si~{\sc xii} abundance is extremely 
sensitive to temperature variations, we cannot rule out the hypothesis that 
our results have been dictated by different temperature regimes along the UVCS 
slit and not by changes in the Si abundance.

\subsection{Third and fourth limb transits}
\label{sec:third_t}

The third limb transit of the AR occurred on November 21-22, 1997, when the AR 
was still very active and changing features appeared at different times. The 
shearing motion of the AR gradually brought the negative polarity closer to 
the south polar CH (see Fig.~\ref{fig:wilcox}b), which is the reason for the 
edge-on view of some of the fan-loop system that makes up the western part of 
the AR. In the EIT image shown in Fig.~\ref{fig:tEIT}c, the three-loop 
configuration is now less evident, and the dark lanes are probably partially 
covered by the large central loop system. Nonetheless, in Fig.~\ref{fig:vmap}c 
plasma regions at relatively high outflow speed are distinguishable at 
latitudes $-20$ and $-40~\deg$, supporting results obtained in the preceding 
limb transits. Another somewhat less evident outflow can be distinguished just 
below the east equator, associated with a few open lines. It reaches latitudes 
of about $-10~\deg$. The $I_{\rm O~{\textsc{VI}}-1031.9}/I_{\rm O~{\textsc{VI}}-1037.6}$ 
profiles in Fig.~\ref{fig:ovirat}b confirm the system of three outflows from 
the AR (two low-latitude and one intermediate-latitude channels), at their 
positions derived from the map in Fig.~\ref{fig:vmap}c.

\cite{Green2002a} made linear force-free extrapolations of the coronal field 
to match low-lying Yohokh SXT loops for five rotations of AR~8100. In their 
Fig.~5 the authors present the extrapolation for November 29, 1997 and show 
that an $\alpha$ value of 0.94 ($10^{-2}Mm^{-1}$) best reproduces the X-ray 
Yohkoh loops. Although obviously different from the potential field we 
computed, the three-loop system is still apparent in the \cite{Green2002a} 
extrapolation, which supports the results of our technique, which compares UV 
features at high heliocentric altitudes with potential coronal fields.

The fourth limb transit of the AR, on December 3-4, 1997, occurred at a time 
when its configuration was quite clear in EIT maps. Figure~\ref{fig:tEIT}d 
shows the EIT~171~\AA\ map on December 4, 1997, where the three-loop system 
configuration is clearly visible. It is obvious that the AR consists of a 
northern, southern, and central loop system. The central system originates in 
the western part of the AR and is now seen edge-on, leaving a clearly defined 
dark lane from the site of the AR in the foreground and a less evident lane 
from the AR-CH boundary. Two bright ribbons, oriented along the N-S direction, 
whose separation increases with increasing latitude, are easily distinguished 
in the EIT image. The dark lane seems to originate from these two ribbons. 
They are visible in the first two west transits in Figs.~\ref{fig:tEIT}b and d 
and are probably covered by the central loops when they are at the east limb.

The coronal PFSS extrapolation for the same day (see Fig.~\ref{fig:vmap}d) 
shows open fied lines originating from latitude $-25~\deg$ and from the 
boundary of the AR with the south CH, and extending in the intermediate corona 
approximately along the radial directions at $-25$ and $-35~\deg$. A partial 
overlay of open magnetic lines from the south CH with those coming from the AR 
affects the intermediate-latitude outflow. The agreement between the 
occurrence of the obvious dark lanes in the EIT image with the high-speed 
lanes in the UVCS outflow speed maps and the occurrence of open field lines in 
Fig.~\ref{fig:vmap}d appears here more strongly than in previous limb 
transits.

The corresponding magnetic field configuration is detected in the Wilcox Solar 
Observatory synoptic map for CR 1930, where the northernmost polarities of the 
AR are adjacent (see Fig.~{\ref{fig:wilcox}b). The map also accounts for the 
occurrence of large-scale trans-equatorial loops joining the south AR we 
analyzed and the AR at the limb in the northern hemisphere, which is partially 
visible in the EIT map. This configuration results in face-on transequatorial 
loops and edge-on central loops, separated by darker lanes, giving us a good 
opportunity of identifying the source region of high-speed lanes.

The profile of the ratio $I({\rm Si~\textsc{xii}})/[n_{\rm e}^2]$ during the fourth 
transit is shown in Fig.~\ref{fig:fip}b, where we can recognize two peaks at 
$-20$ and $-30~\deg$, more or less corresponding to the channel positions in 
Fig.~\ref{fig:vmap}d, and a higher maximum between $-40$ and $-45~\deg$, which 
corresponds to open lines from latitudes higher than $-60~\deg$, however. We 
must add that a CME occurred on December 5, 08:27~UT, $PA=275~\deg$, spanning 
98 degrees, and PFSS extrapolation might not accurately represent the magnetic 
field topology during the fourth limb transit.

\subsection{Fifth and sixth limb transit}

The fifth passage occurred on December 19-20, 1997, when the AR configuration 
appeared to evolve toward a morphology consisting of only two loop systems, as 
can be deduced from Fig.~{\ref{fig:tEIT}e. This is probably to be ascribed to 
the AR progressive rotation, which oriented the central loop system to a more 
face-on angle. In addition to the evident channels at $-32$ and $-45~\deg$, a 
hint of a third channel is visible in Fig.~\ref{fig:vmap}e at latitude 
$-25~\deg$, which is probably associated with positive open lines from the 
northeast side of the AR and is plotted as red lines in the same figure. In 
the EIT images the two ribbons of the AR, which during the fourth transit 
where N-S oriented, now appear to rotate (see Fig.~\ref{fig:tEIT}e), following 
the AR polarity evolution (see also Fig.~\ref{fig:wilcox}c, showing the AR 
polarity rotation).

The sixth AR limb transit occurred on January 2, 1998, and has been the 
subject of the \cite{Zangrilli2012a} paper. We refer to this paper for 
detailed information on that limb transit. Here it suffices to recall that the 
authors concluded that a narrow channel at a latitude of $-45~\deg$ was the 
site of high-speed outflows, originating from the southernmost edge of the AR. 
This is visible in Fig.~\ref{fig:vmap}f (solid triangles), where we also show 
a second intermediate-latitude outflow from the AR (open triangles at 
$-35~\deg$) and a low-latitude outflow, between $-20$ and $-30~\deg$ (solid 
squares). The three channels are associated with three distinct bundles of 
open lines, two originating from latitudes between $-30$ and $-40~\deg$ and 
one from the AR-CH boundary. We point out that in the $-45~\deg$ channel open 
field lines both from the AR and from the south polar CH contribute to the 
observed outflow, analogously to the outflows we found in similar 
circumstances in other limb transits. We recall that the speed map in the 
present work has been made by averaging over a latitudinal width of three 
degrees, while \cite{Zangrilli2012a} averaged over the original data to obtain 
a constant spatial resolution of 105 arcsec along the slit for all the four 
slits of the dataset. This may account for minor differences between plots in 
this and the previous paper.

The configuration of the AR in previous transits and its configuration in this 
passage is changed remarkably: this is illustrated in Fig.~\ref{fig:wilcox}c, 
where the WSO magnetogram of AR~8100 for CR~1931 shows the AR polarity 
rotation at its early stages, and Fig.~\ref{fig:tEIT}f, where we show a large 
face-on loop system, with little evidence of the loop families imaged in 
earlier transits.

In Fig.~\ref{fig:fip}c we show two maxima in the 
$I({\rm Si~\textsc{xii}})/[n_{\rm e}^2]$ profile, as expected on the basis of the 
well-defined low-latitude outflow, which is now quite distinct from the 
channel visible at $-45~\deg$. A small peak at $-38~\deg$ possibly accounts 
for the third outflow, as shown in the speed map of Fig.~\ref{fig:vmap}f.

\subsection{Seventh and eighth limb transits}

The seventh and eighth AR limb transits occurred on January 16-17 and 29-30, 
1998, respectively. The AR configuration is similar to that of the preceding 
sixth transit (see Figs.~{\ref{fig:tEIT}g and h), and it is dictated by the 
E-W alignement of the neutral line, as shown in Fig.~\ref{fig:wilcox}d. 
Figure~\ref{fig:tEIT}h shows that the AR ribbons now appear almost E-W 
oriented. This is an especially quiet time for AR~8100: no flare originated in 
the region until January 22. The outflow speed maps are shown in 
Figs.~\ref{fig:vmap}g and h.

In the seventh transit we probably have only a low-latitude channel (the 
intermediate-latitude channel is largely covered by open lines from the south 
polar CH). The red positive open lines, partially visible in the equatorial 
side of Fig.~\ref{fig:vmap}g, pertain to AR~8141, which is in the northern 
hemisphere, and possibly account for the humps and dips in the speed map at 
equatorial latitudes. In the eighth passage the intermediate-latitude outflow 
is perhaps recognizable again, and only partially to be ascribed to open lines 
originating from the south polar CH, as demonstrated by the PFSS extrapolation 
in Fig.~\ref{fig:vmap}h. This also agrees with the absence of a clear 
intermediate-latitude peak of the ratio $I({\rm Si~\textsc{xii}})/[n_{\rm e}^2]$ in 
Fig.~\ref{fig:fip}d; analogously, the low-latitude maximum is barely 
distinguishable in a globally flattened profile.

\subsection{Last two transits}

The last two transits of AR~8100 occurred on 13-14 and 27-28 February, 1998. 
After that time, the dispersal of the magnetic field in the network between 
growing nearby young ARs, AR~8162 on the east side and AR~8156 on the west 
side, makes AR~8100 barely identifiable (see Fig.~\ref{fig:wilcox}e). In 
Figs.~\ref{fig:vmap}i and j the outflow maps pertaining to transits 9 and 10 
are shown. The last period of the AR life is characterized by two remarkable 
facts: the flattening of the speed profiles in both the low- and 
intermediate-latitude outflows, and the flattening of the 
$I({\rm Si~\textsc{xii}})/[n_{\rm e}^2]$ profile (see Fig.~\ref{fig:fip}e). Although 
the intermediate-latitude outflow still persists (two channels can be 
distinguished), the magnetic field of the AR negative polarity is weakening, 
as demonstrated by the map shown in Fig.~\ref{fig:wilcox}e.

\section{Discussion and conclusions}
\label{sec:discussion}

The purpose of this work is to analyze the evolution of AR outflows detected 
by the SOHO/UVCS spectrometer in the intermediate corona throughout the 
lifetime of NOAA~AR~8100. The observations covered a period of four months 
from the end of October 1997 to the end of February 1998. Previous works about 
AR outflows evolution (e.g., \citealt{Demoulin2013a}; \citealt{Culhane2014a}) 
mainly focused on the time interval of the AR solar disk crossing, or took 
into account only a single AR limb transit (\citealt{Zangrilli2012a}).

Our present study provides direct evidence that channels of coronal outflow 
originate from AR~8100, which appear at the very beginning of the AR life, and 
persist throughout the entire AR evolution. We observed at least two spatially 
distinct types of outflow channels: low-latitude outflows, associated with 
bundles of open magnetic field lines radially extending roughly from within 
the AR, and intermediate-latitude outflows, which, although detected close to 
the CH boundary, are often clearly associated with open field lines whose 
footpoints lie in the AR or in its immediate neighborhood. The evolution of 
the morphology and outflow locations in the corona is illustrated in 
Fig.~\ref{fig:vmap}, from which we conclude that a two-channel pattern roughly 
persists throughout the whole AR evolution, with minor changes in the 
latitudes at which the channels appear.

Our data do not allow us to show how far the outflows can be identified: do 
they really reach outer interplanetary space? ACE slow wind observations 
confirm what we proved out to $\approx 2.5~{\rm R_\odot}$: AR~8100, during 
CR~1929, has been recognized as source of low-speed solar wind and as a site 
of open magnetic field lines by \cite{Rust2008a} (see the synoptic chart shown 
in Fig.~6 therein), who traced back the slow wind observed by ACE at 1~AU to 
AR~8100.

The speed vs. heliocentric distance profiles in the channels are shown for 
each limb transit in Fig.~\ref{fig:profv}. The speed profiles appear 
relatively flat early in the AR life; conversely, during the central part of 
the AR evolution, they reveal a more pronounced acceleration and finally tend 
to level off with the AR age. Both outflow speed and electron density appear 
to have different profiles in the two channels in several transits of the AR, 
as shown in Fig.~\ref{fig:profv}. Electron densities in the low-latitude 
channels often tend to be higher than densities in the high-latitude channel, 
while the opposite occurs for the outflow speed profiles.

However, the exceptions represented by the second and possibly, although with 
a lower degree of evidence, by the first limb passage (see 
Figs.~\ref{fig:profv}a and b), where the density profiles are almost 
indistinguishable, lead us to hypothesize that in some case, data may have 
been contaminated at intermediate latitudes by plasma from the south polar CH. 
It is difficult to estimate how large this contamination might be, hence to 
draw quantitative conclusions about physical differences between the two types 
of outflow channels, that, as we pointed out, are likely to share a common 
origin (see, e.g., the PFSS extrapolations).

As we discussed in Sect.~\ref{sec:analysis}, in some cases (e.g., during the 
second limb transit) the PFSS extrapolation suggested that the two 
types of outflows we found (low-latitude and intermediate-latitude) might 
originate from a single portion of the AR, as some of the open field lines 
from the AR sharply bend toward the south polar CH boundary and then expand 
radially at the locations of intermediate latitude outflow channels. In some 
cases, open field lines from the south polar CH are clearly detected, for 
instance, during limb transits 6 and 8, which project onto the plane of the 
sky above the intermediate-latitude channel. The speed and density values 
deduced from the DD analysis in these cases might be an average of the two 
outflow regimes, that from the AR and that from the CH. It is possible that 
the south polar CH open field lines physically come close together in the open 
bundles from the AR, as during the AR evolution the magnetic field topology 
changes and the AR rotation brings the CH boundary closer to the area of the 
AR from which the open field lines originate.

In this context, it is interesting to note that when in the 
intermediate-latitude outflow channels there is less superposition of CH and 
AR open field lines, we find that the outflow speed and density profiles 
appear indistinguishable from the low-latitude channels (see, e.g., 
Fig.~\ref{fig:profv}b, which summarizes the results of the second limb 
transit). This result supports the hypothesis that all outflow channels we 
found have the same origin in the AR, and differences in the speed profiles 
originate from a contamination with CH outflows.

As we mentioned in Sect.~\ref{sec:second_t}, the AR structure from SOHO/EIT 
images and from its magnetic field topology, as inferred from PFSS 
extrapolation, suggest that the origin of the outflows we detect in the 
intermediated corona is inside the negative AR polarity, at the boundary with 
a fan-loop system. This configuration is similar to other cases described in 
the literature (see \citealt{Schrijver2010a}, \citealt{Baker2009a}), in which 
QSLs in the vicinity of fan-loops and open magnetic field lines have been 
associated with AR outflows detected in the low corona.

A direct comparison of the speed values we derived here with measurements of 
AR outflow speeds in the literature cannot be done because we analyzed data at 
intermediate corona altitudes, whose speed regime has only been probed by UVCS 
with the DD technique. On average, the speed profiles radially accelerate from 
about $50$ up to $150~{\rm km/s}$ from $1.5$ to $2.5~{\rm R_\odot}$. This 
suggests that the AR~outflows we observed have initial speeds lower than 
$50~{\rm km/s}$, which is lower than the values we could expect for the 
high-speed component seen by Hinode/EIS (see \citealt{Tian2011b}) at lower 
altitudes. We may fail to observe the high-velocity component described by 
Tian et al. either because the high-speed AR outflows are episodic, transient, 
or quasi-periodic (see \citealt{De_Pontieu2010a}, \citealt{Tian2011a}) and 
UVCS only measures an average speed value over a two-day interval or because 
the three-degree spatial averaging we adopted is too coarse and they are 
smeared out.

The purpose of the analysis of the second-order Si~{\sc{xii}}~520.7~\AA\ line 
was to determine whether it might support the hypothesis that the outflows 
detected by UVCS in the intermediate corona are a component of the slow wind. 
The profiles of the quantity $I({\rm Si~\textsc{xii}})/[n_e^2]$ along the UVCS slit, shown in 
Fig.~\ref{fig:fip}, are characterized, especially during the early stages of 
the AR life, by two peaks at latitudes that fairly well match those of the AR 
outflows. This seemingly supports our claim; the strong dependence of the 
Si~{\sc xii} abundance on temperature and the lack of information on the 
precise profiles of temperature vs. height in the channels and in the AR allow 
us only to state that our results are consistent with the hypothesis of an 
enhancement of low FIP elements in slow wind, however. We found a general 
tendency of levelling-off of the $I({\rm Si~\textsc{xii}})/[n_e^2]$ profiles across the UVCS 
slit at the end of the AR life, in agreement with the weakened evidence for 
outflow channels during the last two limb transits.

Evidence has been obtained in the low corona by \cite{Doschek2008a}, for 
example, for lower electron temperatures inside the AR outflows 
($1.2-1.4 \times 10^6~{\rm K}$), with respect to the surrounding closed 
corona. The Si~{\sc{xii}} curve of the ionization balance (see, e.g., 
\citealt{Arnaud1985a}) peaks at about $2 \times 10^6~{\rm K}$, well above the 
values estimated by \cite{Doschek2008a}. Under the hypothesis that $T_{\rm e}$ 
inside the AR outflows is maintained at values lower than the surrounding 
corona at the altitudes of UVCS observations, we would expect a decrease in 
the $I({\rm Si~\textsc{xii}})/[n_e^2]$ profiles across the AR, which is the opposite of what 
we obtained. For this reason, we could speculate that we observe an effect of 
silicon abundance enhancement, the electron temperature possibly giving a 
negligible contribution. Moreover, because silicon is a low-FIP element, this 
would be consistent with the expectations for slow wind flows. The limited 
information we can obtain from our data prevents us from drawing any 
conclusion about a possible enhancement of the abundance of low-FIP elements 
in the outflow channels, leaving a clearer interpretation to future studies.

In conclusion, we have shown that ARs contribute throughout their lifetime to 
the slow wind. Whether their contribution is time dependent on short 
timescales cannot be determined from this analysis. On the other hand, our 
work provides strong support to the claim by \cite{Demoulin2013a} that the 
upflow driver acts over a prolonged time, here shown to be on the order of a 
few months.

\begin{acknowledgements}
We would like to thank the anonymous referee for the helpful comments and 
suggestions. The authors acknowledge the use of data of the Kitt Peak 
telescope of the National Solar Observatory; NSO/Kitt Peak data used here are 
produced cooperatively by NSF/NOAO, NASA/GSFC, and NOAA/SEL. We also thank the 
Wilcox Solar Observatory for the use of the synoptic maps of photospheric 
magnetic field distribution. G.P. acknowledges support from ASI I/023/09/0. 
\end{acknowledgements}

\bibliographystyle{aa} 


\end{document}